\documentclass[journal]{IEEEtran}
\ifCLASSINFOpdf
\else
   \usepackage[dvips]{graphicx}
\fi
\usepackage{url}

\usepackage{graphicx}
\usepackage{cite}
\usepackage{cuted}
\usepackage{amsmath,amssymb,amsfonts}
\usepackage{amsmath}
\usepackage{algorithmic}
\usepackage{graphicx}
\usepackage{textcomp}
\usepackage{xcolor}
\usepackage{dsfont}
\usepackage{tkz-euclide}
\usepackage{lipsum}
\usepackage{mathtools}
\usepackage{verbatim} 
\usepackage{cuted}
\usepackage{algorithm}
\usepackage{algorithmic}
\usepackage{comment}
\usepackage{dsfont}
\usepackage[font=small]{caption}
\usepackage{subfig}
\usepackage{stfloats}

\DeclareMathOperator*{\argmin}{argmin}
\DeclareMathOperator*{\argmax}{argmax}
\newcommand{\mb}{\mathbf}
\newcommand{\norm}[1]{\left\lVert#1\right\rVert}
\newcommand*{\tran}{^{\mkern-1.5mu\mathsf{T}}}
\newcolumntype{N}{@{}m{0pt}@{}}

\DeclareMathOperator{\K}{\mathbf{K}}

\DeclareMathOperator{\Khh}{\mathbf{K}_{x,x}}
\DeclareMathOperator{\iKhh}{\mathbf{K}_{x,x}^{-1}}
\DeclareMathOperator{\Kha}{\mathbf{K}_{x,{x}^*}}
\DeclareMathOperator{\Kah}{\mathbf{K}_{{x}^*,x}}
\DeclareMathOperator{\Kaa}{\mathbf{K}_{{x}^*,{x}^*}}

\DeclareMathOperator{\x}{\mathbf{x}}
\DeclareMathOperator{\h}{\mathbf{h}}
\DeclareMathOperator{\g}{\mathbf{g}}
\DeclareMathOperator{\y}{\mathbf{y}}
\DeclareMathOperator{\Y}{\mathbf{Y}}
\DeclareMathOperator{\f}{\mathbf{f}}

\DeclareMathOperator{\GP}{\mathcal{GP}}
\DeclareMathOperator{\N}{N}
\DeclareMathOperator{\E}{\mathbb{E}}

\makeatletter
\newcommand{\mathleft}{\@fleqntrue\@mathmargin0pt}
\makeatother

\begin{document}

\pagestyle{empty} 

\title{Gaussian Process-based Spatial Reconstruction of Electromagnetic fields}


\author{
{Angesom Ataklity Tesfay\textsuperscript{1,2} and Laurent Clavier\textsuperscript{1,2}} \\
\small
\textsuperscript{1}Univ. Lille, CNRS, UMR 8520 - IEMN, F-59000, Lille, France (e-mail: firstname.name@univ-lille.fr)\\
\textsuperscript{2}IMT Lille Douai, France (e-mail: firstname.name@imt-nord-europe.fr)}

\maketitle
\thispagestyle{empty}

\begin{abstract}
These days we live in a world with a permanent electromagnetic field. This raises many questions about our health and the deployment of new equipment. The problem is that these fields remain difficult to visualize easily, which only some experts can understand. To tackle this problem, we propose to spatially estimate the level of the field based on a few observations at all positions of the considered space. This work presents an algorithm for spatial reconstruction of electromagnetic fields using the Gaussian Process. We consider a spatial, physical phenomenon observed by a sensor network.
A Gaussian Process regression model with selected mean and covariance function is implemented to develop a 9 sensors-based estimation algorithm.
A Bayesian inference approach is used to perform the model selection of the covariance function and to learn the hyperparameters from our data set. We present the prediction performance of the proposed model and compare it with the case where the mean is zero. The results show that the proposed Gaussian Process-based prediction model reconstructs the EM fields in all positions only using 9 sensors. 
\end{abstract}

\begin{IEEEkeywords} 
Gaussian Process, Electromagnetic fields, Bayesian inference, sensor network 
\end{IEEEkeywords}

\IEEEpeerreviewmaketitle

\section{Introduction}
Nowadays most of the technological applications apply the concept of electromagnetic field. Electromagnetism has made an important transformation in the area of engineering applications, it also has a great impact on different technologies. It has wide practical implementation that ranges from household appliances to big scientific research applications. In residential applications like kitchen appliances, in near and far field communication systems, in industrial systems like in motors, generators and actuator devices, etc. Therefore there is always an electromagnetic field in our surrounding.
To study the intensity of the electromagnetic fields in the surrounding, it requires the ability of modeling spatial functions using a good observation mechanism. One can use sensor network  to monitor the radiation levels of the electromagnetic fields in the surrounding to provide a way of observation by means of sampling the data and modeling spatial functions. Recently sensor networks were used to monitor different kind of spatial physical phenomenon including some desired aspects like sound intensity, pressure, temperature and pollution concentrations and then send the observations to a fusion center, so it can be used to reconstruct the signal. 
Even though electromagnetic field has many application it can damage our health somehow when we are exposed to high radiation everyday and the electromagnetic equipment can also mutually interfere with one another which have an impact on the performance of the equipment, this is due to the fact that we can not tell the level of electromagnetic fields we are getting in our surrounding. This poses numerous questions about the installation of new equipment as well as protect our own health from the radiation of these fields. The main difficulty is that electromagnetic field is still a concept impossible to visualize normally, which can only be comprehended by some experts. The technician who deploys the equipment, the one who checks the levels of electromagnetic fields to which we are exposed and the individual who fears for his health, all deal with an invisible adversary.
Today, technologies should address this problem. This is the objective of this thesis but several locks remain to be lifted. If we can measure the intensity of the electromagnetic fields in some points using predefined measurement mechanism then we can use these measurements to reconstruct the fields in order to visualize it. A company called Luxondes has developed a slab allowing to display a level of field on the plane of the slab and by the help of Institute for Research on Software and Hardware Components for Advanced Information and Communication (IRCICA), the slab has been replaced by a sensor and augmented reality goggles which allow the measurement points of the field to be displayed in planes. By memorizing these points, we can envisage a spatial reconstruction of the field and work on its representation. Using sensor measurements localized in some specific points, we will reconstruct the field at any point in predefined space. The approach used for the spatial reconstruction is based on the recent works of the Gaussian processes. The confrontation of these tools with the measures will then perhaps require an adaptation of the method to increase the precision. It should also be noted that the proposed statistical methods make it possible to manage sensors of different qualities. It is also possible to orient the mobile sensors to improve the reliability of the reconstruction. Consideration of the imprecision of the location of the sensors and of the temporal evolution also constitute challenges for research.

In this work, we proposed a Gaussian Processes based prediction algorithm that uses a set of sensor measurement points to estimate the field at any point location. The deployed sensors will measure the electromagnetic field intensity in the region of interest and send these measurement values to the algorithm through the network in real-time, then the algorithm will use the received observation points to reconstruct the field. Finally, the algorithm will send its output to the next part, which is the visualization part; all the signal processing and spatial reconstruction is performed in real-time. But, our work is to utilize these sensor measurements to estimate the field intensity using Gaussian process. Now, in order to present the study carried on in this thesis, it is very important to review the researches that has been done so far related to our work.
Gaussian process regression is one of the  common regression technique for data adaptation. The interest in Gaussian processes for regression and classification purposes has been increasing fast in the recent years \cite{MarginalizedNeural,BayesianMultitask,OptimallyPruned}. A Gaussian process regression uses the sampled data to determine the predictive mean function,  predictive covariance function, and  hyperparameters which defines the regression process \cite{Gaussforreg, IntroductionToGaussian, GaussianProcesses}. In \cite{NearOptimal}, Gaussian process regression was applied for controlling the ecological condition of a river for individual sensors. The location of the sensor was determined by maximizing a mutual information gain. In \cite{Adecentralized},  a Bayesian Monte Carlo approach was used to reduce the computation complexity of the Gaussian process regression, and an information entropy was used to assign each sensors. In \cite{Learning}, to construct a gas distribution system and with main goal to reduce the computational complexity a mixture of Gaussian process was implemented. In \cite{OnTraj}, to characterize spatiotemporal functions a Kalman filter was constructed on the top of Gaussian process model. By optimizing a mutual information gain the path planning problem was tackled. 

\begin{figure}
\centering
\includegraphics[scale=0.55]{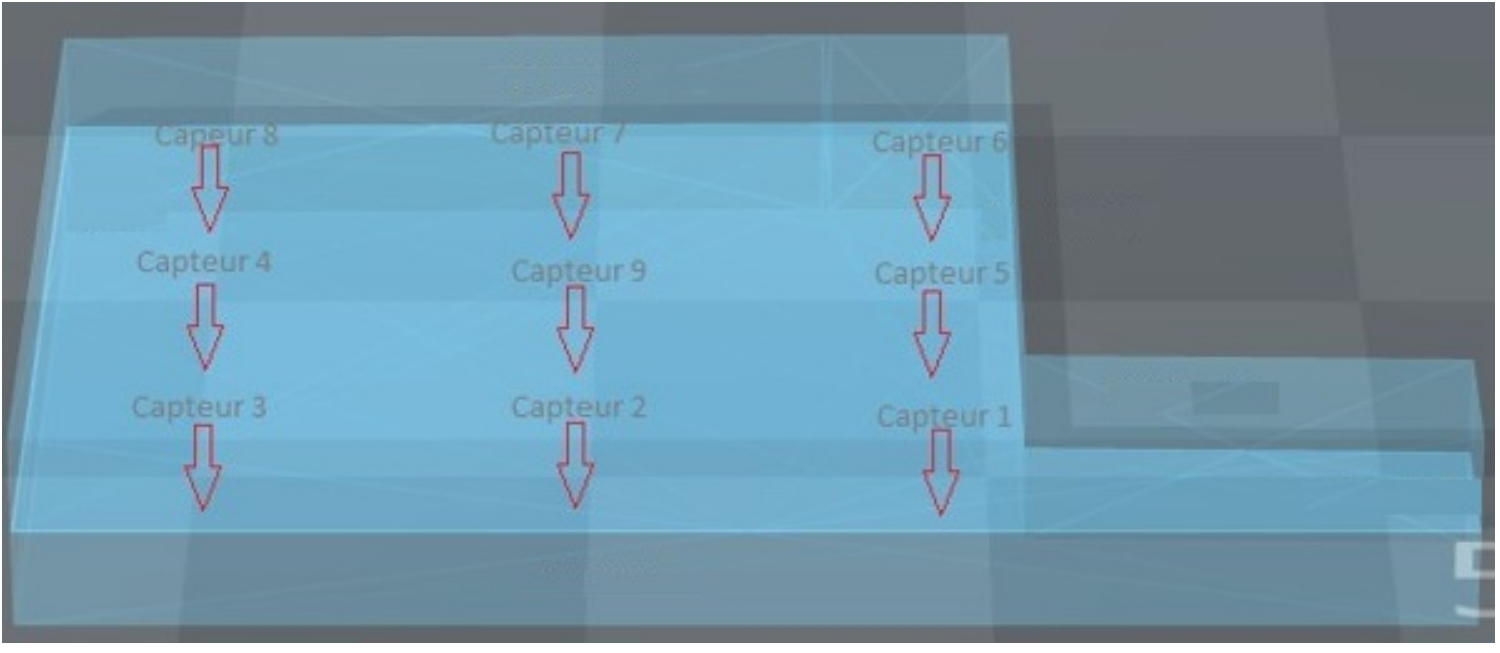}
\caption{The Location of the Sensors in the room}
\label{fig:sen}
\end{figure}

Gaussian process regression provide better prediction of the mean function of the GP but its key advantage comparing to other regression methods is its capability to predict the covariance function of the GP which tells about the uncertainty of the prediction and this is very important information in the special field reconstruction. In \cite{SpatialGauss} GP was used to model spatial functions for mobile wireless sensor networks, they developed a distributed Gaussian process regression technique using a sparse Gaussian process regression approach and a compactly supported covariance function. In \cite{EstimationSpatially} Gaussian Process based algorithms was developed for spatial field reconstruction, exceedance level estimation and classification in heterogeneous Wireless Sensor Networks.  To fuse the different observations from the heterogeneous sensors, an algorithm was developed that is based on a multivariate series expansion approach resulting in a Saddle-point (Laplace) type approximation.
In the research area, so far we have seen that the implemented approaches use large number of sensors points for observation and uses more than 9 points to reconstruct the signal. We proposed  Gaussian process regression method to reconstruct electromagnetic field which is observed using 9 sensors by collecting  data from all sensor nodes to a centralized unit  and then make the computation of the predictive mean function and the covariance function in real time, the spatial reconstruction of the field will be at all positions of the space in the room. The Gaussian processes regression is proposed because it provides prediction uncertainty term, and it is useful to learn the noise and the parameters of our model from the given training data. We will receive the measured values of the electromagnetic field from the sensors deployed in the measurement room in IRCICA at fixed positions and use this data to make the prediction using our algorithm at all position in the room rather than at the location of the sensors. Using our training data we will set the hyperparameters of the model using marginal likelihood approach in which, the maximum a posterior estimation is implemented to compute the optimal hyperparameters that maximize the marginal likelihood. Then we will use the optimal hyperparameter to make spatial reconstruction of the field not only at the location of the sensor but also at all position of the considered room.
We consider a random process represented by spatial function $\h$ and it is modeled as a Gaussian Process $h(x)$ which is defined by a mean function $m(x)$ and a covariance function $Cov(\x, \x')$ and we also consider a noisy observation where the noise is assumed to be Gaussian noise with zero mean and variance $\sigma_\eta^2$. We deploy the sensors in a specific region of interest, a single room of $25 $ square meter area and $2.5$ meter height, which is dedicated for the observation. We used $9$ sensors that are deployed in the measurement room provided by IRCICA, and located in fixed $(x,y,z)$ coordinates, since there is no variation in the z axis at all sensors location we only consider the $(x,y)$ coordinates in the implementation of the algorithm. Top view representation of the measurement room and the location of the deployed sensors is shown in the figure \ref{fig:sen}.
We have collect different set of training data to train our algorithm. The measurement for training sets has been carried out in IRCICA using a modeling tool called SIMUEM3D. The tool was developed by professor Philippe Mariage, it is useful tool for measuring and modeling the radio coverage of confined structure (building, tunnels, dense urban environments) that takes into consideration realistically the influence and shape of the walls on the propagation of electromagnetic waves. By fixing the necessary parameters, this tool will provide us the exact measurement of the electromagnetic field intensity at the specified locations. We begin by modeling the room on SIMUEM3D using the real dimensions of the room, also by setting the permittivity and conductivity properties of the walls. This helps us to set the real parameters of the room in order to collect the observations from all over the room space that considers all the necessary conditions of the room. To take the simulated measurements we start with One reflection and then we increase the number of reflections step by step up-to three reflection, this data will train our algorithm in order to handle different intensity of the field that can occur due to the variation of number of reflections received by the sensors.\\
\begin{figure}
\centering\includegraphics[scale=0.37]{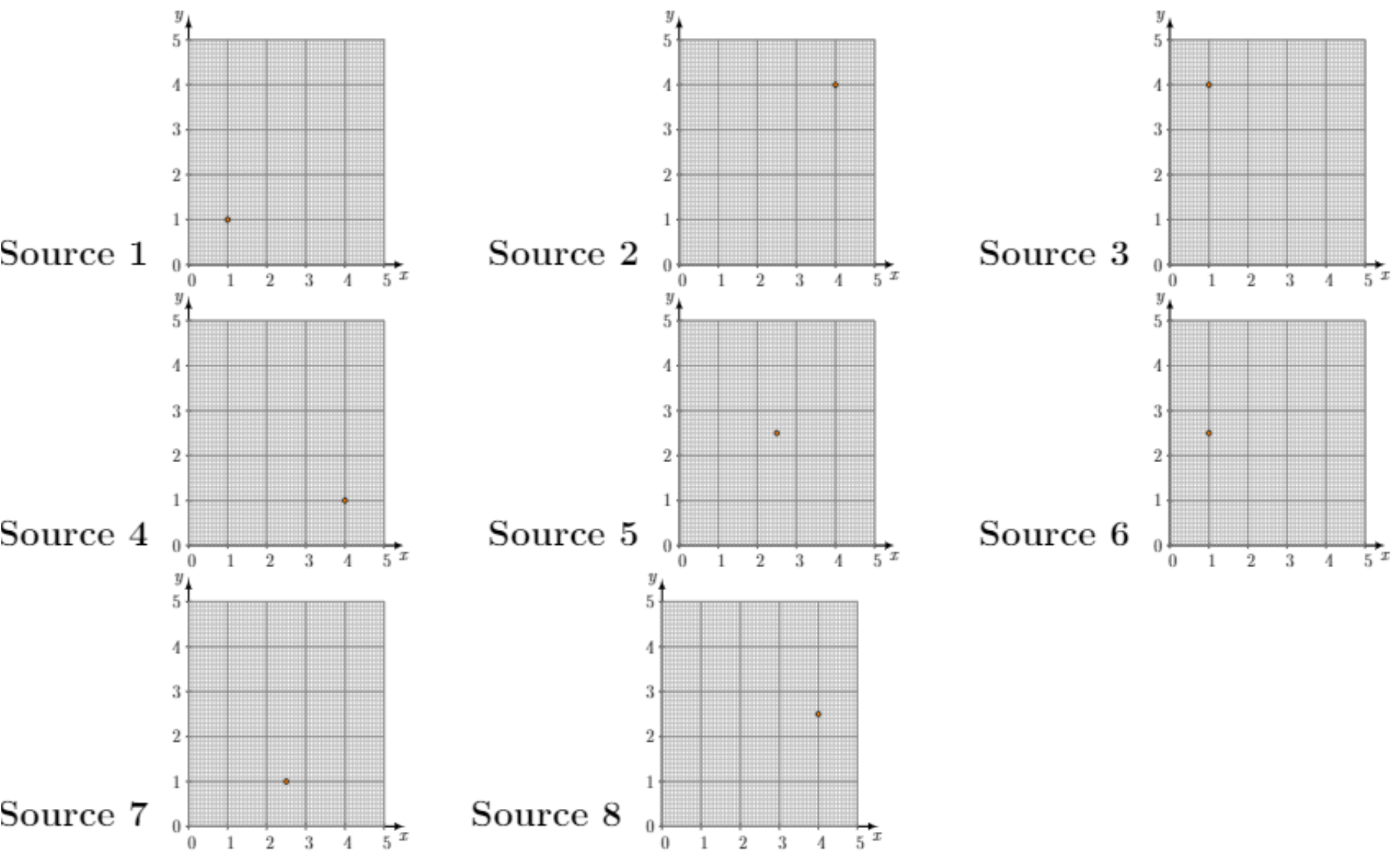}
\caption{EM Source Positions of Source 1 to Source 8}
\label{fig2}
\end{figure}
\begin{figure}
\centering\includegraphics[scale=0.37]{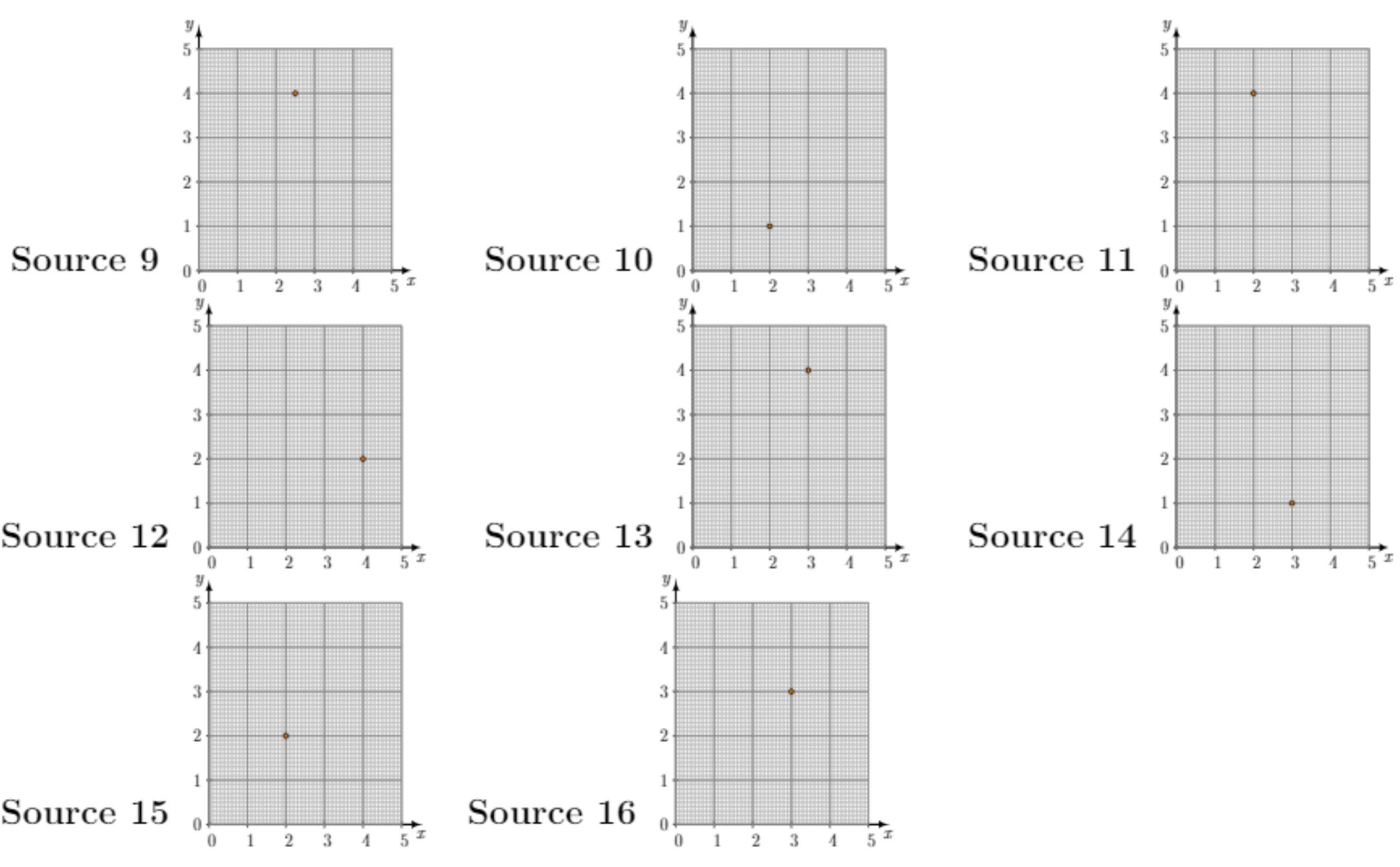}
\caption{EM Source Positions of Source 9 to Source 16}
\label{fig3}
\end{figure}
As it is shown in figures \ref{fig2} and \ref{fig3}  we use 16 different positions of electromagnetic sources in order to cover all the position in the room as much as possible this can help us to test the algorithm capability using different position of the source. Here we set the orientation of the sources along z-axis in all positions. The parameters we use to make the observations are; the relative permittivity $\epsilon_r=5$, frequency $f=900 MHZ$, the conductivity $\sigma=0.001 s/m$ and degree of attenuation, $\alpha=2$. The positions of the observations are set by grid location, which is a grid coordinates starting from $0.1$ up-to $4.9$ and by stepping $0.1$ in both x and y directions. This will be total 49 by 49 gird and we have total of 2401 observation points in our training set. We avoid to locate the observation points and the source on the walls of the room since the tool computes all the reflection angle and this is not possible if we put the observation points and the source exact on the walls of the room.
The algorithm have two parts, the first one is the training phase and the second one is the prediction phase. In the training phase, we use the dataset that includes the measurement of the electromagnetic field intensity in some positions of the considered room to train our algorithm, to select suitable covariance function for our model and to set the hyperparameters by  computing the optimal hyperparameters based on the dataset we have. The different locations of the source are used accordingly to optimize the computation of the algorithm and to observe the response of the algorithm at all possible positions of the EM source. Using the collected dataset we proceed with the model selection for covariance function based on the training data which would be suitable to the problem we have on hand. Selecting a good model means that we will have a best interpretation of the measured data properties when we make prediction using the selected  model.  We implemented the Bayesian theorem for model selection which is also known as the Marginal likelihood model selection and this is because it includes the calculating the probability of our model given the data set. Bayesian theorem have a good and consistent inference method. We implement Gaussian process regression method with consideration of Gaussian noise so the integral over the parameters will be analytically tractable.
Whereas, in the prediction phase, we use the optimized hyperparameters and only 9 sensor measurement values to make the prediction not only at the location of the sensors but also at any location in the room. We also make prediction using 30 points and 300 points to compare the difference between these predictions. This helps on the measurement of the performance of our algorithm. We compute the normalized mean square error of each predictions to measure the accuracy of the prediction. We also develop a network interface between the sensors and the input of our algorithm and on the other side between the output of our algorithm and the visualization part using the open sound control network (OSC) protocol.

The rest of the paper is organized as follows, section II revises the most important theoretical concepts that are helpful for this work, namely those regarding to linear regression model, Bayesian linear regression and Gaussian process model. Section III describes the implementation of Gaussian processes regression studied in this work, focusing on the model specification used for our prediction algorithm. Section IV presents the simulation setup and experimental results. Finally, section V presents the conclusion of the work and the possible future implementations and developments.
\section{Linear Parametric Model}
\label{sec:lin_Para}
For a given training data-set $\{(x^n,y^n), n=1,...,N \}$ a linear parameter regression model is defined by:
    \begin{equation}\label{eq:2.1}
    {y(\mathbf{x})= \mathbf{w^T} \mathbf{\phi(x)}} 
    \end{equation}
where $\mathbf{\phi(x)}$ is a vector valued function of the input vector $\mathbf{x}$ and $\mathbf{w}$ is the weighting parameter. The main concept of linear model is each input variable in the function is multiplied by an unknown weighting parameter, and all of the individual terms are summed to produce the estimated function value. Here the model is linear in the weighting parameter  $\mathbf{w}$ not necessarily linear in $\mathbf{x}$. To determine the parameters  we consider a loss function which is a measure of the discrepancy between the observed outputs and the linear regression. In the literature the widely applied loss function for linear parameter regression model is the sum squared differences method or least squares method. In the least squares method the unknown parameters are estimated by minimizing the sum of the squared deviations between the observed outputs and the estimations under the linear model. Therefore the loss function can be formulated  as:
\begin{equation}\label{eq:2.2}
L(\mathbf{w})=\sum_{n=1} ^{N} \Big(y^n - \mathbf{w^T} \mathbf{\phi(x^n)}\Big)^2
\end{equation}
 And the objective is to find the best possible weighting parameter $\mathbf{w}$ that minimize the loss function:
\begin{equation}\label{eq:2.3}
\min_{{\mathbf{w}}}L(\mathbf{w}) 
\end{equation}
Based on the dimension of the features space of the training set, the parameter vector $\mathbf{w}$ that minimizes  $L(\mathbf{w})$ can be determine by using two widely used learning algorithms. Algorithm based on \textit{Gradient Descent} which is an iterative algorithm, it repeats the equation \eqref{eq:2.4} until convergence and this algorithm is preferable when the dimension of feature space is large.
\begin{equation}\label{eq:2.4}
\mathbf{w}^{new}=\mathbf{w}-\eta \frac{\partial L(\mathbf{w})}{\partial \mathbf{w}} 
\end{equation}
where $\mathbf{\eta}$ is the learning rate. Algorithm based on {Normal Equation} that compute the parameter vector $\mathbf{w}$ analytically by differentiating $L(\mathbf{w})$ with respect to $\mathbf{w}$, and set to zero. 
\begin{equation}\label{eq:2.5}
\mathbf{w}=\Bigg(\sum_{n=1} ^N \mathbf{\phi(x^n)} \mathbf{\phi(x^n)^T}  \Bigg)^{-1} \sum_{n=1} ^N y^n \mathbf{\phi(x^n)}
\end{equation}
   
In linear regression the number of parameters related to the amount of training data should be considered carefully and there should be a mechanism to control it. Consider the problem of predicting $y(\mathbf{x})$ from $\mathbf{x}$. If the prediction function $y(\mathbf{x})$ uses too few parameters then the fit is not very good and it is known as underfitting. In the other hand if the prediction function $y(\mathbf{x})$ uses too many parameters relative to the amount of training data then $y(\mathbf{x})$ fits the available data but does not  {generalize} in a good way to predict new data and it is called  {overfitting}. To overcome the issue of {overfitting} it is possible to use two options; the first option is to reduce the number of features either manually or using a model selection algorithm. The second one is to use  {regularizing term} to the loss function $L(\mathbf{w})$ by keeping all the features, but reduce the magnitude of parameters.
To control the overfitting we have to introduce a complexity penalty term in the loss function. The regularized loss function is given by:
\begin{equation}\label{eq:2.52}
    L(\mathbf{w})=\sum_{n=1} ^{N} \Big(y^n - \mathbf{w^T} \mathbf{\phi(x^n)}\Big)^2 - \mathbf{w^T}\lambda \mathbf{w}
    \end{equation}
where the first term is the model fit term that is used to fit the data in the model and the second term is the complexity penalty term that is used to control the overfitting problem. And where $\lambda$ is regularization term and we can set it using cross validation in which we train the parameters with different settings of $\lambda$ and the one with the lowest error on the validation set is selected to learn $\lambda$. By differentiating the regularized loss function with respect to the parameter, we can compute the optimal $\mathbf{w}$ as:
\begin{equation}\label{eq:2.53}
   \mathbf{w}=\Bigg(\sum_{n=1} ^N \mathbf{\phi(x^n)} \mathbf{\phi(x^n)^T} - \lambda \Bigg)^{-1} \sum_{n=1} ^N y^n \mathbf{\phi(x^n)}
   \end{equation}
In linear parametric model regression, it is possible to use a linear combination of non-linear basis function  to fit the data. In case the data has non-trivial behavior over some region in $x$, then the  the region of $x$ space should be covered densely with bump type basis functions; this means \textbf{k} basis function is needed for one dimension space. Therefore; increasing the dimensionality of the feature space exponentially increases the basis functions need and this cause computational complexity. In [1] this problem is addressed by using the {covariance function}, or a {positive kernel} that produces a positive definite matrix for any inputs $\mathbf{x,x'}$. 
The classification problem is similar the regression problem, except that  now the values to be predicted take on only a small number of discrete values. For binary classification problem given training data $D=\{(\mathbf{x}^n,c^n), n=1,...,N\} $ in which $c  \in \{0,1\}$. Starting from the concept of Linear regression model and reconstruct the classification model that can satisfy the prediction function $0 \leq y(\mathbf{x^n}) \leq 1$. The probability that an input $\mathbf{x}$ belongs to class 1 can be written as;
\begin{equation}\label{clas}
y(\mathbf{x^n})= p(c=1|\mathbf{x})=f(\mathbf{w^T x})
\end{equation}
where $0 \leq f(x) \leq 1$. In \cite{BayesianInference}, $f(x)$ is termed a {mean function}. The most used $f(x)$ is the \textit{Logistic Function}. It can be written as $\sigma(x) $ and is given by;
$$f(x)=\frac{e^{x}}{1+ e^{x}}= \frac{1}{1+ e^{-x}}$$
In [1] logistic regression model is described as;
$$p(c=1|\mathbf{x})= \sigma(b+\mathbf{w^T x})$$ where $b$ is bias and $\mathbf{w}$ is weighting vector. The line that separates the area where $c = 0$ and where $c = 1$ is the decision boundary . It is the set of $\mathbf{x}$ for which $p(c=1|\mathbf{x})=p(c=0|\mathbf{x})=0.5$ and is given by the hyperplane, $b+\mathbf{w^T x}=0$. If  $b+\mathbf{w^T x}>0$ the inputs $\mathbf{x}$ are classified as $1's$ otherwise they are classified as $0's$. The bias $b$ controls the shift of the decision boundary and $\mathbf{w}$ controls the orientation of the decision boundary.

Given a data set $D$, and assuming that  each data point are i.i.d, the logistic regression loss function can be defined using the principles of maximum likelihood estimation:
\begin{align}
p(D|b, \mathbf{w})&=\prod_{n=1} ^N p(c^n|\mathbf{x}^n,b,\mathbf{w})p(\mathbf{x}^n) \nonumber \\
&=\prod_{n=1} ^N p(c=1|\mathbf{x}^n,b,\mathbf{w})^{c^n}\nonumber \\
&\times\big(1-p(c=1|\mathbf{x^n},b,\mathbf{w}) \big)^{1-c^n} \mathbf{w})p(\mathbf{x}^n)
\end{align}
\normalsize
Considering the log likelihood of the output class variables conditioned on the training inputs, the loss function for logistic regression is given as:
\begin{align}
L(\mathbf{w},b)&= \sum _{n=1}^N c^n \log{p(c=1|\mathbf{x}^n,b,\mathbf{w})} \nonumber \\
&+ (1-c^n)(1-p(c=1|\mathbf{x}^n,b,\mathbf{w})) 
\end{align}
To learn the parameters weight vector $\mathbf{w}$ and the bias $b$ in-order to have good classification model, numerically it is difficult to obtain closed form solution to the maximization of log likelihood $L(\mathbf{w}, b)$. Therefore {gradient ascent} is used to update weight vector $\mathbf{w}$ and the bias $b$:
\begin{align}
\mathbf{w}^{new} &= \mathbf{w} + \eta \frac{\partial L(\mathbf{w},b)}{\partial \mathbf{w}}, \\  \nonumber \\ 
b^{new}&= b+ \eta \frac{\partial L(\mathbf{w},b)}{\partial b} 
\end{align}

In case the provided data is linearly separable the updated weights will continue to increase and the classification model will be overfitted. Just like the regularized linear regression model {regularization} technique can be used to avoid this problem. This can be prevented by adding a penalty term to the objective function \cite{bayesianreasoning}:
\begin{align}
L'(\mathbf{w},b)= L(\mathbf{w},b)- \alpha \mathbf{w^T}\mathbf{w}, \hspace{1cm} \alpha > 0
\end{align}
By considering a non-linear mapping of the inputs $\mathbf{x}$ to $\mathbf{\phi(x)}$, it is possible to extend the logistic regression method to more complex non-linear decision boundaries:
\begin{align}
p(c=1|\mathbf{x})=\sigma(\mathbf{w^T}\phi(\mathbf{x})+b) 
\end{align}

The Kernel Trick can be used to address the problem of computational complexity that can be arise due to the increase of dimensionality. Support Vector Machines are also a form of kernel linear classifier  that uses more advanced and have good generalization performance \cite{IntroductionToSvm}. in the next section we will briefly discuss about the Bayesian model. 

\subsection{Bayesian Model}
\label{sec:bayesian}
Considering regression with additive Gaussian noise $\eta$, $\eta \sim \mathcal{N} (\eta|0,\sigma^2)$, and give observed data $D=\{(\mathbf{x}^i,y^i), i=1,...,N \}$, where the input-output pair $(\mathbf{x}^i,y^i)$ are independent and identically distributed, given all parameters and assume that the output is generated with out noise from a model $h(\mathbf{x;w})$, where the parameters $\mathbf{w}$ are unknown. The final output $y$ is generated by addition of noise $\eta$ to $h(\mathbf{x;w})$:
\begin{equation}\label{eq:2.6}
y =h(\mathbf{x;w})+ \eta, \hspace{1cm} h(\mathbf{x;w})=\mathbf{w^T}\phi(\mathbf{x})
\end{equation} 
Since the considered noise is Gaussian distributed, the probability of generating $y$ from input $\mathbf{x}$ is formulated as:
\small
\begin{align}\label{eq:2.7}
p(y|\mathbf{w},\mathbf{x},\sigma^2)&=
\mathcal{N}(y|\mathbf{w^T}\phi(\mathbf{x}),\sigma^2)\nonumber \\
&=|2\pi \sigma^2|^{-1/2} exp\bigg(-\frac{1}{2\sigma^2}[y - \mathbf{w^T} \phi(\mathbf{x})]^2 \bigg)
\end{align}
\normalsize
The likelihood function which is the probability of generating the data given all parameters is:
\begin{equation}\label{eq:2.8}
 p(D|\mathbf{w},\sigma^2)=\prod_{i=1}^N  p(y^i|\mathbf{w},\mathbf{x}^i,\sigma^2)p(\mathbf{x}^i)
\end{equation}
 
In the above equation note that the likelihood function is only a density over the output $y^i$, and it is because the model is conditional and the input point $\mathbf{x}$ is available at prediction. The estimate for $\mathbf{w}$ can be found with maximum likelihood estimation,$\mathbf{w'}=\argmax \big(p(D|\mathbf{w},\sigma^2)\big)$ but if $\mathbf{w}$ is high-dimensional as compared to the number of data points \textbf{N} the maximum likelihood estimation will have problems of overfitting and also the distributions of the parameters is not specified.The Bayesian linear model uses the {prior} weight distributions, $p(\mathbf{w})$, to measure knowledge of each parameters. The {posterior} weight distribution of $\mathbf{w}$, after having seen the data, is given by:
\begin{equation}\label{eq:2.9}
p(\mathbf{w}|D, \sigma^2)=\frac{p(D,\mathbf{w}|\sigma^2)}{p(D|\sigma^2)}=\frac{p(D|\mathbf{w},\sigma^2)p(\mathbf{w})}{p(D|\sigma^2)} 
\end{equation}
where $p(D|\sigma^2)=\int p(D,\mathbf{w}|\sigma^2)d\mathbf{w}$, the above expression is known as Bayes formula and Bayesian apply this formula to insure coherent and consistent inference.

The posterior quantifies the uncertainty in $\mathbf{w}$ given the data and define a region that $\mathbf{w}$ lies in with high confidence. Using the likelihood function and the Gaussian noise assumption the log-posterior weight distribution is:
\begin{align}\label{eq:2.10}
\log p(\mathbf{w}|D, \sigma^2)&=-\frac{1}{2\sigma^2}\sum_{i=1}^N [y^i - \mathbf{w^T}\phi(\mathbf{x}^i)]^2 \nonumber \\
&+ \log p(\mathbf{w})- \frac{N}{2}\log\sigma^2 + constant
\end{align}

In equation \eqref{eq:2.10} the regularizing term that used in linear model is embedded in the prior weight distribution $p(\mathbf{w})$. The posterior weight distribution depends on the prior distribution, in \cite{bayesianreasoning} a natural weight prior distribution is used as:
\begin{equation}\label{eq:2.11}
p(\mathbf{w}|\alpha)=\mathcal{N} (\mathbf{w}|0,\alpha^{-1}\mathbf{I})=\bigg(\frac{\alpha}{2\pi} \bigg)^{\frac{d}{2}}e^{-\frac{\alpha}{2}\mathbf{w^T}\mathbf{w}}
\end{equation}
where $\alpha =\frac{1}{\sigma_w^2}$ is the inverse variance and \textbf{d} is the dimension of the weight $\mathbf{w}$. The posterior distribution can be written as:
\begin{align}\label{eq:2.101}
\log p(\mathbf{w}|D, \sigma^2)&=-\frac{1}{2\sigma^2}\sum_{i=1}^N [y^i - \mathbf{w^T}\phi(\mathbf{x}^i)]^2 \nonumber \\
&{-\frac{\alpha}{2}\mathbf{w^T}\mathbf{w}}+ constant
\end{align}
The posterior weight distribution is described hyperparameter set $\mathbf{\Lambda}$, where $\mathbf{\Lambda=\{\alpha, \sigma}^2\}$ which are parameters that determine the prior distribution and the Gaussian noise respectively. Parameters that determine the functions can also be included in the hyperparameter set. With a Gaussian prior and Gaussian noise assumption, the posterior is also a Gaussian distribution with covariance \textbf{C} and mean \textbf{m}:
\begin{equation}\label{eq:2.12}
p(\mathbf{w}|D,\mathbf{\Lambda})=\mathcal{N}(\mathbf{w}|\mathbf{m},\mathbf{C})
\end{equation}
The covariance and mean of the posterior is described as:
\begin{equation}\label{eq:2.13}
\mathbf{C}=\bigg( \alpha\mathbf{I}+\frac{1}{\sigma^2} \sum_{i=1}^N \phi(\mathbf{x}^i)\phi^T(\mathbf{x}^i)\bigg)^{-1}  
\end{equation}

\begin{equation}\label{eq:2.14}
\mathbf{m}=\frac{1}{\sigma^2}\mathbf{C}\sum_{i=1}^N y^i \phi(\mathbf{x}^i)
\end{equation}

By using the posterior distribution it is now possible to describe the {mean prediction} for an input \textbf{x} as:
\begin{equation}\label{eq:2.15}
\bar{h}(\mathbf{x})\equiv \int h(\mathbf{x;w})p(\mathbf{w}|D,\mathbf{\Lambda})d\mathbf{w}=\mathbf{m^T}\phi(\mathbf{x})
\end{equation}

In the same procedure the variance of the estimated clean function $h(\mathbf{x;w})$ is given by:
\begin{equation}\label{eq:2.16}
Var\big(h(\mathbf{x})\big)=\bigg \langle[\mathbf{w^T\phi(x)}]^2 \bigg \rangle- \bar{h}(\mathbf{x})^2= \mathbf{\phi^T(x)C\phi(x)}
\end{equation}

Note that $Var\big(h(\mathbf{x})\big)$ depends only on the input variables. Since the additive noise is $\eta \sim \mathcal{N} (\eta|0,\sigma^2)$; the {variance prediction} or the variance of the noisy output for an input x is given by:
\begin{equation}\label{eq:2.17}
Var\big(y(\mathbf{x})\big)=Var\big(h(\mathbf{x})\big) + \sigma^2
\end{equation}
Therefore; the primary parameters $\mathbf{w}$ and hyperparameters $\mathbf{\Lambda=\{\alpha, \sigma}^2\}$ are useful in organizing a problem into a chain of tasks. Consider generating the data: to begin with, set hyperparameters, then estimate the parameters given hyperparameters, then estimate data given all parameters. Such progressive models are a Bayesian response to the obvious problem of specifying models and priors which are ambiguous using hard model assumptions and concise distribution families with clear semantics like the Gaussian. The hyperparameter posterior distribution is defined as $p(\mathbf{\Lambda}|D)\propto p(D|\mathbf{\Lambda})p(\mathbf{\Lambda})$ and the optimal $\mathbf{\Lambda'}$ is estimated by {maximum a posteriori (MAP)} approximation. By {marginalizing} over the parameters the predictive distribution can be described as:
\begin{equation}\label{eq:2.18}
p(D|\mathbf{\Lambda})= \int p(D|\mathbf{\Lambda , w})p(\mathbf{w}|D, \mathbf{\Lambda}) d\mathbf{w}
\end{equation}
The quantity $p(D|\mathbf{\Lambda})$ is called \textit{marginal likelihood}. The empirical Bayes estimate of hyperparameters $\mathbf{\Lambda}$ can be also obtained as a maximum of marginal likelihood $p(D|\mathbf{\Lambda})$ and this method of setting hyperparameters  is known as {'ML-II'} \cite{statisticaldecision}.
An advantageous computational technique to a find the maximum of marginal likelihood is to interpret the $\mathbf{w}$ as inactive variables and apply the {Expectation Maximization (EM)} algorithm and it is also possible to use the {Gradient} approach for the hyperparameter optimization \cite{bayesianreasoning}. The best hyperparameters at explaining the training data are those founded by ML-II. Fundamentally, this is not quite the same as those that are best for prediction. Therefore, in practice it is reasonable to set hyperparameters also by validation techniques that is to set hyperparameters by minimal prediction error on a validation set and also to set hyperparameters   by their likelihood on a validation set \cite{statisticaldecision}.
\subsection{Gaussian Process Regression Model}
\label{sec:GP} 
So far in Bayesian learning applications the knowledge about prior weight distribution is the information that the model depends on to make the prediction. However the knowledge about the true latent method behind the data generation process is limited and this is a big problem. For the prior, there are an many infinite set of possible distributions to consider, but the problem is how to deal with the computationally complexity of the infinite dimensionality. Most applied solution is to use Gaussian process prior to inference of continuous values. The main advantage of the Gaussian processes is that the mathematical smoothness properties of the functions are well understood, giving confidence in the modeling the prediction. We will give the definition of Gaussian process and the we will implement Gaussian process for regression.

\textbf{Definition 2.3.1} \textit{(Gaussian Process)  A Gaussian Process is a collection of random variables with the property that the joint distribution of any of its subset is joint Gaussian distribution.}

This means that there should be a consistency, it also can be termed as a marginalization property. We assume that  $ \mb{X} \subset \Re^n$ and $h(\x):\mb{X} \mapsto \Re $ where $\x \in \mb{X}$ The Gaussian process is described by a mean function $m(\x;\theta)$ which is parameterized by $\theta$ and covariance function $Cov(\x,\x';\beta)$ which is the spatial dependence between any two points and parameterized by $\beta$, there is detailed explanation in \cite{GaussianProcesses} \cite{RandomFields}. Therefore a Gaussian process $h(\x)$ is defined as:

\begin{equation}\label{eq_GP}
h(.) \sim \GP(m(.;\theta),Cov(.,.; \beta))
\end{equation}
\small
\begin{equation}\label{eq_GP_mean}
m(\x;\theta):= \E[f(\x)]: \mb{X} \mapsto \Re
\end{equation}
\begin{equation}\label{eq_GP_cov}
Cov(\x,\x';\beta):= \E[(h(\x)-m(\x;\theta))(h(\x')-m(\x';\theta))]: \mb{X} \times \mb{X} \mapsto \Re
\end{equation}
\normalsize
Equation \eqref{eq_GP} tells that the function $\h$ is distributed as a Gaussian process with mean function $m(\x)$ and covariance function $Cov(\x, \x'; \beta)$. We need to fix the mean and the covariance function in order to define an individual Gaussian Process. 
\subsubsection{Prior Information}
Most of the time, there is no prior knowledge about the mean function, $m(x)$ of a given Gaussian Process  because GP is a linear combination of random variables with normal Distribution and this assumed to be zero \cite{patternReco} this means that our initial best predict for the function output at any input is zero. But if there is enough information about the process to be modeled the mean function should be different from zero, we can apply a non-zero constant mean function for all inputs.
First, let's consider the case where there is noise free observations, our training data is given by 
$\mathcal{D} = (\mb{X}, \Y)$ where $\y=[Y_1,...,Y_N]^T$, $\mb{X}=\{\mb{x}_n=[x_{n,1},...,x_{n,d}]^T; n=1,...,N\}$ and $h(\x)$ is a latent function with value $ h_n=h(\mb{x}_n), \h=[h_1,...,h_N]^T$. To model the observations, we put prior information on the function. The Gaussian process prior over our function $h(\x)$ shows that any group of function values $\h$, with an input variable $\mb{X}$, have a multivariate Gaussian distribution:
\begin{equation}\label{eq_GP_prior}
p(\h|\mb{X},\beta) = \N(\h|\mb{0}, \Khh),
\end{equation}
where $\Khh$ is the covariance matrix. The covariance function generates the covariance matrix,  $[\Khh]_{i,j} = k(\mb{x}_i, \mb{x}_j|\theta)$, and this describes the correlation between different points in the process. We can choose any type of covariance function under one condition and that is the covariance matrices produced should be symmetric and positive semi-definite ($\mb{v}^{\text{T}}\Khh\mb{v}\geq 0, \forall \mb{v}\in \Re^n$). A common example of covariance function is the squared exponential which is used widely:

\small
\begin{align}
    \label{k_se}
Cov(h(\x_i), h(\x_j))&=k_{\mathrm{SE}}(\x_i,\x_j|\beta)\nonumber \\
&=\alpha\exp \Big(-{\sum_{l=1}^d (x_{i,l}-x_{j,l})^2/{2\gamma_l^2}}\Big).
\end{align}
\normalsize
where $\beta =[\alpha, \gamma_1,...,\gamma_d]$. Here, $\alpha$ and $\gamma_l$ are known as the \textit{hyperparameters}, where, $\gamma_l$ is the length-scale,that controls the degree of  change in correlation as the distance increases in the direction $l$, and  $\alpha$ is the scaling parameter. Note that most of the time, the performance of Gaussian Process can be considerably influenced by the selection of parameters. Other common covariance functions are discussed in detail by \cite{Geostatistics}. Later, on this chapter we will see some of the common covariance functions and the way to select the best one. In the next subsection, we will see the Gaussian process inference mechanism with and with out noisy observations.
\subsubsection{Inference} 
\label{sec_Inf}
If we want to predict the values $\h ^ *$ at new input $\mb{X}^{*}$. The joint prior distribution for latent variables $\h$ and ${\h}^*$ is given by:
\begin{equation} \label{eg_joint}
\left[ \begin{matrix} \h \\ \h^* \end{matrix} \right] | \mb{X},{\mb{X}^*},\beta
\sim \N\left(\mb{0}, \left[ \begin{matrix} \Khh & \Kha \\ \Kah & \Kaa
    \end{matrix} \right] \right),
\end{equation}
Where $\Khh = k(\mb{X},\mb{X}|\beta)$, $\Kha =
k(\mb{X},{\mb{X}^*}|\beta)$ , $\Kah = k(\mb{X}^*,{\mb{X}}|\beta)  $and $\Kaa =
k({\mb{X}^*},{\mb{X}}^*|\beta)$. 
Here, the covariance
function $k(\cdot,\cdot)$ denotes also vector and matrix valued
functions $k(\x,\mb{X}):\Re^d \times \Re^{d \times n}\rightarrow
\Re^{1\times n}$, and $k(\mb{X},\mb{X}):\Re^{d\times n}\times
\Re^{d\times n}\rightarrow \Re^{n\times n}$. 
From the definition of Gaussian process the marginal distribution of ${\h}^*$ is $p({\h}^*|{\mb{X}}^*,\beta) = \N({\h}^*|\mb{0}, \Kaa)$ which is the same with equation \eqref{eq_GP_prior}. The posterior distribution over functions can be obtained by limiting the joint prior distribution to contain those functions that consider the observed data points and we can perform that by applying conditional probability concept on the joint Gaussian prior distribution. Therefore, the conditional distribution of ${\h}^*$ given $\h$ is expressed as: 
\begin{equation}\label{eq_conditional_ftilde_given_f}\footnotesize
{\h}^*|\h, \mb{X},
{\mb{X}}^*, \beta \sim \N(\Kah\iKhh\h, \Kaa - \Kah\iKhh\Kha),
\end{equation}
\normalsize
Where the mean and covariance of the conditional distribution are
functions of input vector ${\mb{X}}^*$ and $\mb{X}$. We can generalizes the above distribution to Gaussian process with mean function $m(\mb{X}^*|\beta) = k({\mb{X}}^*,\mb{X}|\beta)\iKhh\h$ and covariance function $k({\mb{X}}^*, {{\mb{X}}^*}'|\beta) = k({\mb{X}}^*, {{\mb{X}}^*}'|\beta) - k({\mb{X}}^*,\mb{X}|\beta)\iKhh
k(\mb{X},{{\mb{X}}^*}'|\beta)$. The conditional distribution of the latent function $h({\x}^*)$ is define by the above mean and covariance functions.

Now let's consider noisy observations, our training data is  $\mathcal{D} = \{ \mb{X}, \mb{Y} \}$ where $\mb{Y}= h(\x) + \eta$  and  $\Y=[Y_1,...,Y_N]^T$, are set of noisy observations and $\mb{X}=\{\mb{x}_n=[x_{n,1},...,x_{n,d}]^T; n=1,...,N\}$ are inputs. We assume that each component in the set of observations $\Y$ are a sample from a Gaussian distribution, that is the value of the observations affected by Gaussian noise $\eta \sim \N(0,\sigma_N ^2)$, since there are no assumptions taken about the distributions of the observations, it is better to take it as Gaussian Processes. The prior on the noisy observations will be changed and based on \cite{GaussianProcesses} the covariannce function becomes;
\begin{equation}\label{cov_noise}
 Cov(Y_i,Y_j)= k(\mb{x}_i, \mb{x}_j)+ \sigma_N ^2 \delta_{ij}=\Khh + \sigma_N ^2 \mb{I}
\end{equation}
Where $\delta_{ij}$ is Kronecker delta function which is equal to 1 if the variables are equal, and 0 otherwise. And the \textit{hyperparameters} of the covariance function will include the variance of the noise $\sigma_N^2$ which will be $\beta =[\gamma_l,...,\gamma_d, \alpha,\sigma_N^2]$. The Bayes' theorem is the main foundation of the Bayesian inference that involves the conditional probability of the latent function and parameters after observing our data, which is known as the posterior distribution. All information about the latent function and parameters are included in our posterior distribution, since our model transfer all the information from our training data. It is difficult to compute the posterior distribution but we can use approximation techniques to approximate it. Therefore the conditional posterior of the latent variables given the parameters is given by:
\begin{equation}\label{eq_conditional_posterior_of_f}
p(\h|\mathcal{D}, \beta) =
\frac{p(\Y|\h,\mb{X},\sigma_N^2)p(\h|,\mb{X}, \beta)}{\int p(\Y|\h,\mb{X},\sigma_N^2)p(\h|\mb{X},\beta) d\h},
\end{equation}
We can marginalize  the conditional posterior over the parameters to get the marginal posterior distribution, and can be written as;
\begin{equation}
p(\h|\mathcal{D}) = \int p(\h|\mathcal{D}, \beta) p(\beta|\mathcal{D}) d\beta 
\end{equation}
We can follow the same steps to find the posterior predictive distribution by computing the conditional posterior predictive distribution, $p({h}^*|\mathcal{D}, \beta, {\x}^*)$, and marginalizing it over the parameters. The marginal predictive posterior distribution is given by: 
\begin{equation}\label{eq_margin_ posterior_predictive_distribution}
p(\h^*|\mathcal{D}) = \int p({h}^*|\mathcal{D}, \beta, {\x}^*) p(\beta|\mathcal{D}) d\beta 
\end{equation}
Given the above information, to predict $\h^*$ given the new input $\x^*$ we need to define joint distribution. From equation \eqref{eg_joint} that is the joint prior distribution, we can define the joint distribution of the observation and the function value at the new test inputs using the prior information as:
\begin{equation} \label{eg_joint_y}
\left[ \begin{matrix} Y \\ {\h}^* \end{matrix} \right] 
\sim \N\left(\mb{0}, \left[ \begin{matrix} \Khh + \sigma_N^2 \mb{I}& \Kha \\ \Kah & \Kaa
    \end{matrix} \right] \right)
\end{equation}

The main task in the regression process is to estimate the mean value and the variance of ${\h}^*$.
Our first interest is to determine conditional distribution of ${\h}^*$ given the observed data or the posterior predictive distribution and we can represent it as:
\begin{equation}\label{eq_condi_y}
{\h}^*|D,\beta,{\x}^* \sim \GP \Big(m^p({\x}^*) ,Cov^p({\x}^*, {{\x}^*}')\Big)
\end{equation}
where the mean value of the prediction and the covariance of the  prediction are given respectively as:

\footnotesize
\begin{align}\label{eq_mean_eq_cov}
m^p({\x}^*) &=\E[{\h}^*|D,\beta,{\x}^*]= k({\x}^*,\mb{X})(\Khh + \sigma_N^2 \mb{I})^{-1}\Y \\
k^p({\x}^*, {{\x}^*}') &= k({\x}^*, {{\x}^*}') - k({\x}^*,\mb{X})(\Khh + \sigma_N^2 \mb{I})^{-1}
k(\mb{X},{{\x}^*}')
\end{align}
\normalsize
Therefore, our best estimate for $\h^*$ is given by the mean of the prediction $m^p({\x}^*)$ and the uncertainty of our estimation is described by the covariance of the  prediction. We can conclude that the mean is described as linear combination of the observations $Y$ and covariance depends only in the inputs.

There is no closed form that can represent the predictive posterior distribution $p({h}^*|\mathcal{D},\beta,\x^*)$, but we can approximate the predictive mean and covariance functions, by approximating the posterior mean and covariance functions. The denominator of equation \eqref{eq_conditional_posterior_of_f} is known as the marginal likelihood $ p(Y|\mb{X},\beta)$ which is integration of the likelihood function $p(Y|\h,\mb{X},\sigma_N^2)\sim \N(\h, \sigma_N^2\mb{I})$ over the prior function values $p(\h|\mb{X},\beta)$ gives the marginal likelihood and this is known as marginalization. The marginal likelihood is given as:
\begin{equation}\label{eq_ml}
p(Y|\mb{X},\beta)=\int p(Y|\h,\mb{X},\sigma_N^2)p(\h|\mb{X},\beta)d\h
\end{equation}
In the Gaussian process model the, we assume that the prior is Gaussian, $\h|\mb{X},\beta \sim \N(\mb{0}, \Khh)$, for simplicity we can use logarithm arithmetic:

\small
\begin{equation}
\log p(\h|\mb{X},\beta)= -\frac{1}{2}{\h}^T{\Khh}^{-1}\h  -  \frac{1}{2}\log|\Khh| - \frac{n}{2}log2\pi
\end{equation}
\normalsize
Therefore, to compute the integration over $\h$ we can apply the method used in \cite{GaussianProcesses} and the result of the log marginal likelihood is given by:
\begin{align}
\label{eq_log_marginal_likelihood}
\log p(\Y|\mb{X},\beta) &=  -\frac{1}{2} \Y^{\text{T}} (\Khh + \sigma_N^2\mb{I})^{-1} \Y, \nonumber \\
&- \frac{1}{2}\log |\Khh + \sigma_N^2\mb{I}| -\frac{n}{2}\log(2\pi)
\end{align} 
Since both the prior and the likelihood are Gaussian, we can have exact inference, but if the prior and the likelihood are not Gaussian the integration in equation \eqref{eq_ml} could be intractable. When the likelihood is not Gaussian we have to approximate the marginal likelihood. We can use Laplace and EP approximation to approximate the marginal likelihood, there is more detail in \cite{laplace_Appx} \cite{EP_Apprx}.

In equation \eqref{eq_log_marginal_likelihood}, there are different terms that have different impact on the marginal likelihood. The first term, $-\frac{1}{2} \Y^{\text{T}} (\Khh + \sigma_N^2\mb{I})^{-1} \Y$, involves the observations $\Y$, and it is known as the data-fit term, the second term $- \frac{1}{2}\log |\Khh + \sigma_N^2\mb{I}|$, only depends on the covariance matrix, which is similar to the regularization term used in linear regression, and it is used as penalty term for the complexity \cite{GaussianProcesses}.
It is simple to make predictions of new test points, if we know the covariance function, since it only depends on the ability to solve matrix calculation. But in real life practical applications it is difficult to apprehend what kind of covariance function to use for the problem in hand. Therefore, the performance of our regression model will depend on goodness of the selected  parameters of the  chosen covariance function \cite{predictionWithGauss}. As we describe it before that $\beta$ is the set of hyperparameters used in the given covariance function. Assume the given covariance function is squared exponential function with noisy observations which is the combination of equation \eqref{k_se} and  equation \eqref{cov_noise} and it is given by:

\small
\begin{align}\label{k_se_new}
Cov(h(\x_i), h(\x_j))&=k_{\mathrm{SE}}(\x_i,\x_j|\beta),\\
&=\alpha\exp \Big(-{\sum_{l=1}^d \frac{(x_{i,l}-x_{j,l})^2}{2\gamma_l^2}}\Big) +\sigma_N ^2 \delta_{ij}.\nonumber 
\end{align}
\normalsize
where $\gamma_l$ is the length-scale, $\alpha$ the signal variance and $ \sigma_N$ the noise variance. And $\beta =[\gamma_1,...,\gamma_d, \alpha,\sigma_N  ]$ are the set of \textit{hyperparameters}. The main task after we choose  the covariance function is to assign an appropriate values to the hyperparameters that can suit in the covariance function. The distance in input space before the function value can change considerably is characterized by length-scale. If the length-scale is short the predictive variance can change fast away from the data points, therefore the predictions will have low correlation between each other. The signal variance can be termed as the vertical length-scale. Since the noise that affects the process is random, we don't expect to have correlation between different inputs, therefore we can only find the noise variance on the diagonals of the covariance matrix. The empirical search approach for assigning the suitable values for each parameters is not sufficient enough. Additionally, someone can choose complex covariance function based on the problem in hand, therefore, in this case the number of hyperparameters needed can be large. Therefore, to set the hyperparameters we have to find the set of parameters that maximizes the marginal likelihood. Recall equation \eqref{eq_log_marginal_likelihood} and since the likelihood function or our noise model is Gaussian distributed, $p(Y|\h,\mb{X},\beta)\sim \N(\h, \sigma_N^2\mb{I})$ and after marginalizing over the latent variables the marginal likelihood is given as $p(\Y|\mb{X},\beta)\sim \N(\Y|\mb{0}, \Khh + \sigma_N^2\mb{I})$. Therefore, the log marginal likelihood is written as:

\small
\begin{equation}\label{eq_log_marginal_likelihood_new}
\log p(\Y|\mb{X},\beta) =  -\frac{1}{2} \Y^{\text{T}} \K^{-1} \Y - \frac{1}{2}\log |\K| -\frac{n}{2}\log(2\pi) 
\end{equation}
\normalsize
Where $\K=\Khh + \sigma_N^2\mb{I}$, to set our hyperparameters first we have to put a prior information $p(\beta)$ for our hyperparameters, then by using the  maximum a posterior (MAP) estimation we can compute ${\tilde{\beta}}$  that maximising the log hyper-posterior \cite{ModellingLocal} or we can compute ${\tilde{\beta}}$ that minimizes the negative log hyper-posterior. The hyper-posterior is defined as $p(\beta| \Y,\mb{X} ) \propto p(\Y|\mb{X},\beta)p(\beta)$ and the MAP estimation is given by:
\begin{align}
\tilde{\beta} &= \argmax_{\beta} \log p(\beta| \Y,\mb{X} ), \nonumber \\
&= \argmin_{\beta}
\left[ -\log p(\Y|\mb{X},\beta) - \log p(\beta) \right].
\end{align}

Therefore, by setting the hyperparameters using the MAP estimation we specify our Gaussian Process model. In other words, to train the hyperparameters we need to compute the partial derivatives of the marginal likelihood with respect to the hyperparameters, $\beta$, in order find the hyperparameters that maximize the marginal likelihood or minimize the negative marginal likelihood . Using equation \eqref{eq_log_marginal_likelihood_new} and by applying the matrix derivatives described in \cite{GaussianProcesses}, we have:
\begin{equation}\label{eq_log_marginal_likelihood_derivative}
\frac{\partial}{\partial{\beta_i}}  \log p(\Y|\mb{X},\beta) =  \frac{1}{2} \Y^{\text{T}} \K^{-1}\frac{\partial{K}}{\partial{\beta_i}}\K^{-1} \Y - \frac{1}{2} tr\Big( \K^{-1}\frac{\partial{K}}{\partial{\beta_i}}\Big)
\end{equation}
After this approximation, the values of the parameter are given to the posterior distribution, and the marginal predictive posterior distribution in equation \eqref{eq_margin_ posterior_predictive_distribution} is approximated as $p(\h^*|\mathcal{D}) \approx p(\h^*|\mathcal{D},\tilde{\beta})$. Both the log marginal likelihood, and its approximations, are differentiable with respect to the hyperparameters \cite{GaussianProcesses}. This implies that, the log posterior is also differentiable, that permit us to use gradient based optimization.
Gradient descent is widely used technique to find the set of optimal hyperparameters that maximize the log marginal likelihood. 
A detail discussion about the techniques used to estimate values for hyperparameters can be
found in \cite{GaussianProcesses} \cite{predictionWithGauss}. In the next subsection, we will discuss about the types of covariance function widely implemented in Gaussian process applications.
\subsubsection{Covariance Functions for GP}
\label{covfuncion}
The covariance function is the main building block of the Gaussian process, since it contains the assumptions about the function which we want to model. Therefore, in this section we will discuss some popular covariance functions that are used in Gaussian process and we will define the covariance function used in our algorithm. A covariance function , also called a kernel, kernel function, or covariance kernel, is a positive-definite function of two inputs $\x$ and $\x'$. When we select a specific covariance function, we are selecting the property of the solution function, that means we select if our solution functions should be smooth, linear, periodic and polynomial. If we want to totally change the algorithm's functionality, then we just have to change the covariance function we used before. Additionally, inferring distributions over the hyperparameters of the covariance function, we can also understand the property of the data we have, properties including rate of variation, smoothness and periodicity. To define the prior covariance between any two function values in Gaussian process models Kernel is used. The covariance matrix is generated from the covariance function and it should be symmetric positive semi-definite matrix, $\mb{v}^{\text{T}}\K\mb{v}\geq 0, \forall \mb{v}\in \Re^n$, see equation \eqref{cov_matrix}.
  \begin{figure*}
  $$Cov[f(\x),f(\x')]= k(\x, \x')$$
  \begin{align}\label{cov_matrix}
 k(\x_N, \x_N) & =
 \begin{bmatrix}
 Cov(\x_1, \x_1) &  Cov(\x_1, \x_2) & \dots  & Cov(\x_1, \x_n) \\
 Cov(\x_2, \x_1) &  Cov(\x_2, \x_2) & \dots  & Cov(\x_2, \x_n) \\
 \vdots       &  \vdots          & \ddots & \vdots \\
 Cov(\x_n, \x_1) & Cov(\x_n, \x_2) & \dots  & Cov(\x_n, \x_n)
 \end{bmatrix} \in \Re^{N\times N} 
  \end{align}
  \vspace{0.2cm}
  $$k(\x^*,\x_N)=\E[f(\x^*) f(\x_N)] \in \Re^{d}\times \Re^{d \times N} \to \Re^{1 \times N} $$
  \end{figure*}
  
 There are different type of covariance function used in Gaussian process, and we have to select the covariance function carefully because some covariance function may not be good choice based upon the process being observed or the training data we have on hand. The covariance function describes which functions are possibly under the Gaussian process and which holds the generalization properties of the model. We can construct a new covariance function from other covariance function by using different techniques like addition of kernels, product of kernels, exponent of kernels, etc. For better understanding of different kernel function families and the kinds of structures representable by Gaussian process we will make kernel implementation on a some of prior functions by using some frequently used kernel functions. For every implementation of covariance functions there are different set of assumptions considered based upon the function or process we want to model. For instance, applying square exponential covariance function infers that the function we are modeling has infinitely many derivatives. There are many different alternatives of local kernel like the squared exponential covariance function, but each will encode somewhat different assumptions about the smoothness of the function to be modeled. Covariance Function parameters: Each covariance function contains many parameters that describes the properties of the covariance function like, variability and the shape of the covariance function. These parameters are also known as hyperparameters since they are describing a distribution over function rather than describing directly a function. For example, we can see the parameters of squared exponential covariance function, the  length-scale  $l $, which controls the correlation scale in the input and defines the width of the kernel and the magnitude or scaling parameter $\sigma^2$ , which governs the overall variability of the function.
 Stationary Covariance Function: A covariance function is stationary if it is translation invariant in the input space, that means these covariance function depends on the separation between the two input points $\x- \x'$. This indicates that even if we interchange  all the $\x$ input values by the same scale, the probability of observing the dataset will remain the same, therefore they are invariant to translation in our input space. we can mention an Isotropic kernels as examples of stationary covariance function. If a kernel is only a function of the norm of the distance between the inputs, $\norm {\x-\x'}$, and not in their direction, we call this Isotropic covariance function. 
 In contrast,a non-stationary covariance function is a kernel that vary with translation, which means if our input points are moved, the equivalent Gaussian process model will give us totally different predictions even all the covariance function parameters remains the same, these functions vary with translation. Linear covariance function is one example of non-stationary covariance function. Another way to represent covariance function of a stationary process is using Bochner's theorem \cite{bochner} \cite{interpolation} that uses Fourier transform of positive finite measure.\\
 \textbf{Theorem 2.2.} \textit{(Bochner) A complex-valued function k on $\mathbb{R}^B$ is the covariance
 	function of a weakly stationary mean square continuous complex-valued random
 	process on $\mathbb{R}^B$ if and only if it can be represented as:}
 \begin{equation}
 k(\tau)= \int_{\mathbb{R}^B} e^{2\pi iS.\tau} d\beta(S) 
 \end{equation}
 \textit{where $\beta$ is a positive finite measure.}
 
 If the positive finite measure $\beta$ has a density $S(s)$, then we call $S$ the power spectrum or specteral density of $k$ and of course both $k$ and $s$ are Fourier dual \cite{timeseries}. This means that the property of the stationarity of the covariance function is governed by  the spectral density and this is due to the more interpretability of spectral density than covariance function. If we apply Fourier transform of a stationary covariance function, the obtained spectral density indicates the distribution of support for many frequencies.
 \begin{equation}
 k(\tau)=\int S(s) e^{2\pi iS.\tau} ds
 \end{equation}
 \begin{equation}
 S(s)=\int k(\tau) e^{2\pi iS.\tau} d\tau
 \end{equation}
 
 The stationarity property has an important inductive bias, as we will discuss it in the next sections the most common covariance functions like squared exponential, rational quadratic and Mat\'{e}rn  covariance functions are examples of isotropic stationary kernels. In the next section we will discuss about the squared exponential, Matern, neural network, rational quadratic, periodic and linear covariance functions. We will see the covariance functions with some prior functions on one dimensional case, this makes the understanding much easy. We will plot the covariance function $k(\x_i, \x_j)$ and the function $f(\x)$ sampled from Gaussian process prior.
  \subsubsection*{Squared Exponential Covariance Function}   
\begin{figure}
 		\centering
 		\includegraphics[scale=0.5]{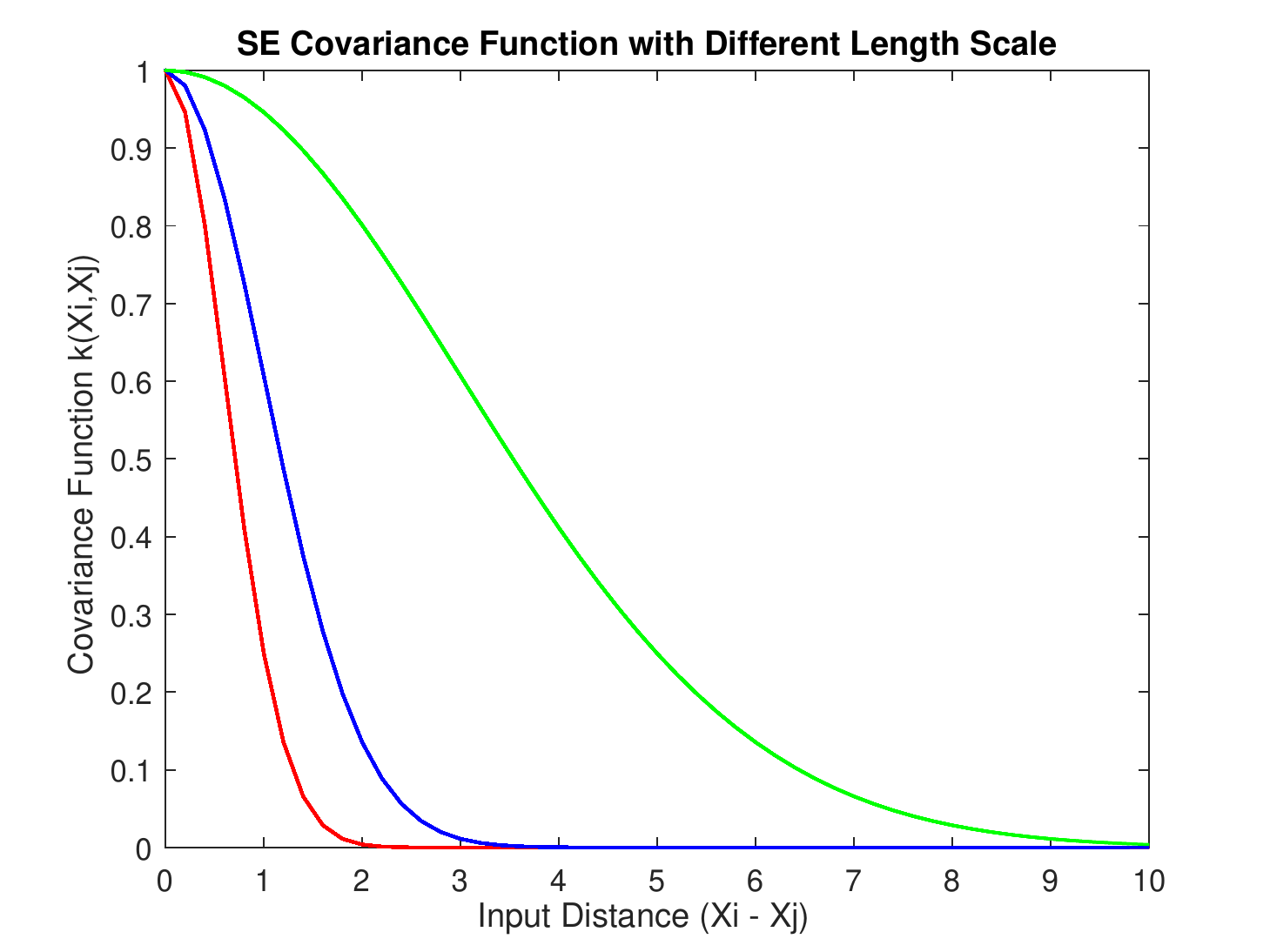}
 		\caption{SE Covariance Function with three different
 			settings of  length scale}
 		\label{fig:se_l}
 	\end{figure}

\begin{figure}
 		\centering
 		\includegraphics[scale=0.46]{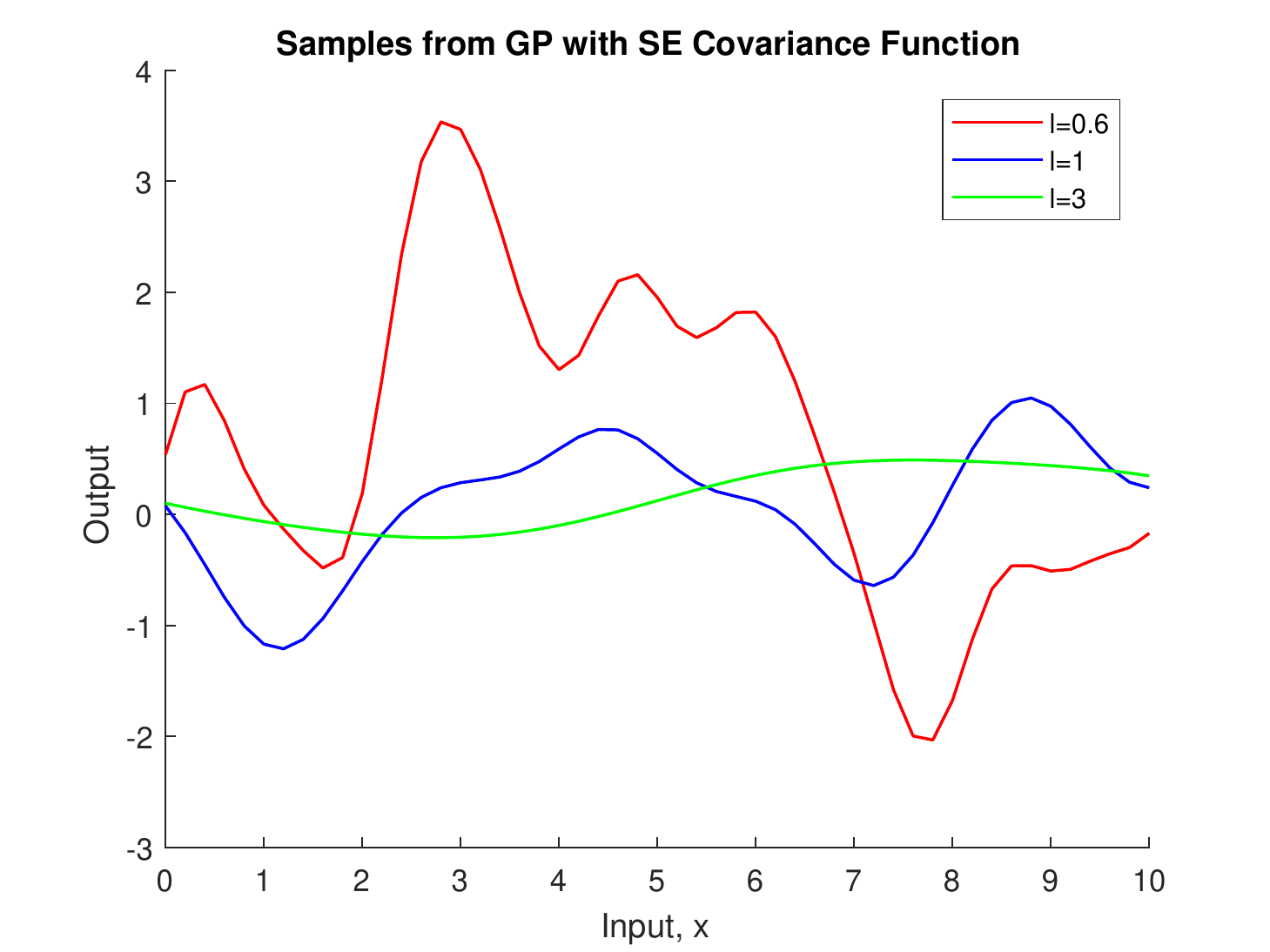}
 		\caption{Random Sample Functions drawn from Gaussian Process with Squared Exponential Covariance Function, for different values of $\ell$ and with $\sigma_{se}=1$.}
 		\label{fig:se_2}
 		
 	\end{figure}

 The Square exponential covariance function is also called radial basis function (RBF) and it is probably the most
 widely-used covariance function within the kernel machines field \cite{GaussianProcesses}. This covariance function is known for its infinitely differentiable property, which implies that the Gaussian process with this specific covariance function includes mean square derivatives of all possible orders, and due to this it is very smooth. \\
 \begin{equation}\label{k_se1}
 k(\x_i, \x_j)=\sigma_{se}^2 \exp \Big(-\frac{|x_{i}-x_{j}|^2}{2{\ell}^2} \Big)
 \end{equation}
 The length scale $\ell$ controls the correlation between the inputs and the variance of the function $\sigma_{se}^2$ governs the overall variability of the function. Figures \ref{fig:se_l} and \ref{fig:se_2}  illustrates the squared exponential covariance function with respect the separation between the inputs with different length scale and the random sample functions drawn from Gaussian process with these covariance function
 	%
  \subsubsection*{Rational Quadratic Covariance Function}
 The squared exponential covariance function considers that the inputs are changing only at one specific length scale, but in reality the inputs could change on different scales. In \cite{apractical}, they showed that the variance on returns on equity indices includes patterns that change over different scales. Therefore, to model the unpredictability of the equity index returns,  It is be reasonable to use a sum of squared exponential covariance function that have different length scales learned from the data set. Therefore we can define the rational quadratic covariance function as a combination of infinite sum of squared exponential covariance function that have different length scales. The rational quadratic covariance function can be expressed as:
 \begin{equation}
 k(\x_i,\x_j) = \left(1+
 \frac{(x_{i}-x_{j})^2}{2 \alpha\ell^2} \right)^{-\alpha}
 %
 \end{equation}
 	\begin{figure}
 		\centering
 		\includegraphics[scale=0.5]{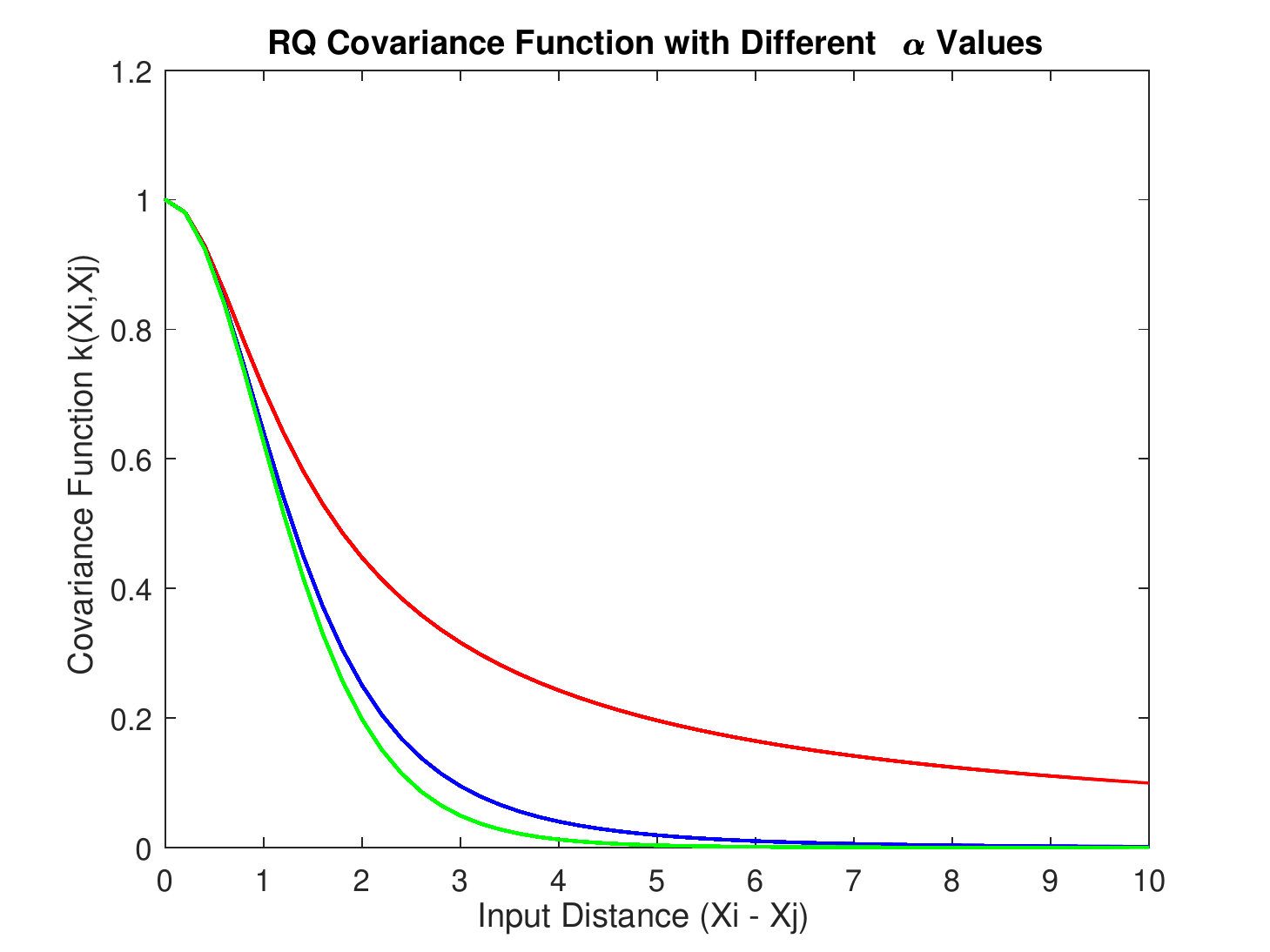}
 		\caption{Rational Quadratic Covariance Function with three different settings of $\alpha$, each with the length-scale $\ell=1$,}
 		\label{fig:rq_l_1}
 	\end{figure}%
 	\begin{figure}
 		\centering
 		\includegraphics[scale=0.5]{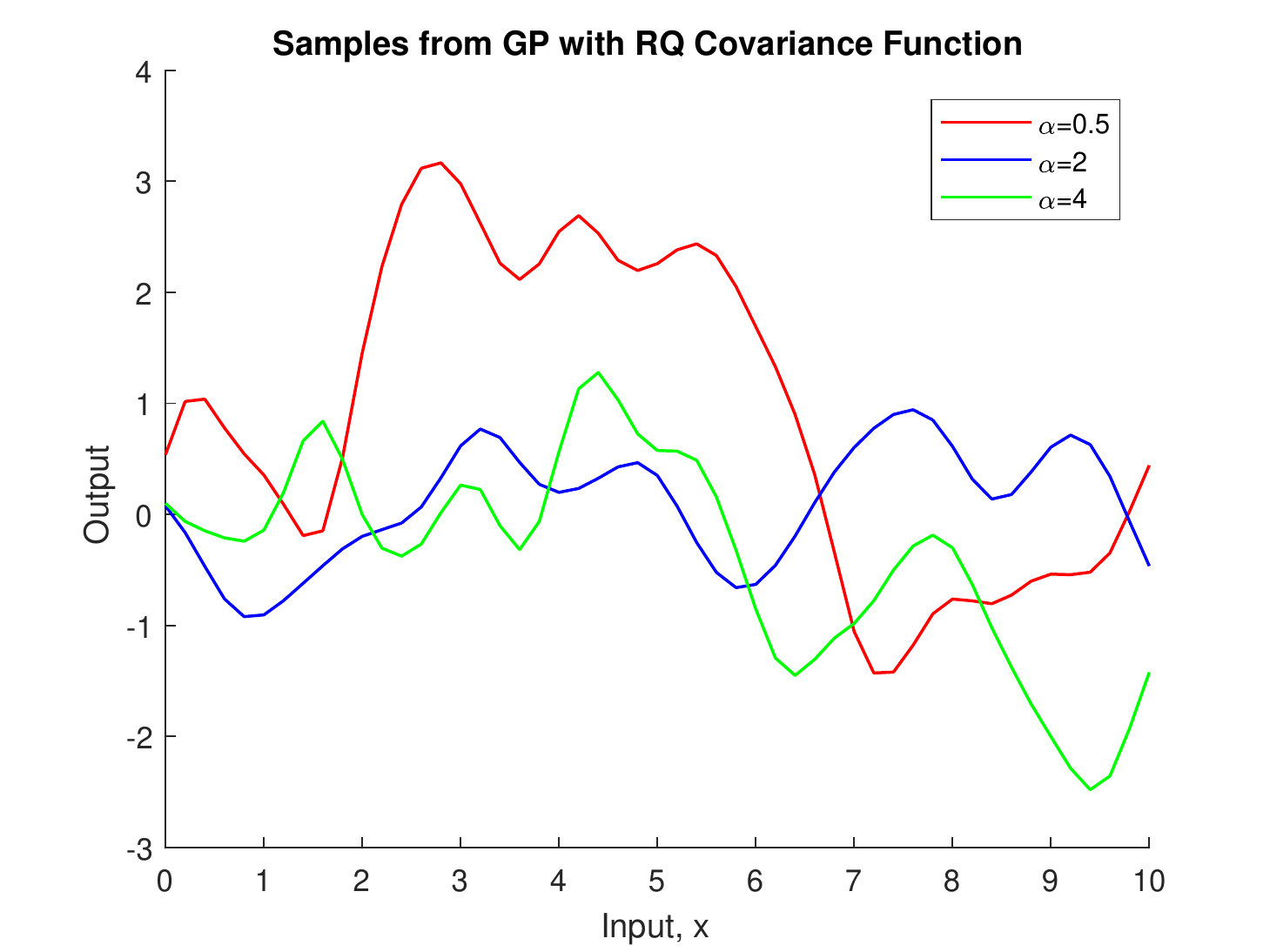}
 		\caption{Random Sample Functions drawn from Gaussian Process with each of these RQ Covariance Function. }
 		\label{fig:rq_l_2}
 	\end{figure}%

The rational quadratic covariance function is proposed to model multi scale inputs. Figures \ref{fig:rq_l_1} and \ref{fig:rq_l_2} illustrates the rational quadratic covariance function with respect the distance between the inputs and functions drawn from Gaussian process with rational quadratic covariance function  with different values of $\alpha$ and when $\alpha$ approaches to infinity the rational quadratic covariance function becomes square exponential covariance function.
 
 \subsubsection*{Neural Network Covariance Function}
 The neural network covariance function is possibly the most outstanding contribution for Gaussian process researches in the research community. The Bayesian models do not suffer with overfitting problem, thus Neal in \cite{bayesianlearn} proposed that we should use more expressive models which have better capability of describing the complicated real world physical process. The data set we have could also be helpful for improving the modelling performance even if we have more expressive model. Therefore, in \cite{bayesianlearn} Neal followed the limits of large models, and indicated that Bayesian neural network becomes a Gaussian process with neural network covariance function when the number of the units becomes infinity. Williams and Rasmussen \cite{Gaussforreg} was inspired by this observation to explore Gaussian process models. A nonstationary neural network covariance function is given as:
 
 \small
 \begin{equation}\label{cf_neuralnetwork}
 k(\x_i,\x_j)=\frac{2}{\pi}\sin^{-1}\left(\frac{2\mathbf{\hat{x}}_i^{\text{T}}\Sigma_{nk} \mathbf{\hat{x}}_j}{(1+2\mathbf{\hat{x}}_i^{\text{T}}\Sigma_{nk}
 	\mathbf{\hat{x}}_i)(1+2\mathbf{\hat{x}}_j^{\text{T}}\Sigma_{nk}
 	\mathbf{\hat{x}}_j)}\right)
 \end{equation}
 \normalsize
 
 Where $\mathbf{\hat{x}}=(1,x_1,\ldots,x_B)^{\text{T}}$  is the input vector increased with 1 and
 $\Sigma_{nk}=\text{diag}(\sigma_0^2,\sigma_1^2,\ldots,\sigma_B^2)$ is a diagonal weight prior, where $\sigma_0^2$ 
 is a variance for bias parameter governing the function offset from the origin. The variances for the weight parameters are $\sigma_1^2,\ldots,\sigma_B^2$, and with small values for weights, the neural network covariance function gives smooth and rigid functions. If the values for the weight variance are large, the covariance function produces more flexible functions.
 \subsubsection*{Periodic Covariance Function}
\begin{figure}
 		\centering
 		\includegraphics[scale=0.4]{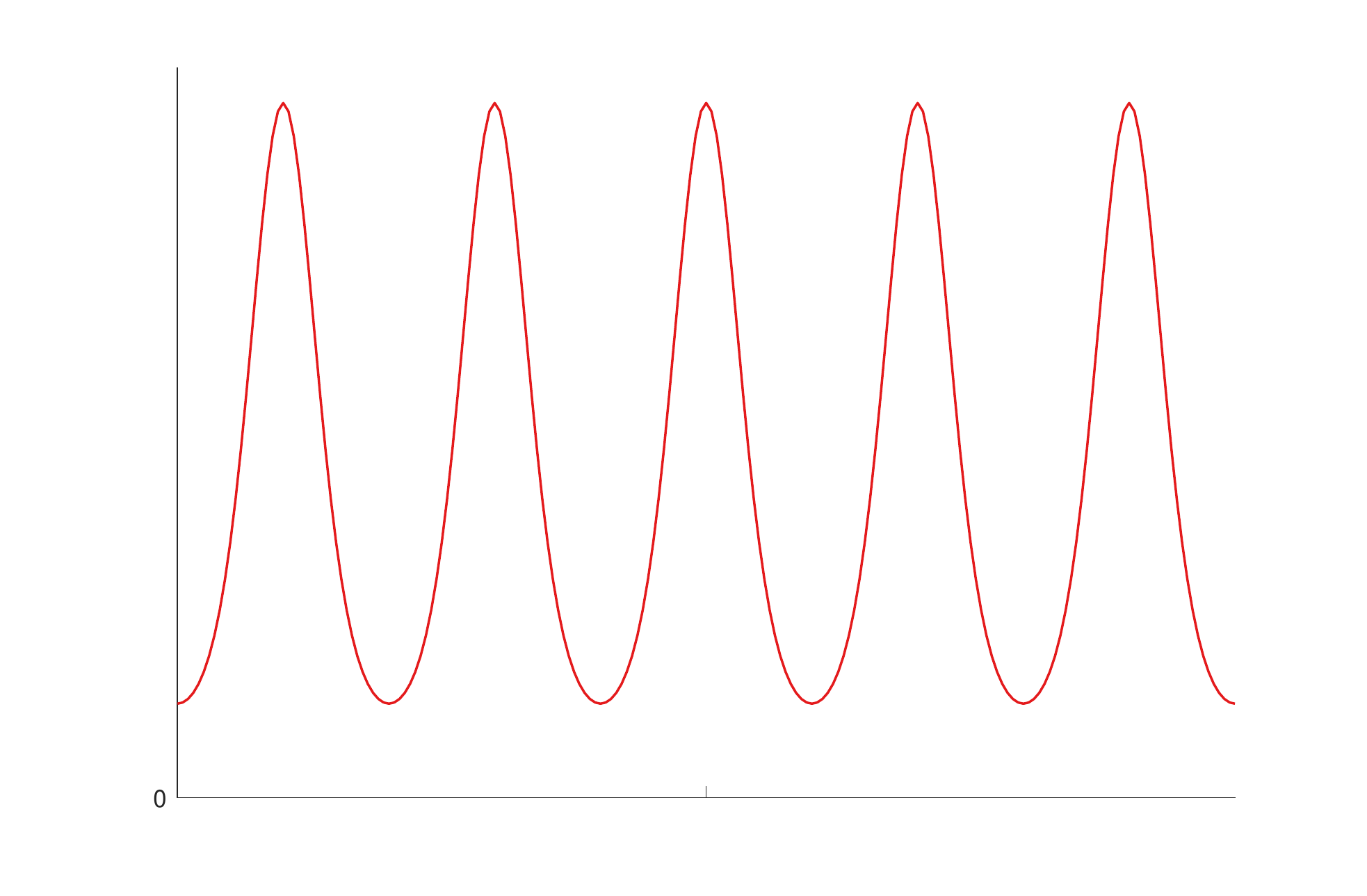}
 		\caption{Periodic covariance Function with respect the distance between the inputs $p=4$, each with the length-scale $\ell=1$ and $\sigma_p=2$,}
 		\label{fig:pe_l_1}
 	\end{figure}

\begin{figure}
 		\centering
 		\includegraphics[scale=0.4]{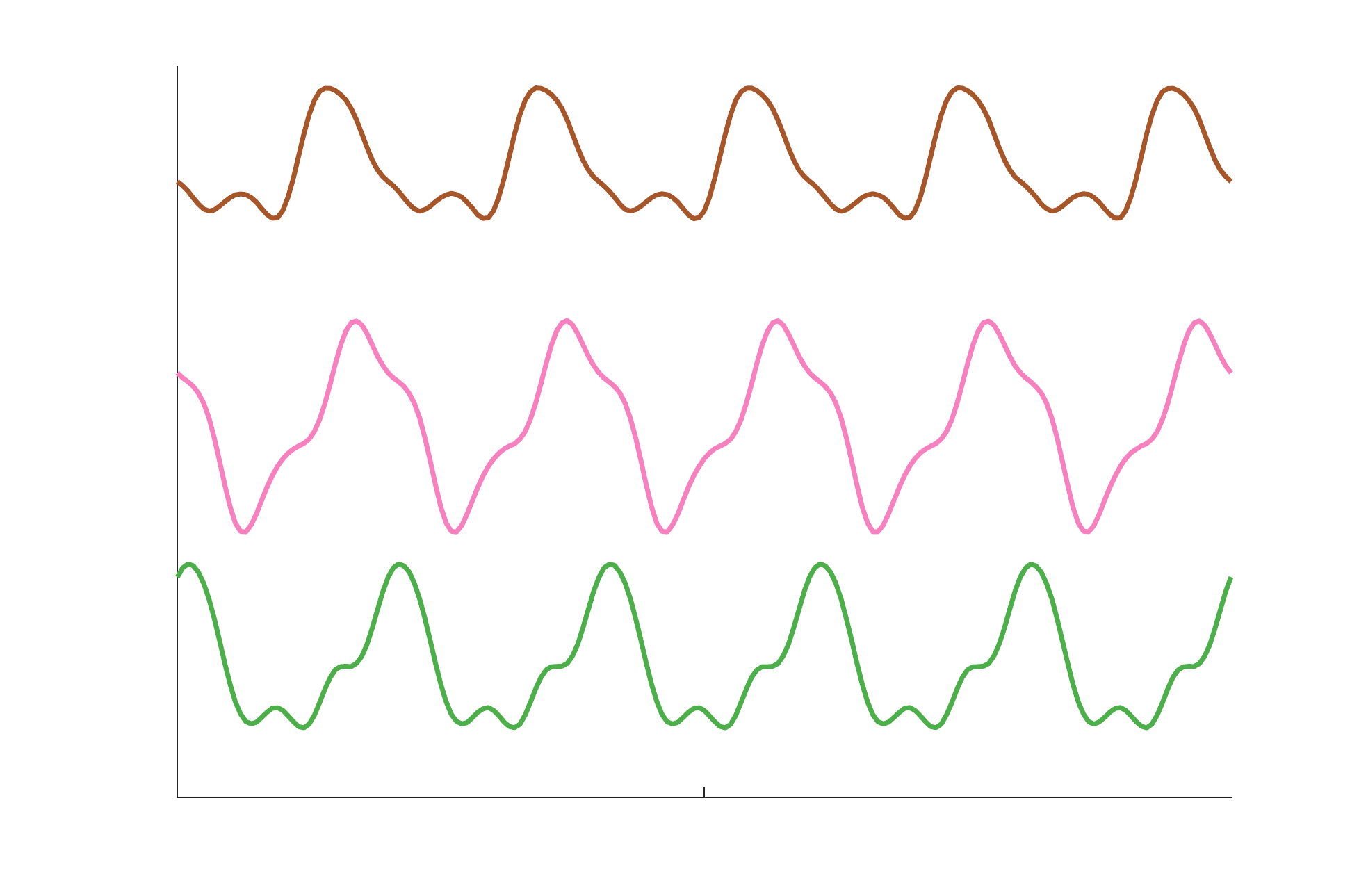}
 		\caption{Random Sample Functions drawn from Gaussian Process with each of these Periodic Covariance Function.}
 		\label{fig:pe_l_2}
 	\end{figure}

 There are many real world processes that shows periodic behavior and we can model this kind of systems with periodic covariance function. Periodic covariance function can be written as:
 \begin{equation}
 k(\x_i, \x_j)=\sigma_{p} \exp\Big(- \frac{2}{\ell^2} \sin^2(\pi \frac{\x_i - \x_j}{p})  \Big)
 \end{equation}
 Where the parameter $\ell$ controls the smoothness of the function and $p$ governs the inverse length of the periodicity. Most of the time periodic covariance function is useful in mixture with other covariance function like squared exponential and rational quadratic covariance function. Figures \ref{fig:pe_l_1} and \ref{fig:pe_l_2} shows the periodic covariance function with respect to the distance between the inputs and the random sample function drawn from Gaussian process. %
 	%

 \subsubsection*{Mat\'{e}rn Covariance Function}
 
 The Mat\'{e}rn Covariance function is also one of the most used covariance function next to square exponential covariance function. Even if square exponential covariance function is the most used one but its strong smoothness assumptions are not more realistic to model many real world physical processes and in \cite{interpolation} Stein recommends the  Mat\'{e}rn class to model real physical process. The Mat\'ern class of covariance functions is given by:
 
 \footnotesize
 \begin{equation}\label{matern}
 k(\x_i,\x_j)=\sigma_{\textrm{m}}^2
 \frac{2^{1-\nu}}{\Gamma(\nu)}\left(\frac{\sqrt{2\nu}|\x_i -\x_j|}{\ell}\right)^{\nu}
 K_{\nu}\left(\frac{\sqrt{2\nu}|\x_i -\x_j|}{\ell}\right)
 \end{equation}
 \normalsize
 The parameter $\nu$ controls the smoothness of the function, and $K_{\nu}$ is a modified Bessel function \cite{Handbook}. The most implemented Mat\'{e}rn covariance functions are the cases when $\nu=3/2$ and $\nu=5/2$,  we can write the two cases as:
 
\small
\begin{align}\label{matern3/2_ARD}
 k_{\nu=3/2}(\x_i,\x_j)& =\sigma_{\mathrm{m}}^2\left(1+\frac{\sqrt{3}|\x_i -\x_j|}{\ell}\right)
 \exp\left(-\frac{\sqrt{3}|\x_i -\x_j|}{\ell}\right),  \\
 k_{\nu=5/2}(\x_i,\x_j) & =\sigma_{\textrm{m}}^2\left(1+\frac{\sqrt{5}|\x_i -\x_j|}{\ell}
 + \frac{5|\x_i -\x_j|^2}{3\ell}\right)\nonumber \\
 &\times \exp\left(-\frac{\sqrt{5}|\x_i -\x_j|}{\ell}\right).\label{matern5/2_ARD}
\end{align}
\normalsize

\begin{figure}
\centering
\includegraphics[scale=0.5]{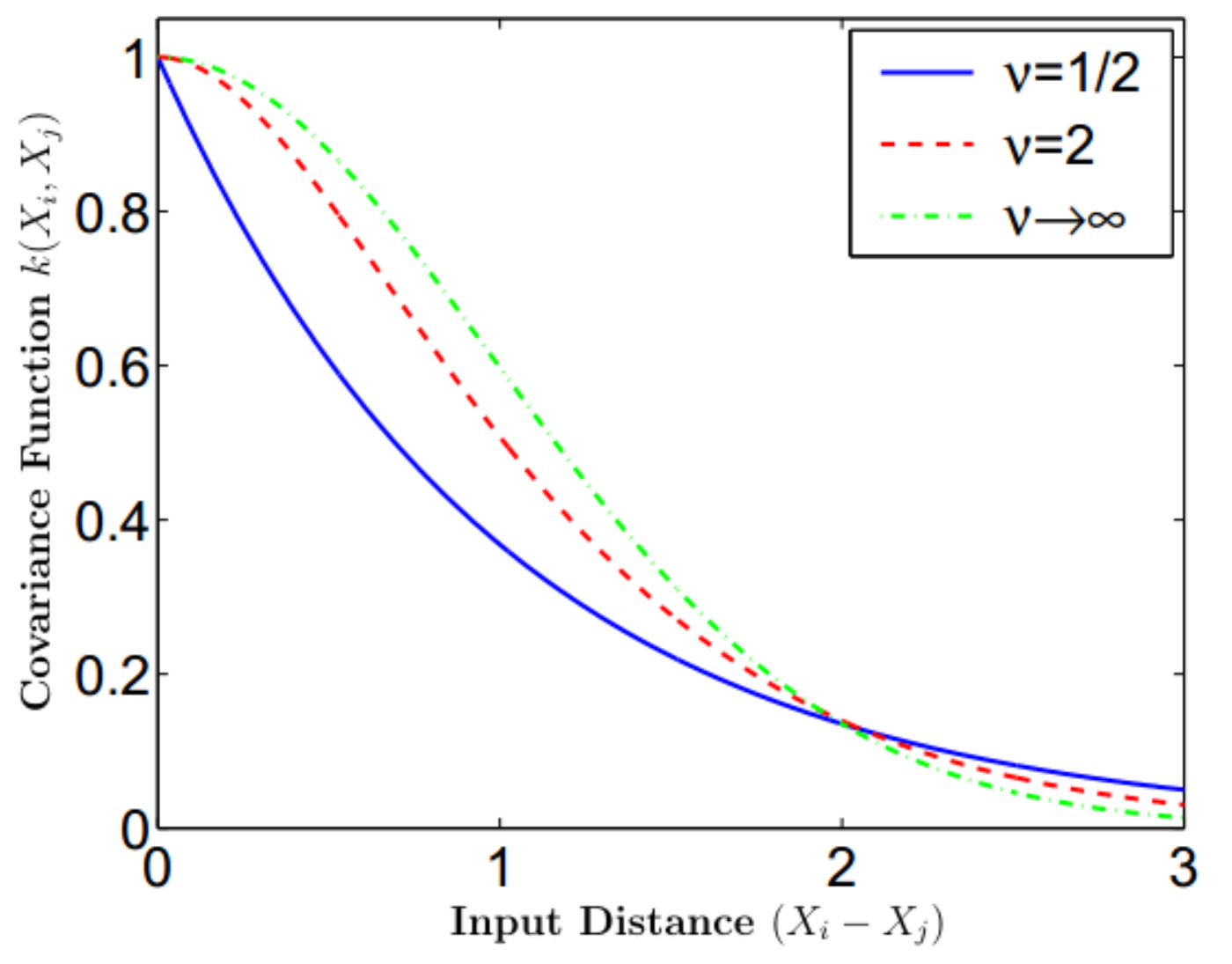}
\caption{Mat\'ern covariance Function with three different settings of  $\nu$, each with the length-scale $\ell=1$,This figure is reproduced from Rasmussen and Williams book \cite{GaussianProcesses}.}
\label{fig:ma_l_1}
\end{figure}%
 	
 	\begin{figure}
 		\centering
 		\includegraphics[scale=0.5]{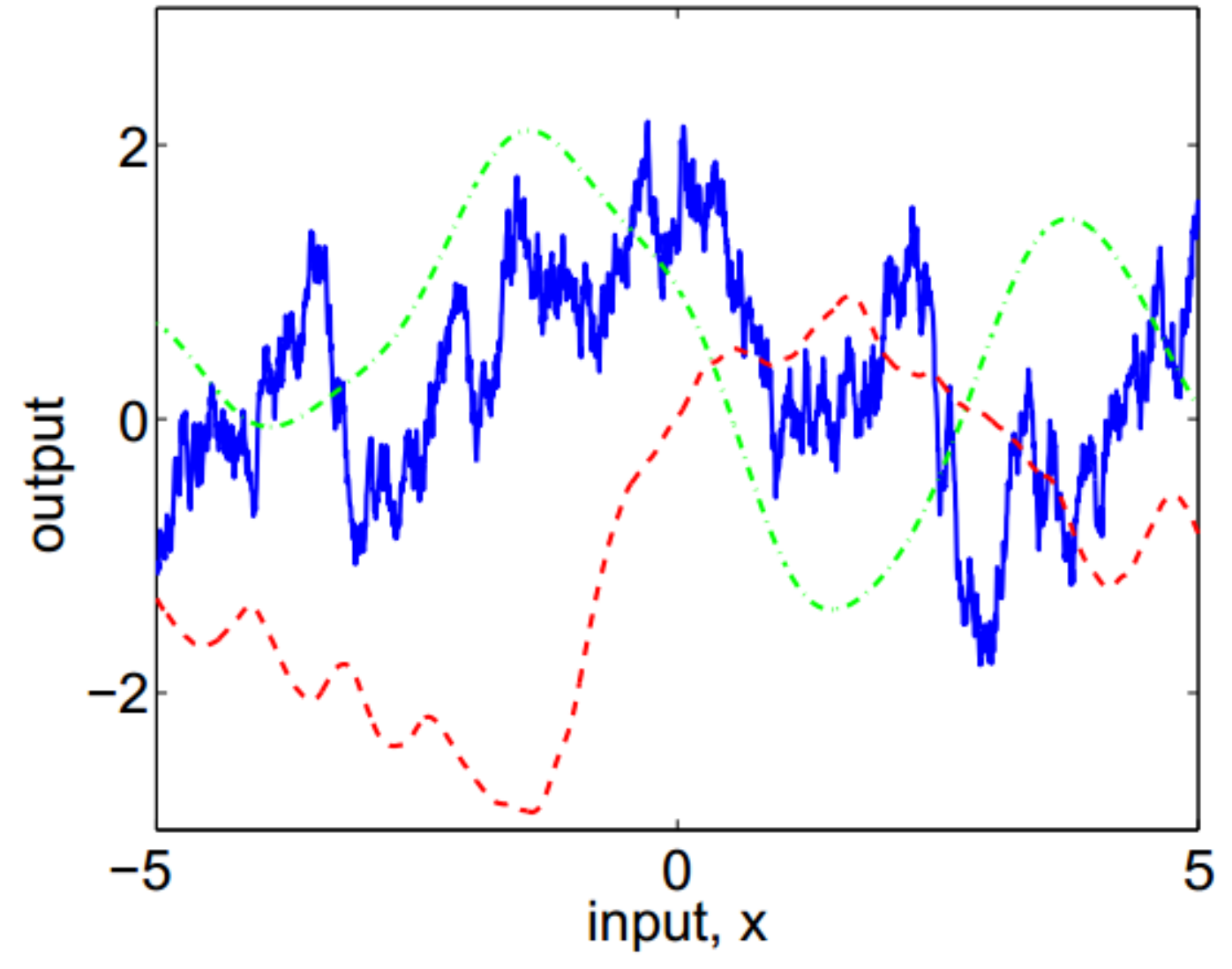}
 		\caption{Random Sample Functions drawn from Gaussian Process with each of these Mat\'ern Covariance Function. This figure is reproduced from Rasmussen and Williams book \cite{GaussianProcesses}. }
 		\label{fig:ma_l_2}
 	\end{figure}%
 The Mat\'{e}rn covariance function depends on the parameter $\nu$ and it changes its behavior based on the values of $\nu$. For example, exponential covariance function $k(\x_i,\x_j)= exp(\frac{-|\x_i -\x_j|}{\ell})$ is generated when $\nu=1/2$. In recent research \cite{fastfood} , it was discovered that the Mat\'{e}rn covariance function has better performance than the squared exponential covariance function on data set with high dimensional inputs, for all inputs $\x \in \mathbb{R}^B$, $B>>1$. This improved performance is achieved due to the Mat\'{e}rn covariance function avoids a 'concentration of measure' effect in high dimensions. Figures \ref{fig:ma_l_1} and \ref{fig:ma_l_2}  illustrates the form of the Mat\'ern covariance function with respect to the separation between the input and random sample functions drawn from Gaussian process with Mat\'ern covariance function for different values of $\nu$.
\subsection{Loss Function In Regression}
\label{sec:loss_fun}
There are a number of distortion measuring criterion used in regression using Gaussian process, but here we will discuss some of them. We will discuss about the widely used criterion like Mean Square Error (MSE), Root Mean Square Error (RMSE) and their normalization.  
\subsubsection{Mean Square Error (MSE)}
Mean square error \textbf{(MSE)} is one of the most used error measure techniques for estimation processes. It simply measures the average of the different between the predicted value and the observed values. Mean square error\textbf{(MSE)} is written as;
\begin{align}
\begin{aligned}
MSE=\E[(\h^* - \tilde{\h}^*)^2]             \\
MSE=\frac{1}{N}\sum_{i=1} ^N (\h^* - \tilde{\h}^*)^2
\end{aligned}
\end{align}

%
Where $\h^*$ is the spatial random process at any location $\x^*$ and $\tilde{\h}^*$ is the predicted value.

\subsubsection{Root Mean Square Error (RMSE)}

The Root Mean Square Error \textbf{(RMSE)} is also one of the commonly used measure of the differences between predicted values and the observed values and it helps to combine these differences into a single measure of predictive power. The Root Mean Square Error \textbf{(RMSE)} of a prediction model with respect to the predicted values $\tilde{\h}^*$ is expressed as the square root of the mean squared error:
\begin{align}
\begin{aligned}
RMSE= \sqrt{MSE}             \\
RMSE=\sqrt{\frac{\sum_{i=1} ^N (\h^* - \tilde{\h}^*)^2}{N}}
\end{aligned}
\end{align}

%
Where $\h^*$ is the spatial random process at any location $\x^*$ and $\tilde{\h}^*$ is the predicted value.

\subsubsection{Normalization}

For comparison with different scales it is very important to normalizing the MSE or RMSE. In the literature there is no constant means of normalization term used, but there are two common approaches most applied normalization term, these are normalizing over the \textbf{mean} of the observed values and normalizing over the \textbf{range} of the observed values. 
 \subsubsection*{Normalized MSE (NMSE)}
 \begin{itemize}
 \item Normalizing Over the \textbf{mean}
 \begin{equation}
 NMSE= \frac{MSE}{\overline{\h^*}} 
 \end{equation}

 \item Normalizing Over the \textbf{range}
 \begin{equation}
 NMSE= \frac{MSE}{\h^*_{max}-\h^*_{min}} 
 \end{equation}

 \end{itemize}

 \subsubsection*{Normalized RMSE (NRMSE)}
 \begin{itemize}
 \item Normalizing Over the \textbf{mean}
 \begin{equation}
 NMSE= \frac{RMSE}{\overline{\h^*}} 
 \end{equation}

 \item Normalizing Over the \textbf{range}
 \begin{equation}
 NMSE= \frac{RMSE}{\h^*_{max}-\h^*_{min}} 
 \end{equation}
 \end{itemize}
 

\section{System Model: Sensor Network}
\label{sensor_net}
The model of the physical phenomenon observed by the sensor network is considered to be a Gaussian random field with a spatial correlation structure. The Gaussian processes model we consider assumes that the special filed observed by the sensors network is a noisy observation where the noise is independent and identically distributed Gaussian Noise. A general system model have been considered in which a sensors network are deployed in one room. The measurements from the sensors is collected in real time by the fusion center in order to make an estimate of the spatial phenomenon at any point of interest in the room space. Here we will discuss all the considered assumption of the proposed system model, the sensors network observation.
We may have a source of electromagnetic fields from different directions and the field intensity captured by the sensors may vary. For that reason we Consider a random spatial phenomenon to be observed over a two dimensional space $\mb{X}\in \Re^2$ in the measurement room. We modeled our system as a Gaussian process since we want the average response of our conditional method to be smooth and persistent spatial function $h(.)$:
\begin{equation}
h(\x) \sim \GP(m(\x),Cov(\x, \x'))
\end{equation}
We have used $9$ sensors  in our measurement room which is 2 - dimensional space $\mb{X}\in \Re^2$ and the location of the sensors is represented by $\x_n \in \mb{X}$, where $n=\{1,...,9\}$.
We have also assume that the observation from each sensor is corrupted by independent and identically distributed Gaussian noise: 
\begin{equation}
 {Y}_n= h(\x_n) + \eta_n   
\end{equation}
where the noise assumption, Gaussian noise is  $\eta \sim \N(0,\sigma_{\eta}^2)$. Therefore the measurement from each sensor is noisy observation.

\subsection{Mean Function}\label{sec:mean}
Most of the time when Gaussian process is implemented the mean function is consider as zero mean to indicate that we do not have any knowledge about the function output at any input, but it would be better to include a mean function if we have some prior information about the behavior of the function output. In this thesis since we are working on spatial reconstruction of electromagnetic fields we extract some information from their properties. The intensity of electromagnetic fields observed by the sensors is inversely proportional to the square of the separation between the positions of the source and sensors. This implies that the EM fields observed by the nearest sensor to the source have higher intensity relative to the distant sensors, the EM fields intensity will decay with distance. We include this information to our model by considering a mean function for the proposed Gaussian process and this increases the expressiveness of our dataset.
In many applications it is common to use deterministic mean function that can be applied by fixing a mean function, but to define a fixed mean function is difficult in practical applications. Therefore, it will be more advantageous if we can define some basis functions with their standard coefficients and learn the coefficients from our dataset. For the proposed algorithm we consider a mean function that is defined by weighted sum of few fixed basis function $\mb{g}$.
\begin{equation}
m=\g(x)^{\tran} \Omega
\end{equation}
where $\g(x)$ are some fixed basis functions and $\Omega$ are the weighting coefficient, they are an additional parameters of the mean function to be learned from the dataset. Modeling of the latent process is assumed as a sum of the considered mean function and a zero mean Gaussian process.
\begin{align}
h(\x)&=  \g(x)^{\tran} \Omega + f(\x) \nonumber\\
h(\x)&=  \g(x)^{\tran} \Omega + \GP(\mb{0}, \K)
\end{align}
where $f(\x)$ is a zero mean Gaussian process, $\K=\Khh + \sigma_{\eta}^2\mb{I}$. From the concept of electromagnetic wave propagation path loss the intensity observed by the sensors is described by:
\begin{equation}\label{pathloss}
\mb{P_r \sim P_e d^{-2}}
\end{equation}
Based on the equation \eqref{pathloss}, we define the basis function to consider this relation on our mean function and are represented by:
\begin{equation}
\g(\x)=\frac{1}{1+\mb{d}(\x,\x_s)^2}
\end{equation}
\begin{align}
\mb{d}(\x,\x_s)=\sqrt{(x_1- x_{s1})^2 + (x_2- x_{s2})^2}. \nonumber
\end{align}
where $\x=[x_1, x_2]^{\tran}$ is the location of the sensors,  $\x_s=[x_{s1}, x_{s2}]^{\tran}$ is the location of the source and $d(\x,\x_s)$ is the distance between the sensors and source. This basis function ensures that we will have a maximum electromagnetic fields intensity at the source location and this is the prior information that we want to include in our algorithm to enhance its prediction performance.

The idea is to learn the weighting coefficient from our data, so we can have a good representation of the parameters using the specified dataset. To do so, first we assumed a prior on the coefficient $\Omega$ to be Gaussian distributed with a mean and variance, $\Omega \sim \N(a,A)$. To fit our model, we implement optimization over the parameters of $\Omega$ together with the hyperparameters of our covariance function. Using the concept of marginalization we can integrate out the weight parameters and the prior for our latent function $\mb{h}$ is also another Gaussian process with a specified mean and covariance function:
\begin{equation}\label{fig:mean_fun}
\mb{h}(x) \sim \GP(\g(\x)^{\tran}\mb{a}, \K + \g(\x)^{\tran}\mb{A}\g(\x))
\end{equation}
From the consider mean function now we have an additional contribution in the covariance function that is related to the uncertainty in the parameters of our mean function. Therefore, we obtain the predictive equation by using the mean and covariance function of our latent prior $\mb{h}(\x)$ in the predictive equation of  zero mean Gaussian process equation \eqref{eq_condi_y}, and it is given by:

\small
\begin{align}
m(\mb{x}^*)&=\mb{G}^*\bar{\Omega} +\K^{*\tran}\K^{-1}(\Y - \mb{G}^{\tran}\bar{\Omega}) =\E[\f^*] + \mb{R}^ {\tran}\bar{\Omega}\\
Cov(\mb{h}^*)&= Cov(\f^*) + \mb{R}^{\tran}(\mb{A}^{-1} +\mb{G}\K^{-1}\mb{G}^{\tran}) \mb{R} \\
\nonumber \\
\bar{\Omega}&=(\mb{A}^{-1} + \mb{G}\K^{-1}\mb{G}^{\tran})^{-1} (\mb{A}^{-1} \mb{a} + \mb{G}\K^{-1}\Y) \nonumber \\
\mb{R}&=\mb{G^*} - \mb{G}\K^{-1}\K^* \nonumber
\end{align}
\[ 
\mb{G}=\left( \begin{array}{ccc}
g_1(\x) \\
g_2(\x) \\
\vdots\\
g_k(\x) \end{array} \right)
\] 
\normalsize
where $\mb{G}$ is a matrix that combines all the vectors of $g(\x)$ for all the datasets and $\mb{G}^*$ for all test points, $\bar{\Omega}$ is the mean of the weighting parameters and it has a trade-off between the data and the prior information, and our predictive mean is the sum of mean linear output and the prediction of zero mean Gaussian process.  
\subsection{Model Selection and Hyperparameter Learning }
\label{sec:modelsele}
In many practical applications, it is not easy problem to specify all characteristics of the covariance function with having any problem. However some properties like stationarity of
the covariance function is not complicated to understand from the setting, we have a tendency to usually have only uncertain information regarding additional properties, for instance the value of the hyper-parameters.
In section \ref{covfuncion}, we have discussed about many examples of covariance functions, with different and large numbers of parameters. Furthermore, the precise form and the free parameters of the likelihood function might be unknown. Therefore, to optimize the advantage of Gaussian process for practical application there should be approaches that overcome the problem of the model selection. Mainly the problem of model selection includes the selection of the functional arrangement of the covariance function and the suitable values of the hyperparameters of the covariance function. We have to make decision about every detail of our specification in order to model for our practical tool.
From many covariance function families that are available like, squared exponential, Mat\'{e}rn, Rational Quadratic it is not an easy task to select on that can interpret the dataset we have on hand. Each covariance function families have many free hyperparameters and we need to determine their values. Therefore, when we select covariance function for our application, we are both setting the values of the hyperparameters of the covariance function and comparing it with other different families. Our main task is to apply inference to determine the type of the covariance function and its hyperparameters based on the training data we have. In other word we are letting the dataset to determine the suitable covariance function and hyperparameters. The smooth shape and variation of the likelihood function using Gaussian process are governed by the covariance function and the included hyperparameters. Therefore, the main idea behind the model selection and hyperparameters learning using Gaussian process is to involve the data we have in the process of  choosing the covariance function and setting the hyperparameters.

\begin{figure}[!htb]
\centering
\includegraphics[scale=0.5]{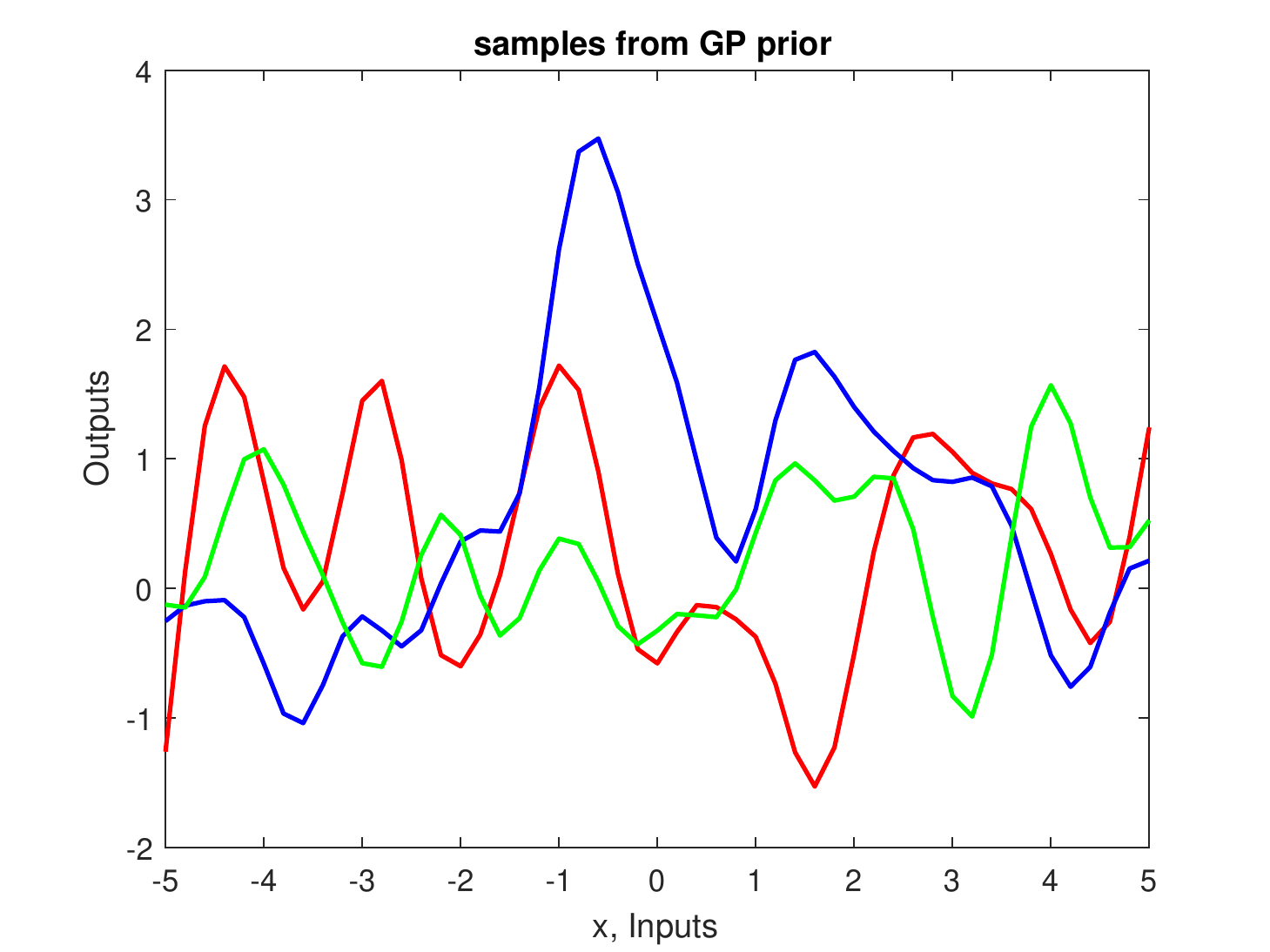}
\caption{GP prior with squared exponential covariance function with small value $\ell=0.3$.}
\label{fig:GP_small_1_prior}
\end{figure}

\begin{figure}[!htb]
\centering
\includegraphics[scale=0.5]{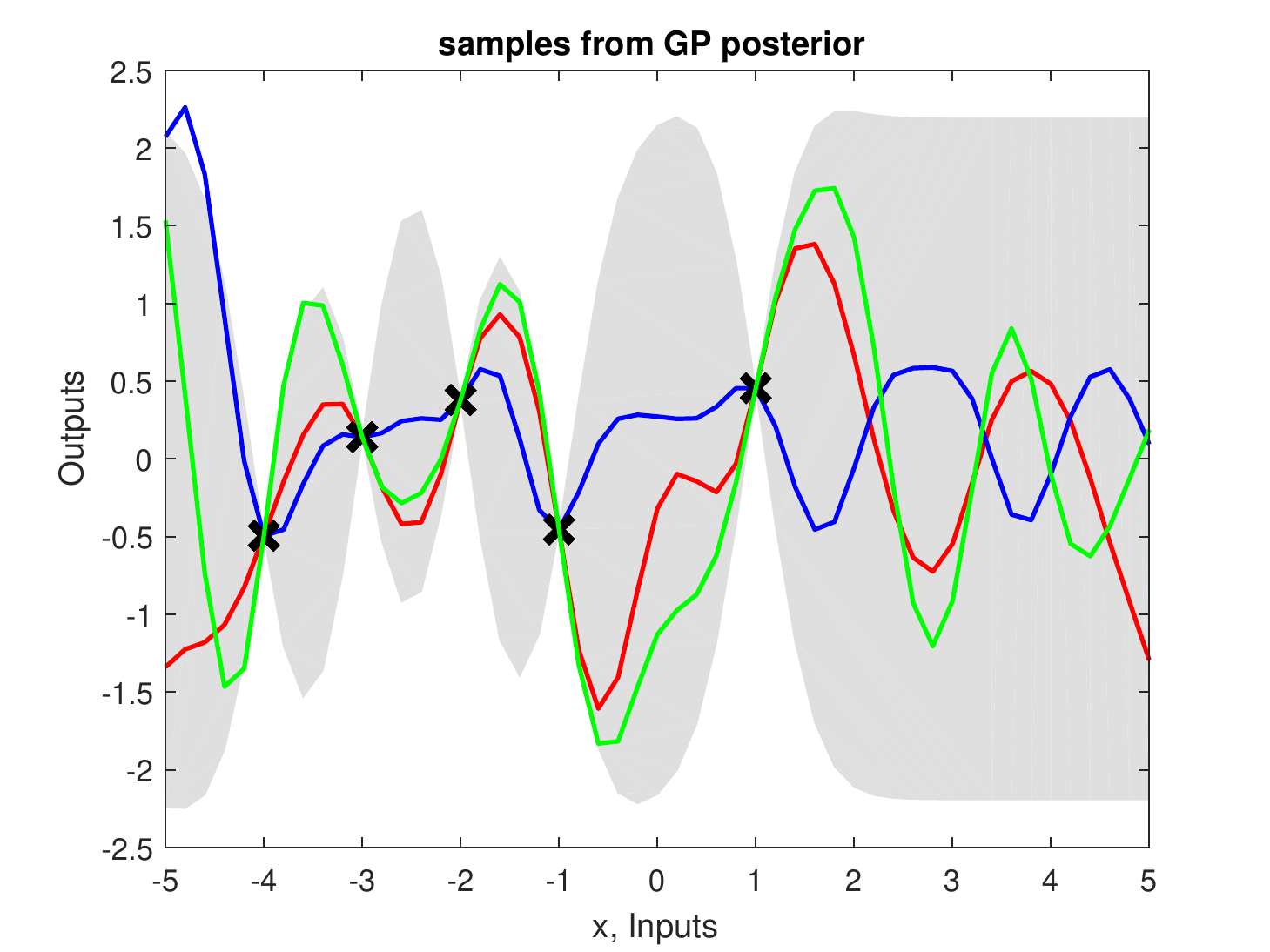}
\caption{Posterior functions drawn from a GP with squared exponential covariance function with small value of length-scale, $\ell=0.3$.}
\label{fig:GP_small_1_post}
\end{figure}

\begin{figure}[!htb]
	\centering
	\includegraphics[scale=0.5]{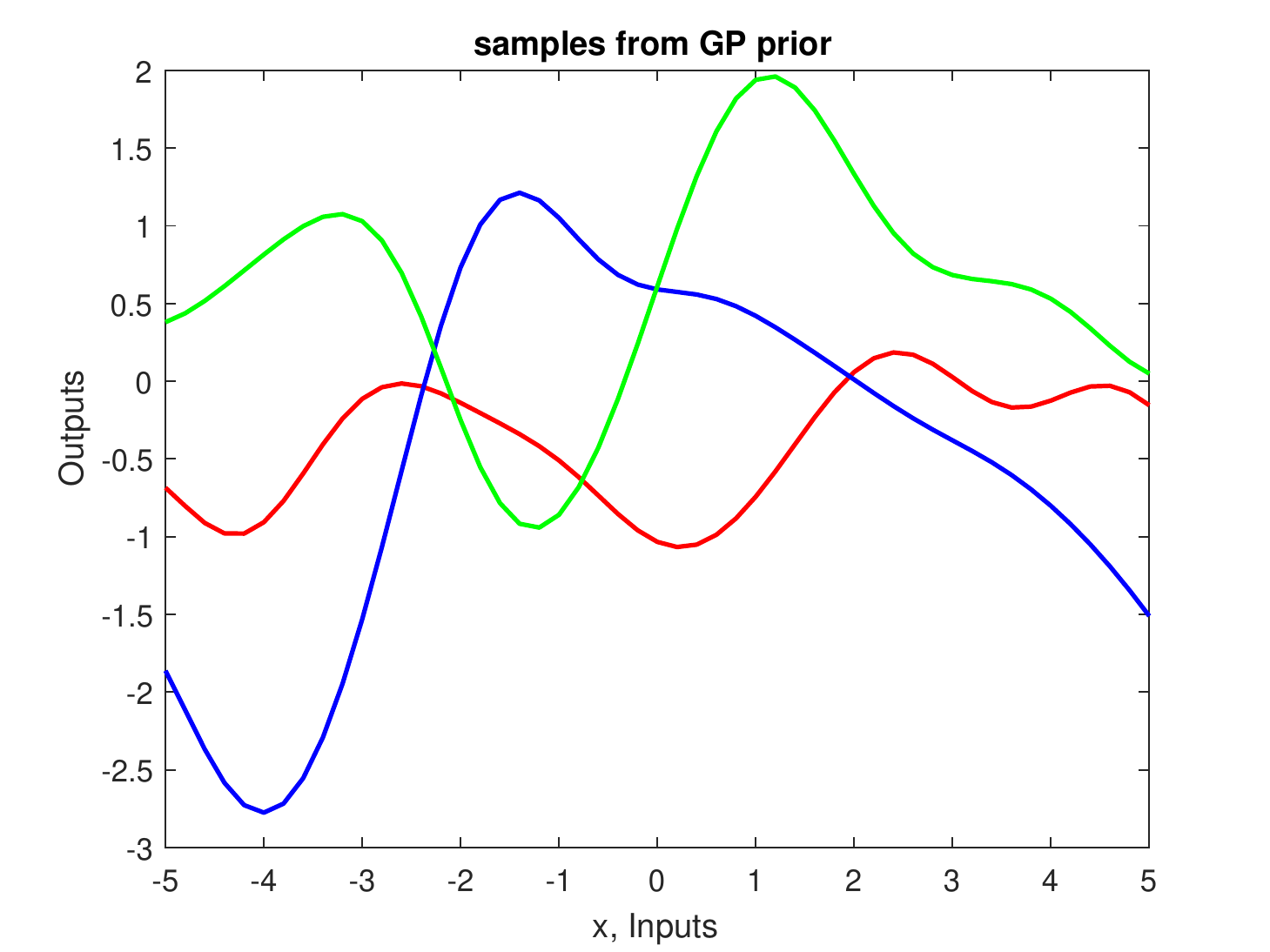}
	\caption{Gaussian Process prior with squared exponential covariance function with appropriate value of $\ell=1.2$}
	\label{fig:GP_small_2_prior}
\end{figure}

\begin{figure}[!htb]
	\centering
	\includegraphics[scale=0.5]{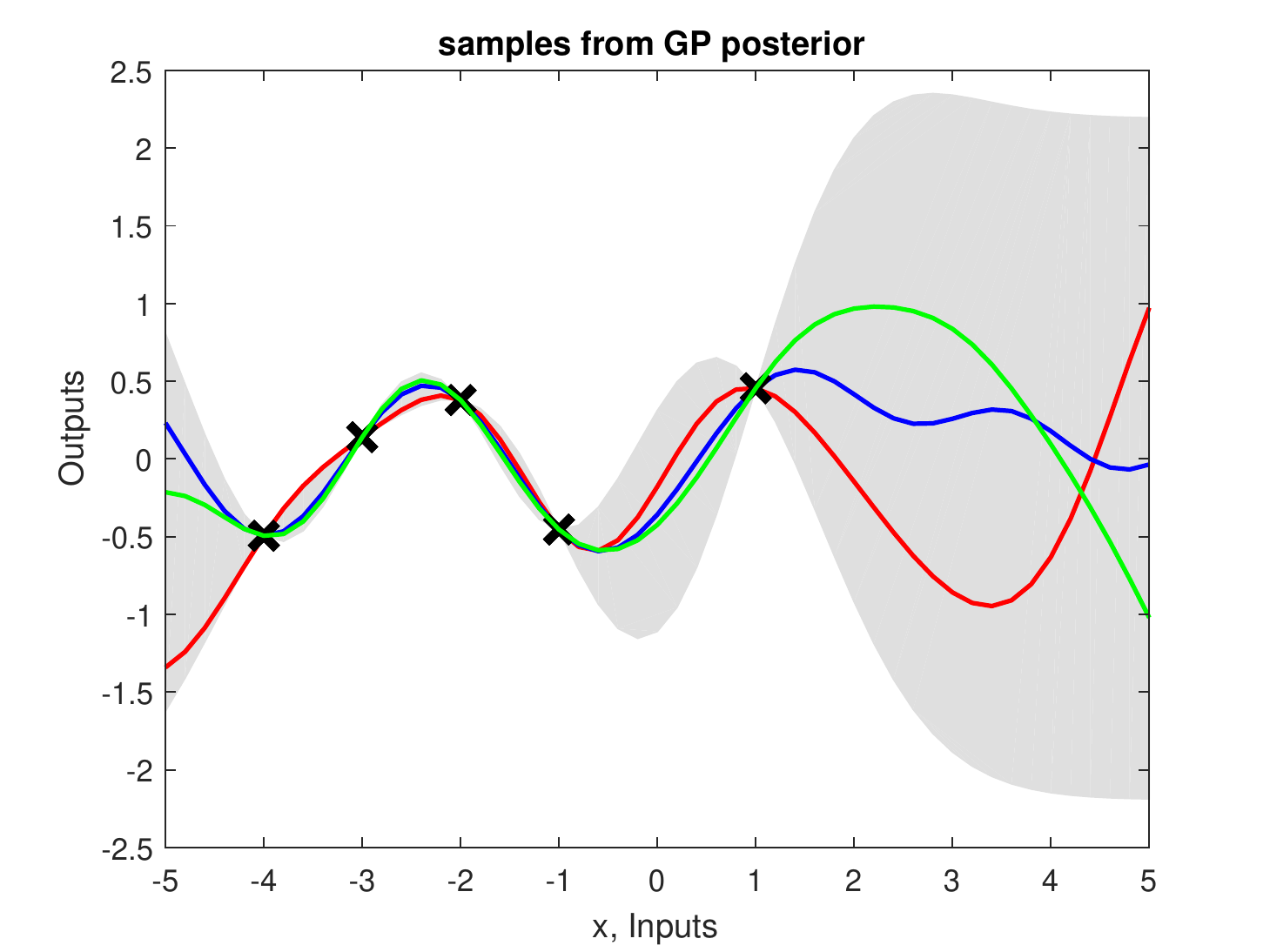}
	\caption{Posterior functions drawn from a Gaussian process with squared exponential covariance function with appropriate value of length-scale, $\ell=1.2$.}
	\label{fig:GP_small_2_post}
\end{figure}

\begin{figure}[!htb]
 	\centering
 	\includegraphics[scale=0.25]{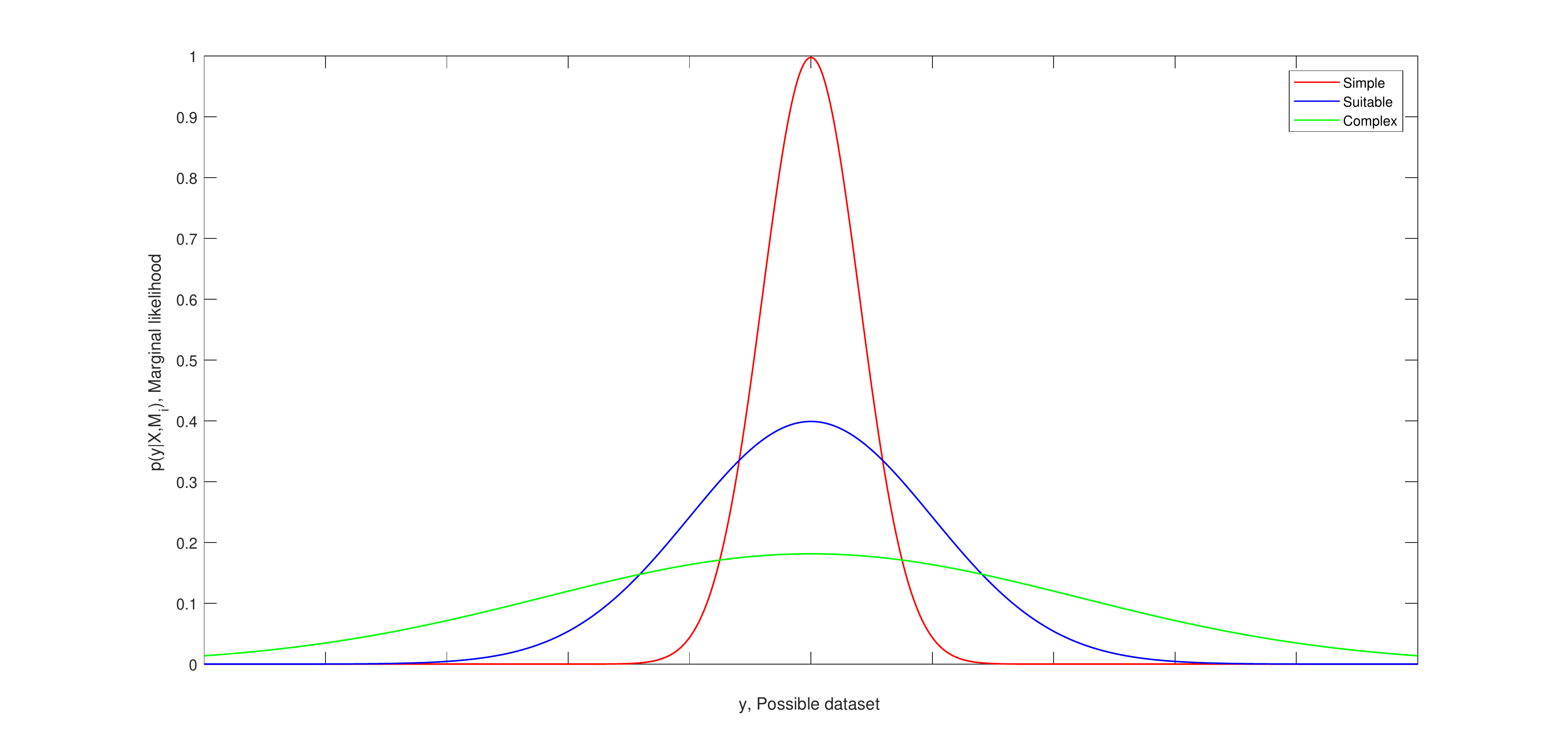}
 	\caption{Model selection using marginal likelihood, for three different model complexities. The probability of the data given the model $p(\y|\mb{X}, M_i)$ is the marginal likelihood. All the possible data-sets are on horizontal axis and the vertical axis shows the marginal likelihood.}
 	\label{fig:ModelSele}
 \end{figure}

In model selection we need to focus on selecting a model that compromise the complexity of the model and the capability to fit the data. For example, the model in Figure \ref{fig:GP_small_1_prior} and \ref{fig:GP_small_1_post} shows functions drawn from Gaussian process with square exponential covariance function and with small value of  hyperparameter, length scale. The model fits the data points in a good way, but the small length scale is making the sample functions to vary quickly and trying to fit every point in the dataset. The scale of the variation of the function and the variety of data required to achieve a good prediction are controlled by the length scale. This means the correlation between the function values of close input points is very low. Therefore, this model is complex model. In Figure \ref{fig:GP_small_2_prior} and \ref{fig:GP_small_2_post} we improve the length scale and draw the sample function from Gaussian process. The functions are varying normally, nether too fast nor too slow. Now we have a good compromise between the complexity of the model and the data-fitting.
%
%
From the demonstration we can observe that the performance of Gaussian process depends on the hyperparameters of a covariance function , changing even a single hyperparameter also changes the performance. Accordingly, it is very important to determine the hyperparameters correctly. Since what we need is a model that have a good compromise between the data-fit and complexity, we want a mechanism that can do this automatically, with out fixing any parameters. Even though there are a number of approaches for model selection, it is always a problem to choose the best one. Gaussian processes implement a good probabilistic properties  that provide automatic model selection. In this discussion we will consider the Bayesian model selection that is computing the probability of a model given data to make the model selection and it is technically based on the marginal likelihood.
Now let's assume that we have a defined model $M_i$ that model the observations $y$ and this model relates our inputs to the observation, considers prior distributions and the model  for noise during the observation. Bayesian model selection approach can be described by using three level inference \cite{bayeMac} . At the lower level, we want to deduce the model's parameters given the data which is also called the posterior of parameters. Applying Bayes' rule the distribution is given by:
\begin{equation} \label{pos}
p(\omega|\y,\mb{X},\beta, M_i)=\frac{p(\y|\mb{X},\omega, M_i) p(\omega|\beta,M_i)}{p(\y|\mb{X},\beta, M_i)}
\end{equation}
Where $\omega $ is the model parameters , $\beta$ is the hyperparameters of the model, the likelihood is given by $p(\y|\mb{X},\omega, M_i)$ and $p(\omega|\beta,M_i)$ is the prior distribution of the parameters. The posterior integrates both the prior information and the data using likelihood. The denominator of equation \eqref{pos} is the normalizing term and it is called the marginal likelihood or evidence. It is given by:    
\begin{equation} \label{marglik}
p(\y|\mb{X},\beta, M_i)= \int {p(\y|\mb{X},\omega, M_i) p(\omega|\beta,M_i)} d\omega
\end{equation}
At the second level of inference, we infer what the hyperparameters of our model might be given the data we have. The posterior over the hyperparameters is expressed by:
\begin{equation} \label{post_hy}
p(\beta|\y,\mb{X}, M_i)= \frac{p(\y|\mb{X},\beta, M_i) p(\beta|M_i)}{p(\y|\mb{X}, M_i)}
\end{equation}
Where $p(\y|\mb{X},\beta, M_i)$, the marginal likelihood of the lower level of inference is used as likelihood and $p(\beta|M_i)$ is the prior information of hyperparameters. We can express the normalizing term of equation \eqref{post_hy} as:
 \begin{equation} \label{marglik_hy}
 p(\y|\mb{X}, M_i)= \int {p(\y|\mb{X},\beta, M_i) p(\beta|M_i)} d\beta
 \end{equation}
At the higher level of inference, we want to infer which model is the most reasonable given the data we have, that means we are applying model comparison. The posterior of the model is given by:
\begin{equation} \label{post_mod}
p(M_i|\y,\mb{X})= \frac{p(\y|\mb{X}, M_i) p(M_i)}{p(\y|\mb{X})}
\end{equation}
 where $ p(\y|\mb{X})=\sum_{i} p(\y|\mb{X}, M_i) p(M_i)$ and by assuming that the prior $p(M_i)$ remains the same for different models, we can perform the model comparison by just evaluating the marginal likelihood or evidence. Many complicated integrals are used in different level of inference and based on the model’s specification the included integrals can be analytical traceable or not. If Gaussian likelihood assumption is not considered, we have to implement analytical approximation approaches. The integral in equation \eqref{marglik_hy} can be evaluated by first estimating hyperparameters, $\beta$ that maximize the marginal likelihood in equation \eqref{marglik} and then we can use the expression around the maximum.
 The main advantage of the marginal likelihood is the ability to automatically ensure the best compromise between the complexity of the model and the data fit. In figure \ref{fig:ModelSele} we present a diagram of the property of the marginal likelihood for three different model complexities. Assume that we fix the number of data-points used; the vertical axis indicates the marginal likelihood $p(\y|\mb{X}, M_i)$ and the horizontal axis shows all the possible observation $y$. A complex model has the capacity to include large number of datasets, but this implies that it has lower values of marginal likelihood. On the other hand, a simple model is capable of including limited number of possible datasets, and the marginal likelihood that is the probability distribution over y has a higher value. For the suitable model we have best trade-off between the data fit and the model complexity. 
 The figure explains that the marginal likelihood does not just select the model that fits the data in a best way, but also considers the model complexity. This principle is called the Occam's razor which states "plurality should not be assumed without necessity" \cite{bayeMac}. Now, we will see the marginal likelihood implemented in the algorithm to select the covariance function and to learn the hyperparameters of the considered mean and covariance function. Since we are using a Gaussian prior and likelihood function the computation of the integral is analytical tractable and we select our specification of the covariance function by comparing the objective function of the considered model with others.
 \subsubsection{Marginal Likelihood}\label{mal}
To specify the marginal likelihood of our model, first let's recall the marginal likelihood function of Gaussian process prior with zero mean that discuss in section \ref{sec_Inf}. The log marginal likelihood from equation \eqref{eq_log_marginal_likelihood} is give by:

\small
\begin{equation}\label{eq_log_m_l_2}
\log p(\y|\mb{X},\beta) =  -\frac{1}{2} \y^{\tran} \K^{-1} \y - \frac{1}{2}\log |\K| -\frac{n}{2}\log(2\pi) 
\end{equation} 
\normalsize

where $\K=(\Khh + \sigma_{\eta}^2\mb{I})^{-1}$ is the covariance function of the noisy target $\y$. This implementation shows that the marginal likelihood has this trick used to apply the automatic compromise between the data fitting process and the model complexity, and this is because if we select hyperparameters of the covariance function in such way to obtain a better data fitting the second term in of the equation \eqref{eq_log_m_l_2} will also scale up and that is the complexity term that regularize the model.

In the proposed algorithm we consider a mean function, that is sum of weighted fixed basis function, and this causes inclusion of additional contribution on the covariance function.
 Therefore, the marginal likelihood will change accordingly. The log marginal likelihood for our model with $\Omega \sim \N(a,A)$ on the hyperparameter of equation \eqref{fig:mean_fun} is computed as:
\begin{align}\label{eq_log_ml_new}
\log p(\y|\mb{X},\boldsymbol{\theta} ,\boldsymbol{\beta})&=  -\frac{1}{2} \mb{U}^{T} \mb{V}^{-1} \mb{U} - \frac{1}{2}\log |\K|  \nonumber \\
&- \frac{1}{2}|\mb{A}| - \frac{1}{2}|\mb{W}| -\frac{n}{2}\log2\pi 
\end{align}  
\begin{align}
\mb{W}&=\mb{A}^{-1} + \mb{G}\K \mb{G}^{\tran} \hspace{1cm} \mb{U}=\mb{G}^{\tran} \mb{a}-\y \nonumber \\
\mb{V}&=\K +\mb{G}^{\tran} \mb{AG} \hspace{1cm} \K=(\Khh + \sigma_{\eta}^2\mb{I})^{-1} \nonumber
\end{align}   
Where $\boldsymbol{\theta}$ and $\boldsymbol{\beta}$ are the hyperparameters of our mean and covariance functions respectively and we implement the matrix inversion equations A.9 and A.10. We set our hyperparameters by maximizing the marginal likelihood and to search the hyperparameters that maximize the marginal likelihood we need to compute the partial derivative of the marginal likelihood with respect to the hyperparameters. Therefore, we use equation \eqref{eq_log_marginal_likelihood_new} to compute the partial derivative as:
\begin{align}
  \frac{\partial}{\partial{\beta_i}} {log(p(\y|\mb{X},\boldsymbol{\theta} ,\boldsymbol{\beta}))}=\frac{1}{2} \mb{U}^{\tran} \mb{V}^{-1} \frac{\partial \K}{\partial{\beta_i}} \mb{V}^{-1} \mb{U}^{\tran} \nonumber \\
  - \frac{1}{2} tr\Big( \K^{-1} \frac{\partial \K}{\partial{\beta_i}}\Big)  - \frac{1}{2} tr\Big( \mb{W}^{-1} \frac{\partial \mb{W}}{\partial{\beta_i}}\Big)  
\end{align}
\begin{equation}
\frac{\partial \mb{W}}{\partial{\beta_i}}= -\mb{G} \K^{-1} \frac{\partial \K}{\partial{\beta_i}} \K^{-1} \mb{G}^{\tran}  \nonumber
\end{equation}
The partial derivative of the marginal likelihood is then  fed to the scaled conjugate gradient optimization function to proceed with the hyperparameter optimization.

\subsubsection{Covariance Function Selection}\label{cov_new}
To select the suitable covariance function for our model, we have implemented the marginal likelihood approach to compare between other covariance function that could fit in a good way. In the process of model selection, we develop two Gaussian process; the first one is based on square exponential whereas the other is based on Mat\'{e}rn covariance function, then we compared the performance of the model by computing and comparing the objective function. 
As we have discuss it in section \ref{covfuncion} the square exponential covariance function given in equation \eqref{k_se_new_new} is widely used because it is infinitely differentiable, meaning it has mean square derivative of all orders and this contributes to its strong smoothness property. But in  real life application this property may not be practical to model physical process. Compared to this, the smoothness of Mat\'{e}rn covariance function stated in equation \eqref{matern3/2_ARD_1} is governable, since it is $\nu -1$ times differentiable. Therefore, we can govern the smoothness by controlling $\nu$. 
\begin{equation}\label{k_se_new_new}
k(\x_i, \x_j)_{SE}=\sigma_{se}^2 \exp \Big(-\frac{|x_{i}-x_{j}|^2}{2{\ell}^2} \Big),
\end{equation}
\begin{align}\label{matern3/2_ARD_1}
k_{\nu=3/2}(\x_i,\x_j)& =\sigma_{\mathrm{m}}^2\left(1+\frac{\sqrt{3}|\x_i -\x_j|}{\ell}\right)  \nonumber \\
&\times\exp\left(-\frac{\sqrt{3}|\x_i -\x_j|}{\ell}\right). 
\end{align}
\\
We used 16 different training set to train the Gaussian process based on the two covariance functions, then we used the optimized hyperparameters to make the prediction. The electromagnetic source is located at 16 different positions to ensure that we have sufficient data points to make the prediction at any point regardless the location of the electromagnetic source.
Three hierarchal steps is applied in the procedure of selecting our covariance function; in the initial step, the training sets are used to compute the model marginal likelihood and its partial derivative with square exponential and Mat\'{e}rn covariance function, then the scaled conjugate gradient optimization algorithm associates the optimized hyperparameters with each covariance function. In the second step, we make the prediction for different number of points using the optimized hyperparameters of each models. In the third step, to evaluate the performance of each models we computed the difference between the real target values and the predicted values separately.

Two different evaluating criteria are implemented to observe and compare the performance of the square exponential and Mat\'{e}rn covariance functions:
\begin{itemize}
	\item The Mean squared error (MSE): computes the prediction error, which measures the difference between the true and the predicted values. Lower value indicates a better performance.
	\begin{equation}
	 NMSE=\frac{\E[(\h - \hat \h)^2]}{\E[\h]} \nonumber
	\end{equation}  
	
	\item The correlation (CR): computes the correlation between the predicted and true values. Higher value shows better performance.
	
\end{itemize}

\begin{table}[!thb]
\caption{Comparison of SECF and MCF for Source 1 - 8}
	\label{table1} 
	\centering
	{
\begin{tabular}{|*{4}{c|}}  
	\hline
	{Datasets} & {Models} & \multicolumn{2}{|c|}{9 Observations}  \\ \hline
	         &   & NMSE & CR  \\ \hline
    Source 1 & SECF &0.5431 & 0.7325      \\ 
             & MCF  &0.3798 & 0.8043  \\ \hline
    Source 2 & SECF &0.5581 & 0.7171      \\ 
             & MCF  &0.3763 & 0.8141      \\ \hline 
    Source 3 & SECF &0.5671 & 0.7202      \\ 
             & MCF  &0.3978 & 0.8011   \\ \hline   
    Source 4 & SECF &0.5212 & 0.7312      \\ 
             & MCF  &0.3783 & 0.8103    \\ \hline    
    Source 5 & SECF &0.5677 & 0.7297      \\ 
             & MCF  &0.3868 & 0.8245     \\ \hline                    
    Source 6 & SECF &0.5902 & 0.7104       \\ 
             & MCF  &0.3712 & 0.8144       \\ \hline          
    Source 7 & SECF &0.5381 &  0.7228    \\ 
             & MCF  & 0.3766 & 0.8132    \\ \hline         
    Source 8 & SECF &0.5881& 0.7023      \\ 
             & MCF  & 0.3774   & 0.8251       \\ \hline 
\end{tabular}}
\end{table}
\begin{table}[!thb]
	\caption{Comparison of SECF and MCF for Source 9 - 16}
	\label{table2} 
	\centering
	\begin{tabular}{|*{4}{c|}}  
		\hline
		{Training sets} & {Models} & \multicolumn{2}{|c|}{9 Observations} \\ \hline
		&   & NMSE & CR  \\ \hline
		Source 9 & SECF  &0.5531 & 0.7125       \\ 
		          & MCF  &0.3898 & 0.8041    \\ \hline   
		Source 10 & SECF &0.5677 & 0.7297      \\ 
		          & MCF  &0.3712 & 0.8144       \\ \hline         
		Source 11 & SECF &0.5881& 0.7023     \\ 
	       	      & MCF  & 0.3771 & 0.8152  \\ \hline 
		Source 12 & SECF &0.5971 & 0.7082       \\ 
		          & MCF  &0.3862 & 0.8132      \\ \hline 
		Source 13 & SECF &0.5719 & 0.7203       \\ 
		          & MCF  &0.3833 & 0.8069     \\\hline
		Source 14 & SECF &0.5831 & 0.7019   \\ 
		          & MCF  &0.3794 & 0.8154   \\ \hline           
		Source 15 & SECF &0.5623 & 0.7206     \\ 
		          & MCF  &0.3853 & 0.8023       \\ \hline 
		Source 16 & SECF &0.5327 & 0.7241       \\ 
		          & MCF  &0.3780 & 0.8193      \\ \hline   
	\end{tabular}
\end{table}
Mat\'{e}rn covariance function also has better performance than the squared exponential covariance function on data set with high dimensional inputs, for all inputs $\x \in \mathbb{R}^B$, $B>1$. This improved performance is achieved due to the Mat\'{e}rn covariance function avoids a concentration of measure effect in high dimensions. An empirical performance comparison of the two models using 16 different training sets and observed by 9 sensors is illustrated in Tables \ref{table1} and \ref{table2}. From the results shown in Tables \ref{table1} and \ref{table2} we can observe that Mat\'{e}rn covariance function performances is better that the squared exponential. This implies that based on our training set the suitable model for our algorithm is the Mat\'{e}rn covariance function.

\subsection{Implementation of Prediction Criteria}
\label{sec:imple}
The main objective of this thesis is to develop low complexity algorithm to overcome the problems related with spatial reconstruction of electromagnetic fields from few sensor observations using Gaussian process. The task is to correctly estimate the spatial random electromagnetic fields at unobserved point locations, $\x^* \in \mb{X}$, from an observations collected by 9 sensors. The estimation of a spatial random physical process at any unobserved point location $\x^*$ is expressed by $\h^*=\h(\x^*)$. To measure the deviation between the estimated and true values, we used mean square error which is the most common approach and it is defined by:
\begin{equation}\label{loss_f}
L(\h,\h^*)=\E[(\h - \h^*)^2]
\end{equation}
The best prediction should be the optimal one that minimizes the loss function stated in equation \eqref{loss_f}. From the Gaussian process regression model discussed in section \ref{sec:GP} the optimal estimation that minimizes the deviation measure is the mean of the posterior predictive distribution of the observation, it is described as:
\begin{equation}\label{l_l}
\h^* = \E[p(\h^*|\mb{X}, \x^*, \Y)]   
\end{equation}
\begin{equation}\label{h_pred}
\h^* = \int  \h^* p(\h^*|\mb{X}, \x^*, \y) d\h^*.
\end{equation}
Therefore, the entire estimation process depends on the accurate computation of the posterior predictive distribution. Since our assumption for the likelihood function and prior distribution is Gaussian, computation of posterior predictive distribution involves solving many analytical tractable integrals. Using the concept of marginalization we can drive the posterior predictive distribution by integrating the conditional predictive prior distribution $p(\h^*| \h, \mb{X}, \x^*)$ over $p(\h| \mb{X}, \Y) $ the posterior distribution of our underlying spatial function at location of the sensors given the observation by the sensors and this is written as:
\begin{equation}
p(\h^*|\mb{X}, \x^*, \Y) = \int \hdots \int_{\Re^{N}} p(\h^*| \h,\x^* , \mb{X}) p(\h| \mb{X}, \Y) d\h.
\end{equation}
where $N$ is the number of sensors used for observation, in our case we used 9 sensors. In the next subsection we develop an algorithm that shows all the necessary procedures to compute the posterior predictive distribution.
\subsubsection*{Computation of Predictive Distribution}
In the proposed algorithm, the computation of the posterior predictive distribution is implemented in three main steps and a detail procedure is presented below:

1) First and foremost the conditional predictive prior distribution $p(\h^*| \h,\x^* , \mb{X})$ is computed. The considered Gaussian process prior is described in equation \eqref{fig:mean_fun}, the joint distribution between the out put of underlying function $\h$ and the output of function $\h^*$ at uncontrolled input location $\x^*$ is given by
   \begin{equation} \label{eg_joint_new}
   \left[ \begin{matrix} \h \\ \h^* \end{matrix} \right] | \mb{X},{\mb{X}^*},\boldsymbol{\theta},\boldsymbol{\beta}
   \sim \N\left(\left[  \begin{matrix} \boldsymbol{\mu} \\\boldsymbol{\mu}^* \end{matrix}\right], \left[ \begin{matrix} \K_1 & \K_2 \\ \K_3 & \K_4
   \end{matrix} \right] \right),
   \end{equation}
   where $\boldsymbol{\mu}=\mb{G}^{\tran} \boldsymbol{\bar{\Omega}}$ and  $\boldsymbol{\mu}^*=\mb{G}^{*\tran} \boldsymbol{\bar{\Omega}}$ are the considered mean functions for $\h$ and $\h^*$ respectively, $\mb{G}=[g_1(\x), g_2(\x),...,g_K(\x)]^{\tran}$ are the fixed basis functions, $\boldsymbol{{\Omega}} \sim \N(\mb{a,A})$ is the weighting parameter, $\boldsymbol{\theta}=[\mb{a},\mb{A}]$ and $\boldsymbol{\beta=[\sigma_{\textrm{m}}^2,\ell_1,\ell_2]}$ are the hyperparameters of the mean and covariance functions respectively. 
    \begin{align}
   \K_1 &= \K_{\x,\x} + \mb{G}^{\tran}\mb{A}\mb{G} \hspace{1cm}
   \K_2 = \K_{\x,\x^*} + \mb{G}^{\tran}\mb{A} \mb{G}^* ,\nonumber \\
   \K_3 &= \K_{\x^*,\x} + \mb{G}^{*\tran}\mb{A} \mb{G}  \hspace{0.75cm}
   \K_4 = \K_{\x^*,\x^*} + \mb{G}^{*\tran}\mb{A} \mb{G}^* .\nonumber 
   \end{align}
  Using Gaussian identities for conditioning and marginalizing, the conditional predictive prior distribution is expressed as:
  \begin{equation}\label{cond_pre_prior}
 p(\h^*| \h,\x^* , \mb{X},\boldsymbol{\theta},\boldsymbol{\beta}) = \N(\mb{m}_{h^*|h},\mb{Cov}_{h^*|h})
   \end{equation}
  where 
  \begin{align}
  \mb{m}_{h^*|h}&= \mb{G}^{*\tran} \boldsymbol{\bar{\Omega}} + \K_{\x^*,\x}^{\tran}\K_{\x,\x}^{-1}(\h- \mb{G}^{\tran} \boldsymbol{\bar{\Omega}}) \nonumber \\
  \mb{Cov}_{h^*|h} &= \K_{\x^*,\x^*} - \K_{\x^*,\x}\K_{\x,\x}^{-1}\K_{\x,\x^*} + \K' \nonumber \\
  \K'&=\mb{R}^{\tran}(\mb{A}^{-1} +\mb{G}\K_{\x,\x}^{-1}\mb{G}^{\tran}) \mb{R} \nonumber\\
  \mb{R}&=\mb{G^*} - \mb{G}\K_{\x,\x}^{-1}\K_{\x^*,\x}^* \nonumber
  \end{align}
  The properties of conditional Normal distribution is used to drive equation \eqref{cond_pre_prior} which originated from the Gaussian process prior assumption on $\h$. Therefore, the conditional predictive prior distribution  $p(\h^*| \h,\x^* , \mb{X},\boldsymbol{\theta},\boldsymbol{\beta})$  is a multivariate Normal distribution. Now in the next step we will compute the conditional posterior distribution of the latent function $p(\h| \mb{X}, \Y,\boldsymbol{\theta}, \boldsymbol{\beta})$. 
  
2) In the second step, we drive the conditional posterior distribution of our underlying spatial function $p(\h| \mb{X}, \Y,\boldsymbol{\theta}, \boldsymbol{\beta})$ at location of the sensors given observations then we marginalize over the hyperparameters. By applying Bayes law, we have:

 \small
 \begin{equation}\label{eq_posterior_of_f}
 p(\h|\mb{X}, \Y, \boldsymbol{\theta}, \boldsymbol{\beta}) =
 \frac{p(\Y|\h,\mb{X},\sigma_{\eta}^2)p(\h|\mb{X},\boldsymbol{\theta},\boldsymbol{\beta})}{p(\Y|\mb{X},\boldsymbol{\theta}, \boldsymbol{\beta})},
 \end{equation}
 \normalsize
where \small{$p(\Y|\mb{X},\boldsymbol{\theta}, \boldsymbol{\beta})=\int p(\Y|\h,\mb{X},\sigma_{\eta}^2)p(\h|,\mb{X},\boldsymbol{\theta},\boldsymbol{\beta}) d\h$} \normalsize is the marginal likelihood. Since both the likelihood function $p(\Y|\h,\mb{X},\sigma_{\eta}^2)$ and our prior $p(\h|,\mb{X},\boldsymbol{\theta},\boldsymbol{\beta})$ are Gaussian distributed, the marginal likelihood is also Gaussian which is described in equation \eqref{eq_log_ml_new}.

Therefore, substituting equation \eqref{eq_log_ml_new} in the denominator of equation \eqref{eq_posterior_of_f} gives a Gaussian distributed conditional posterior of the underlying spatial function $\h$:
 \begin{equation}\label{post_latent}
 p(\h|\mb{X}, \Y, \boldsymbol{\theta}, \boldsymbol{\beta})= \N(\mb{m}_{h},\mb{Cov}_{h}),
 \end{equation}

\small
\begin{align}
 \mb{m}_{\h}&= \mb{G}^{\tran} \boldsymbol{\bar{\Omega}} + \Khh\K^{-1}(\Y- \mb{G}^{\tran} \boldsymbol{\bar{\Omega}}) \nonumber , \\
 \mb{Cov}_{\h} &= \K_{\x,\x} - \K_{\x,\x}\K^{-1}\K_{\x,\x} + \mb{R}^{\tran}(\mb{A}^{-1} +\mb{G}\K^{-1}\mb{G}^{\tran}) \mb{R} \nonumber.
  \end{align}
 \begin{align}
 \boldsymbol{\bar{\Omega}} &=(\mb{A}^{-1} + \mb{G}\K^{-1}\mb{G}^{\tran})^{-1} (\mb{A}^{-1} \mb{a} + \mb{G}\K^{-1}\Y) \nonumber, \\
   \mb{R}&=\mb{G^*} - \mb{G}\K^{-1}\K_{\x^*,\x}^* \nonumber \\ 
 \K&=\Khh + \sigma_{\eta}^2\mb{I} \nonumber 
 \end{align}
\normalsize
To marginalize the conditional posterior of the underlying spatial function over the hyperparameters, we need to integrate $p(\h|\mb{X}, \Y, \boldsymbol{\theta}, \boldsymbol{\beta})$ over the posterior of the hyperparameter $p(\boldsymbol{\theta}, \boldsymbol{\beta}|\mb{X},\Y)$,
\begin{equation}\label{marginal}
p(\h|\mb{X}, \Y) = \int p(\h|\mb{X}, \Y, \boldsymbol{\theta}, \boldsymbol{\beta}) p(\boldsymbol{\theta}, \boldsymbol{\beta}|\mb{X}, \Y) d\boldsymbol{\theta} d \boldsymbol{\beta}.
\end{equation}
\normalsize
\begin{equation}
p(\boldsymbol{\theta}, \boldsymbol{\beta}|\Y, \mb{X}) \propto p(\Y| \mb{X}, \boldsymbol{\theta}, \boldsymbol{\beta}) p(\boldsymbol{\theta}, \boldsymbol{\beta}) \nonumber
\end{equation}
Solving the posterior distribution of the hyperparameter is complex due to the lack of prior information. Therefore, the integration in equation \eqref{marginal} is analytical intractable.
The marginal latent function is approximated using maximum a posterior (MAP) estimate by setting a joint prior distribution $ p(\boldsymbol{\theta}, \boldsymbol{\beta}) $ for the hyperparameters. The MAP estimation is obtained by:

\footnotesize
\begin{align}
\{ \hat{\boldsymbol{\beta}}_{MAP}, \hat{\boldsymbol{\theta}}_{MAP} \}&= \argmax_{\boldsymbol{\beta,\theta}} p(\boldsymbol{\theta}, \boldsymbol{\beta}|\Y, \mb{X}) \nonumber \\
  &= \argmax_{\boldsymbol{\beta,\theta}} p(\Y| \mb{X}, \boldsymbol{\theta}, \boldsymbol{\beta}) p(\boldsymbol{\theta}, \boldsymbol{\beta}) \nonumber \\
  &=\argmax_{\boldsymbol{\beta,\theta}} \Big(\int p(\Y|\h,\mb{X})p(\h|,\mb{X},\boldsymbol{\theta},\boldsymbol{\beta}) d\h  \Big) p(\boldsymbol{\theta}, \boldsymbol{\beta})
\end{align}
\normalsize
The marginal likelihood is computed in equation \eqref{eq_log_ml_new} and we consider a Gaussian distributed prior for our hyperparameters. Therefore, the solution for the optimization problem is:

\footnotesize
\begin{align}
\{ \hat{\boldsymbol{\beta}}_{MAP}, \hat{\boldsymbol{\theta}}_{MAP} \} &=\argmax_{\boldsymbol{\beta, \theta}} \Big(\int p(\Y|\h,\mb{X})p(\h|,\mb{X},\boldsymbol{\theta},\boldsymbol{\beta}) d\h  \Big) p(\boldsymbol{\theta}, \boldsymbol{\beta}) \nonumber \\
&= \argmax_{\boldsymbol{\beta,\theta}} \Big[ \log  p(\Y| \mb{X}, \boldsymbol{\theta}, \boldsymbol{\beta}) + \log p(\boldsymbol{\theta}, \boldsymbol{\beta}) \Big]  \nonumber \\ 
&= \argmin_{\boldsymbol{\beta,\theta}} \Big[ -\log  p(\Y| \mb{X}, \boldsymbol{\theta}, \boldsymbol{\beta}) - \log p(\boldsymbol{\theta}, \boldsymbol{\beta}) \Big] 
\end{align}
\normalsize
We now  approximate the marginal posterior of the latent function using the optimized hyperparameters as follows:
\begin{align}
p(\h|\mb{X}, \Y) &= \int p(\h|\mb{X}, \Y, \boldsymbol{\theta}, \boldsymbol{\beta}) p(\boldsymbol{\theta}, \boldsymbol{\beta}|\Y, \mb{X}) d\boldsymbol{\theta} d \boldsymbol{\beta}. \nonumber \\
 p(\h|\mb{X}, \Y) & \approx p(\h|\mb{X}, \Y, \boldsymbol{\hat{\theta}}, \boldsymbol{\hat{\beta}}).
\end{align}
We also use the optimized hyperparameters to marginalize the conditional predictive prior distribution over the hyperparameters as: 
\begin{align}\label{mar_pre_prior}
p(\h^*| \h,\x^* , \mb{X}) &= \int p(\h^*| \h,\x^* , \mb{X}, \boldsymbol{\theta}, \boldsymbol{\beta}) p(\boldsymbol{\theta}, \boldsymbol{\beta}|\Y, \mb{X}) d\boldsymbol{\theta} d \boldsymbol{\beta}, \nonumber \\
p(\h^*| \h,\x^* , \mb{X}) & \approx p(\h^*| \h,\x^* , \mb{X},\boldsymbol{\hat{\theta}},\boldsymbol{\hat{\beta}}).
\end{align}
So far we have computed the marginal posterior of the latent function $p(\h|\mb{X}, \Y)$ and the marginal predictive prior distribution $p(\h^*| \h,\x^* , \mb{X})$. To complete the algorithm computation we will combine these results to compute the posterior predictive distribution $p(\h^*|\mb{X}, \x^*, \Y) $.

3) Finally, we compute the posterior predictive distribution $p(\h^*|\mb{X}, \x^*, \Y) $ by integrating the computed conditional predictive prior distribution  over the marginal posterior of the latent function as follows
\begin{align}
p(\h^*|\mb{X}, \x^*, \Y) &= \int  p(\h^*| \h,\x^* , \mb{X}) p(\h| \mb{X}, \Y) d\h, \nonumber \\
 p(\h^*|\mb{X}, \x^*, \Y) & \approx  \int \N(\mb{m}_{h^*|h},\mb{Cov}_{h^*|h}) \N(\mb{m}_{h},\mb{Cov}_{h}) d\h, \nonumber\\
 p(\h^*|\mb{X}, \x^*, \Y) & = \N(\h^*; \mb{m}^p_{h^*},\mb{Cov}^p_{h^*}) .               
\end{align}
where 	
	\begin{align}
	\mb{m}^p_{h^*}&=\mb{G}^{*\tran}\bar{\Omega} +\Kah^{\tran}\K^{-1}(\Y - \mb{G}^{\tran}\bar{\Omega}) \\
\mb{Cov}^p_{h^*}&=  \Kaa - \Kah\K^{-1}\Kha + \mb{R}^{\tran}(\mb{A}^{-1} +\mb{G}\K^{-1}\mb{G}^{\tran}) \mb{R} \\
	\nonumber \\
	\bar{\Omega}&=(\mb{A}^{-1} + \mb{G}\K^{-1}\mb{G}^{\tran})^{-1} (\mb{A}^{-1} \mb{a} + \mb{G}\K^{-1}\Y) \nonumber \\
	\mb{R}&=\mb{G^*} - \mb{G}\K^{-1}\K^*  \\ \nonumber
	\K&=\Khh + \sigma_{\eta}^2\mb{I} \nonumber
	\end{align}

	Now we have complete the all the necessary computation and we define the predictive mean that minimizes the loss function in equation \eqref{l_l} by computing the predictive mean using equation \eqref{h_pred} and it is given by
	\begin{align}
	\h^* &= \int  \h^* p(\h^*|\mb{X}, \x^*, \Y) d\h^*, \nonumber \\
	&= \int \h^* \N(\h^*; \mb{m}^p_{h^*},\mb{Cov}^p_{h^*}) d\h^*, \nonumber\\
	&=  \mb{G}^{*\tran}\bar{\Omega} + \Kah^{\tran}\K^{-1}(\Y -\mb{G}^{\tran}\bar{\Omega} ) \nonumber
	\end{align}

\section{Result and Discussion}
\label{cha:4}
In our study we use the training set from SIMUEM3D, the tool is available for research in IRCICA and takes all the exact consideration as sensors to make the observation. The observation made by the tool have similar values with the sensor measurements. The real measured data by sensors is complete and provides features like the location of the observations on Cartesian grid location and the intensity levels at the observation location. We selected only two source positions to analyze and calibrate the algorithm. These are: source 2 positioned at (4,4), and source 5 positioned at (2.5,2.5), but for the performance comparison we will focus on source 2. Now, we will first make the adjustment of our model setting then we will make the estimation of the electromagnetic field intensity by applying different system configurations and we will compare the results.
\subsection{Proposed Model Adjustment EM field Intensity}
We made model adjustment by fitting the hyperparameters of our model using the Bayesian or marginal likelihood approach which use maximum a posterior (MAP) estimation to estimate the hyperparameters, as we described on section \ref{sec:modelsele} the detail procedure to estimate Gaussian process hyperparameters. In subsection \ref{cov_new} we illustrate the selection of suitable covariance function, and we shaw that we select a 2-D Mat\'{e}r covariance function with $\nu=3/2$. It is given as:%

\small
\begin{align}\label{matern3/2}
k_{\nu=3/2}(\x_i,\x_j)& =\sigma_{\mathrm{m}}^2\left(1+\frac{\sqrt{3}|\x_i -\x_j|}{\ell_1}\right)
\exp\left(-\frac{\sqrt{3}|\x_i -\x_j|}{\ell_2}\right), 
\end{align}
\normalsize
The smoothness of the covariance function is controlled by the fixed parameter $\nu=3/2$, this is suitable value $\nu$  and the MAP estimation of our hyperparameters the length scale $\ell$ and the magnitude scale $\sigma^2_m$ parameters is given in Table \ref{table3}. we will use the optimized hyperparameters to estimate the electromagnetic field intensity.
\begin{table}
	\caption{MAP estimation of the hyperparameters of covariance function}
	\label{table3} 
	\centering
	\resizebox{\columnwidth}{!} {
	\begin{tabular}{| >{\centering\arraybackslash}m{0.8in} | >{\centering\arraybackslash}m{0.8in} | >{\centering\arraybackslash}m{0.8in} | >{\centering\arraybackslash}m{0.8in} |N}
		\hline
		& \multicolumn{3}{|c|}{{ $\beta_{MAP}$}} &\\ \hline
		\textbf{Dataset}    & {$\sigma^2_m$} & {$\ell_1$} & {$\ell_2$}  &\\  [10pt]\hline
	    Source 2 &  5.9420  &   1.418  & 1.2908 &\\ [10pt]\hline   
		Source 5 &  5.8637  &   1.3157 & 1.3156 &\\[10pt] \hline   
	\end{tabular}}
\end{table}
  
\subsection{Estimation of Electromagnetic Field Intensity}
\begin{figure*}[!htb]
\centering
\subfloat[True value]{\includegraphics[scale=0.4]{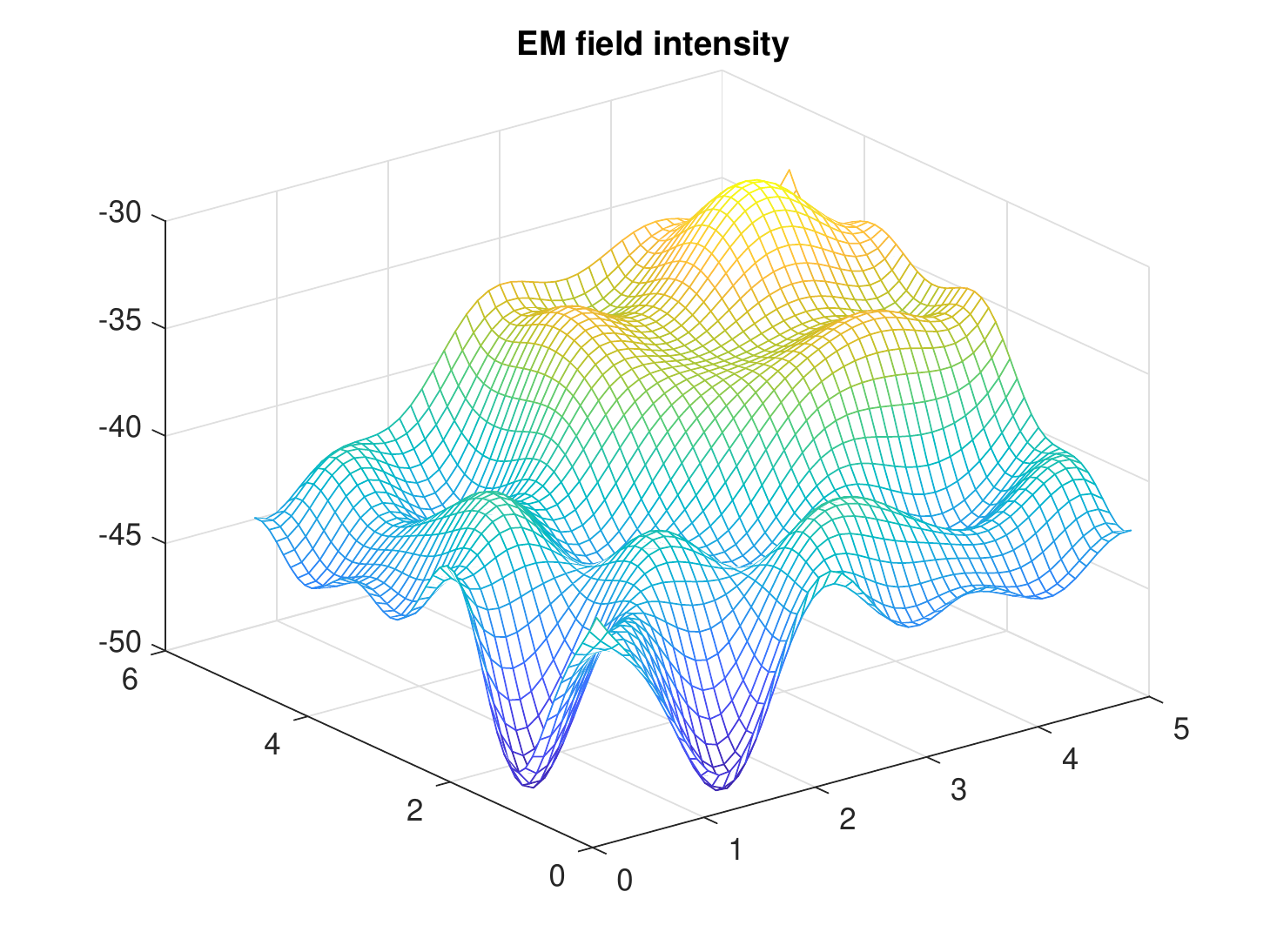}\label{fig2aa}}
\subfloat[Estimated]{\includegraphics[scale=0.4]{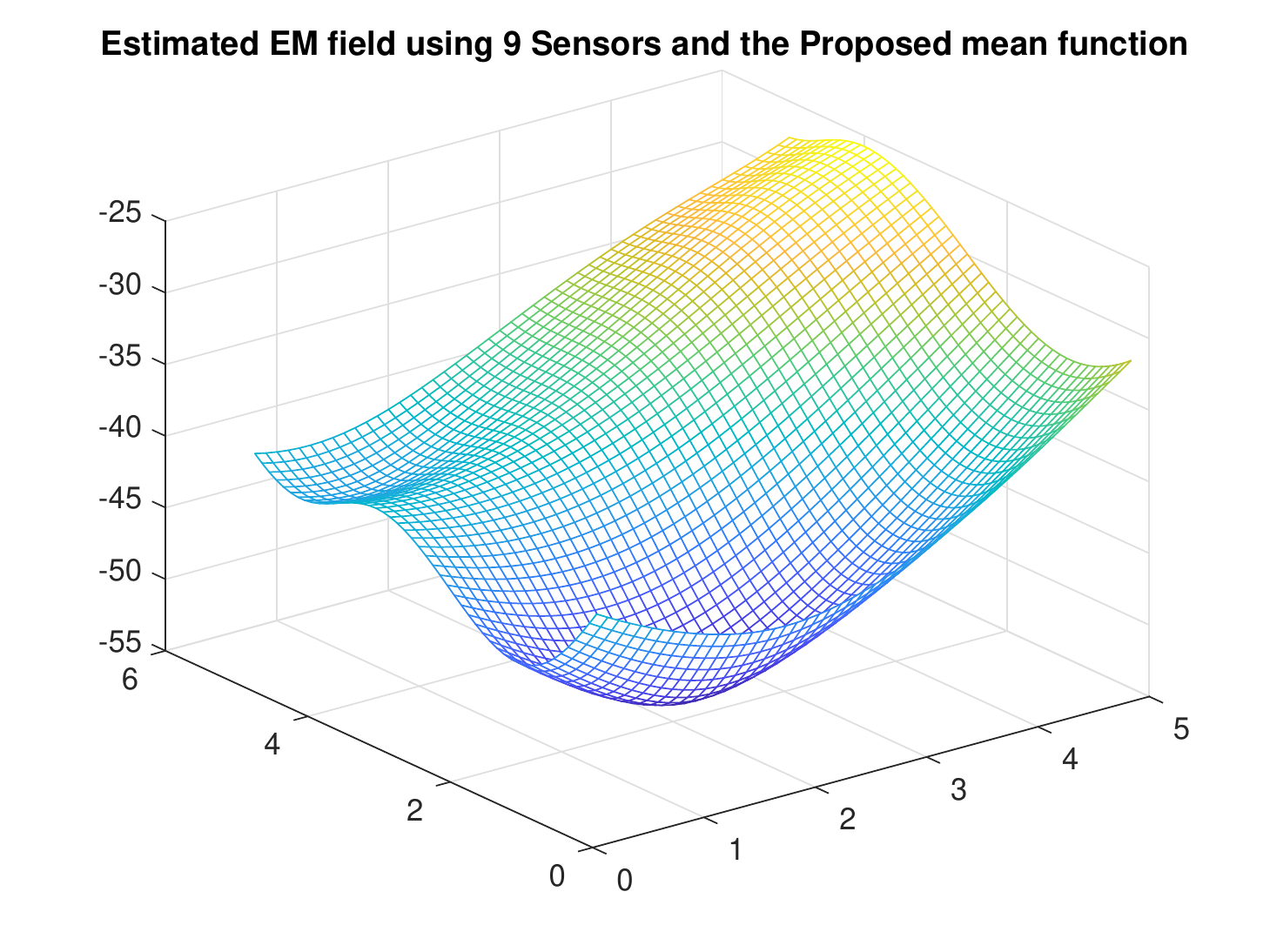}\label{fig2a}}
\subfloat[Difference]{\includegraphics[scale=0.4]{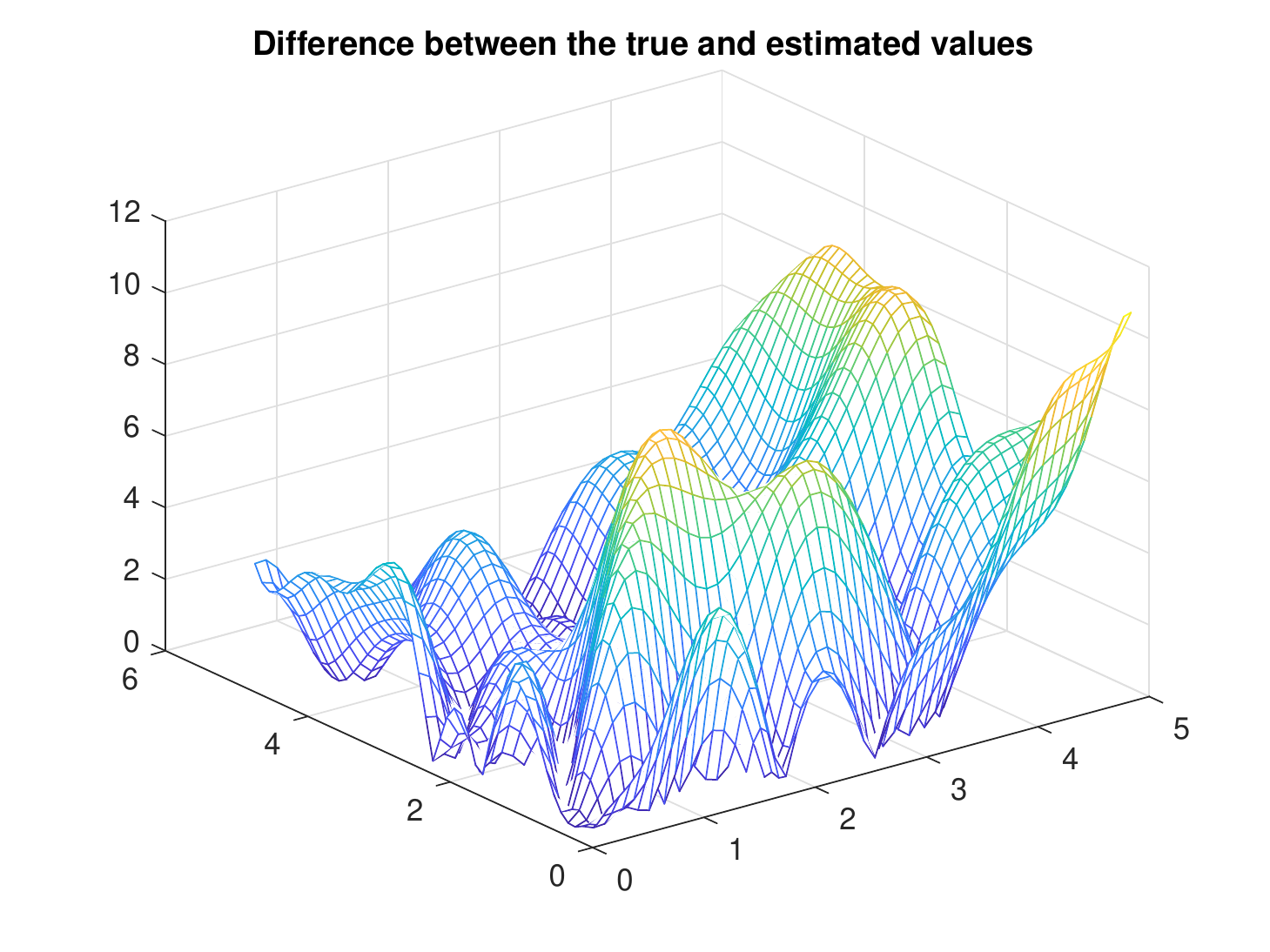}\label{fig2b}}
\caption{EM Field Reconstruction using the Proposed Algorithm and 9 observation points.} 
\label{fig:diff_9sensor}
\end{figure*}

\begin{figure*}[!htb]
\centering
\subfloat[True value]{\includegraphics[scale=0.4]{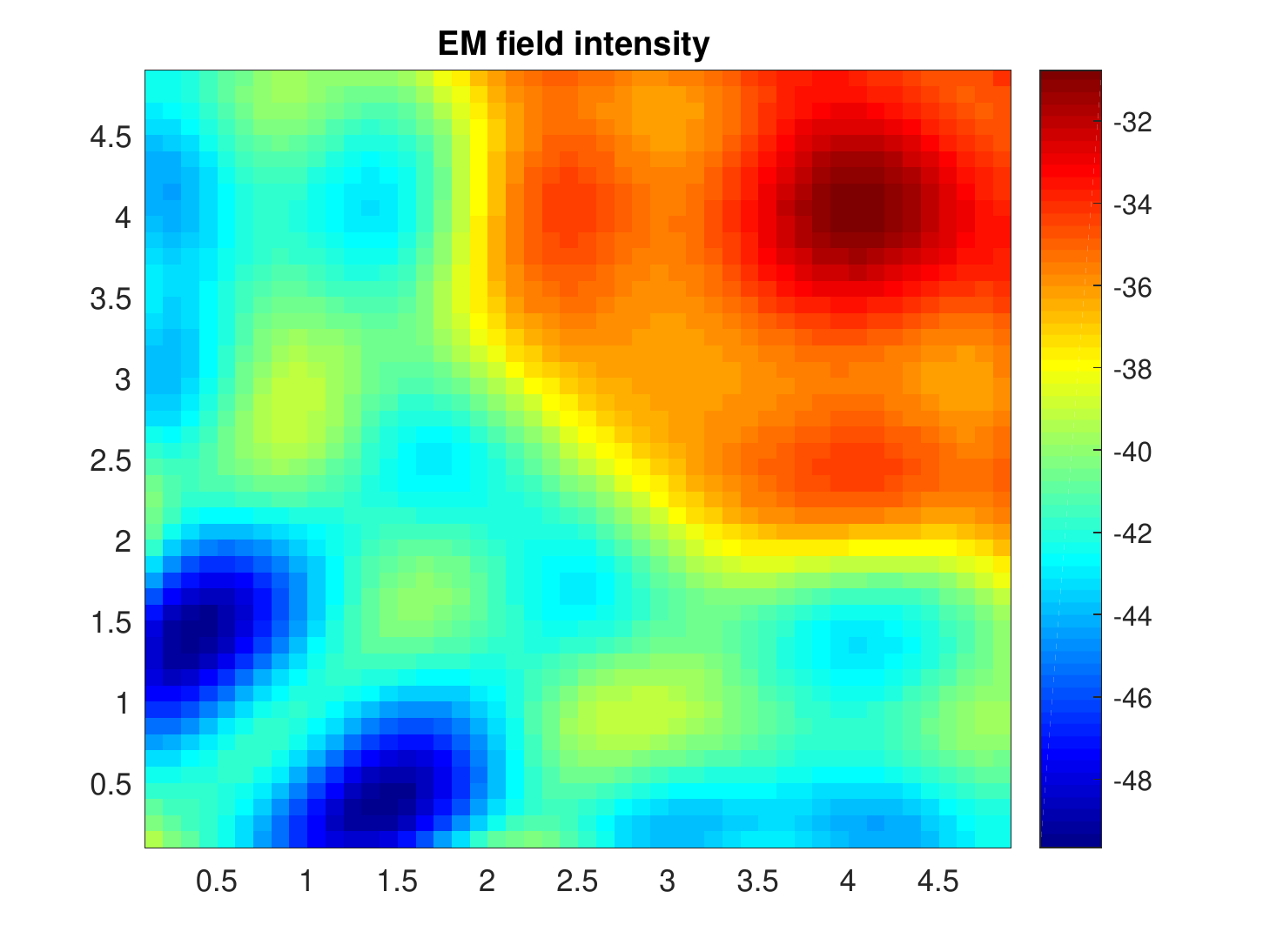}\label{fig3a}}
\subfloat[Estimated]{\includegraphics[scale=0.4]{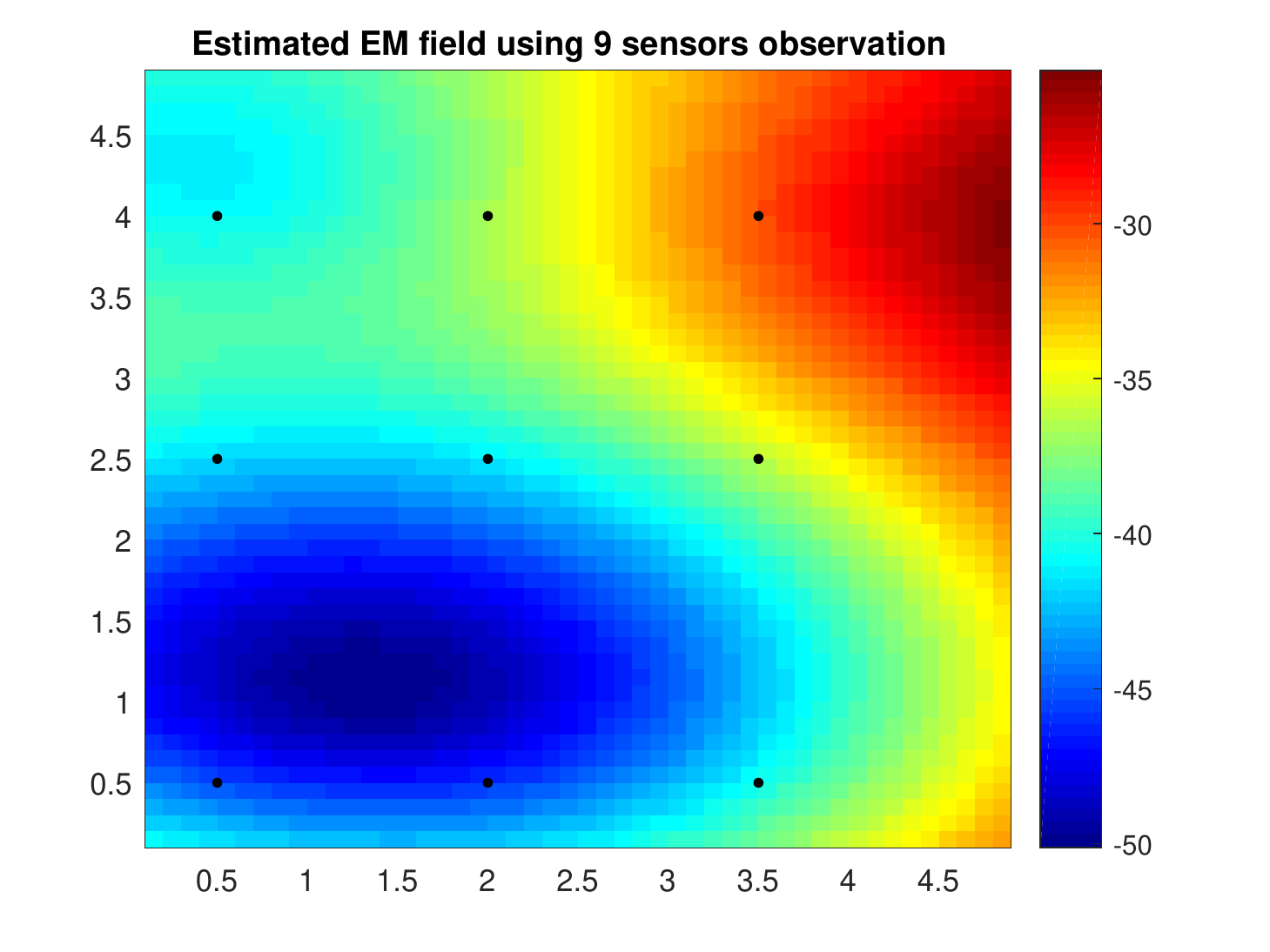}\label{fig3b}}
\subfloat[Difference]{\includegraphics[scale=0.4]{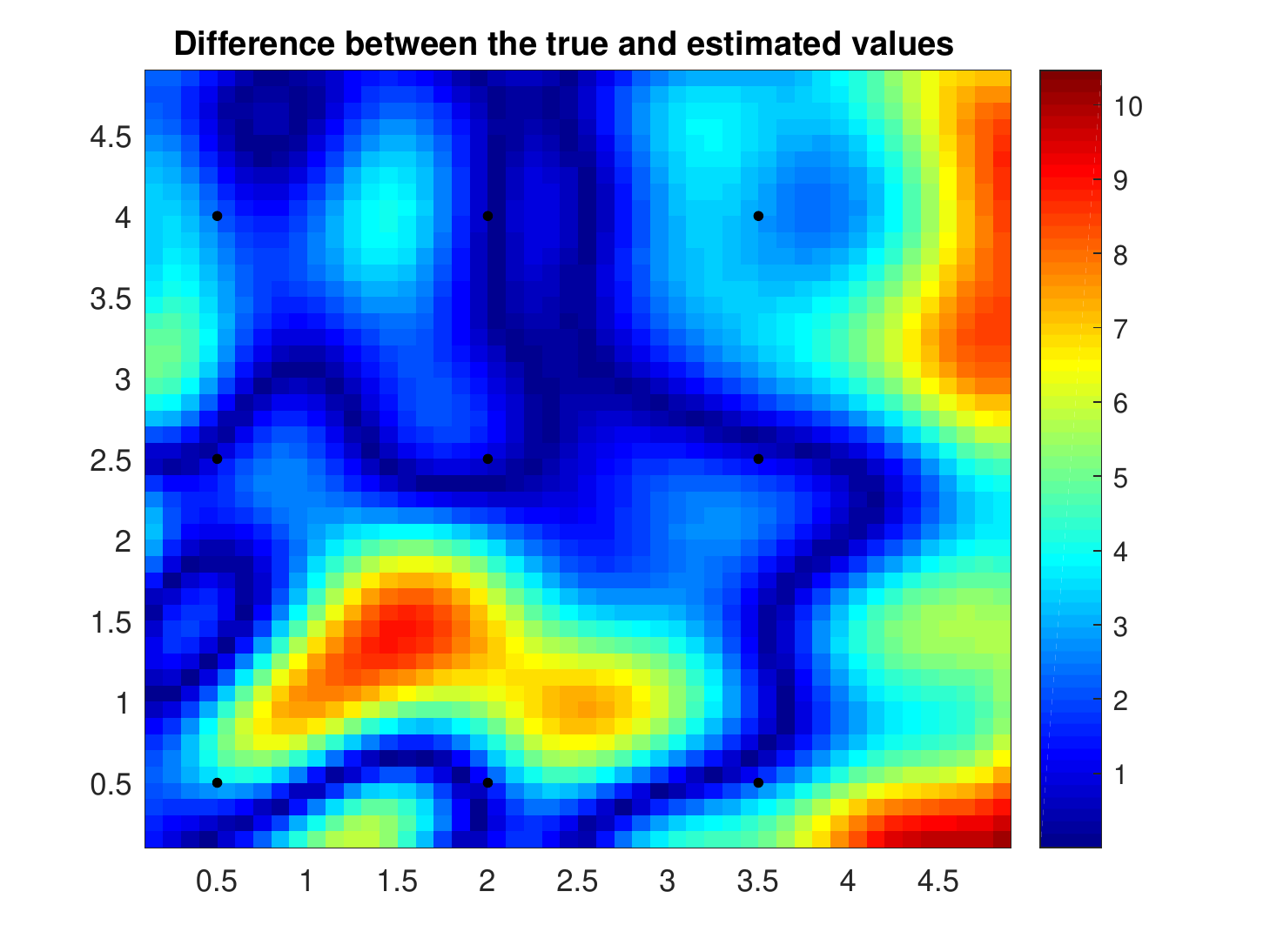}\label{fig3c}}
\caption{Pseudocolor plot of EM Field Reconstruction using the Proposed Algorithm and 9 observation points.} 
\label{fig:pplot_9sensor}
\end{figure*}

 We implement the electromagnetic field intensity estimation using our algorithm for different system configuration and compare the results. Reconstruct the electromagnetic field using the selected covariance and mean function. First of all we performed electromagnetic field intensity estimation using only 9 observation points and the proposed mean and covariance function, then we compare the results to the case where the mean function is zero mean. The results shown in figures \ref{fig:diff_9sensor} and \ref{fig:pplot_9sensor} are the electromagnetic field reconstruction using the proposed mean and covariance function for the electromagnetic field intensities in dB emitted by source 2. The actual data of the electromagnetic field intensities [dB] is shown in figure \ref{fig2aa}. The estimated electromagnetic field intensities based on only 9 sensors is shown in figure \ref{fig2a}, the spatial EM field reconstruction contains total 2401 spatial points. The performance measure is based on the normalized mean square error (NMSE) and we achieve $NMSE=0.3763$. 
 
The figure shows that a good estimation of  electromagnetic field intensities is achieved using only 9 sensors in all the 2401 spatial points. The Pseudocolor plot of the true,  the spatial reconstructed and the difference between these values EM field intensities is shown in figures \ref{fig3a}, \ref{fig3b} and \ref{fig3c}, respectively. The results shown in figures \ref{fig3c} up-to \ref{fig_meanVSzeromean_pplot} are the comparison we made between the estimation performance of the proposed algorithm and to the case where the mean function is zero mean. We compare the performance of the two cases by computing the normalize mean square error of the estimation and the correlation between the true and estimated electromagnetic field intensities for each cases, and this is given in Table \ref{table4}.
\begin{figure*}[!htb]
\centering
\subfloat[Proposed Mean]{\includegraphics[scale=0.3]{Re_Pred_Mean9sensors-eps-converted-to.pdf}\label{fig4a}}
\subfloat[Zero Mean]{\includegraphics[scale=0.3]{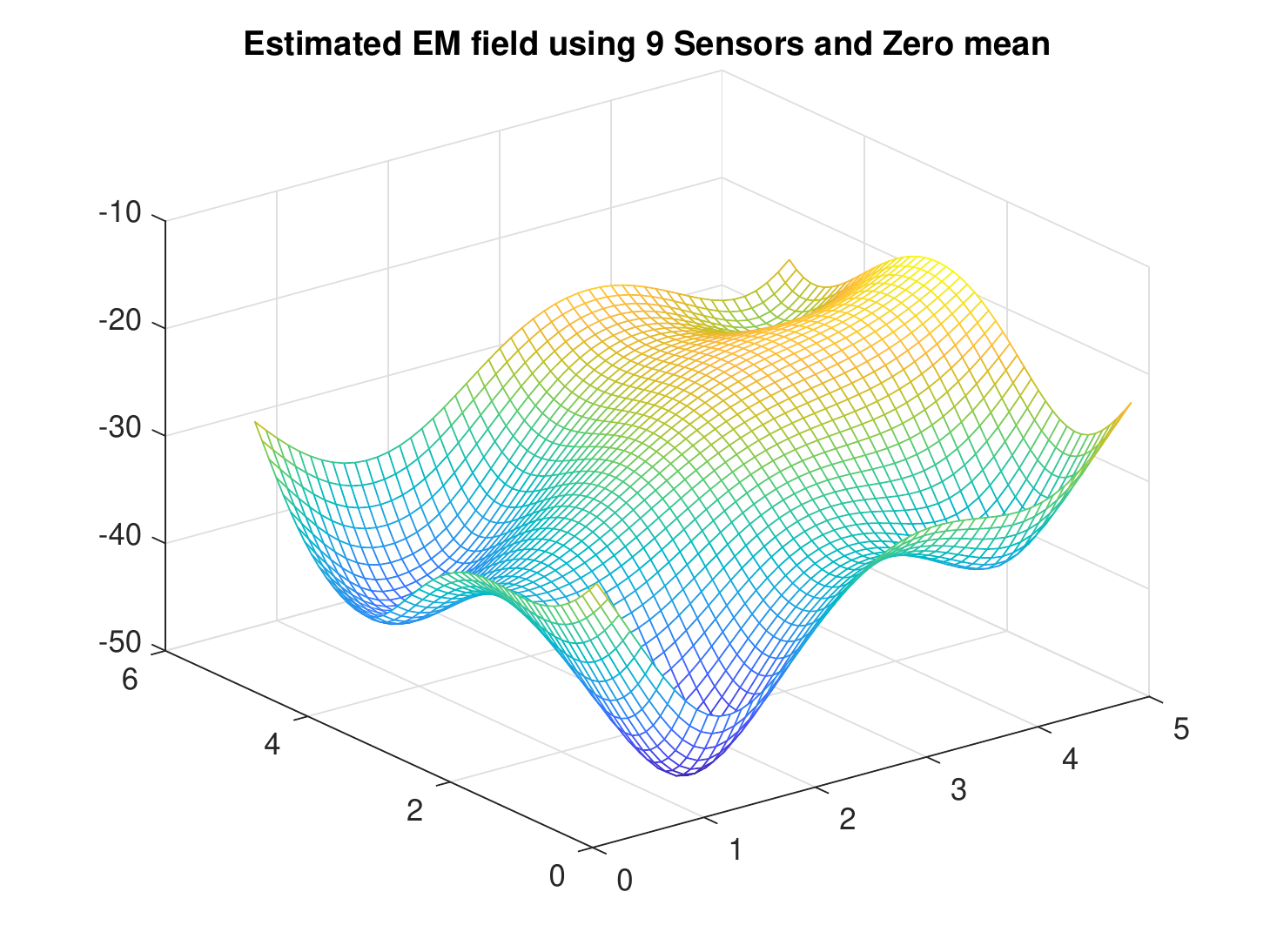}\label{fig4b}}
\subfloat[Difference Proposed Mean]{\includegraphics[scale=0.3]{Re_DiffMean-eps-converted-to.pdf}\label{fig5a}}
\subfloat[Difference Zero Mean]{\includegraphics[scale=0.3]{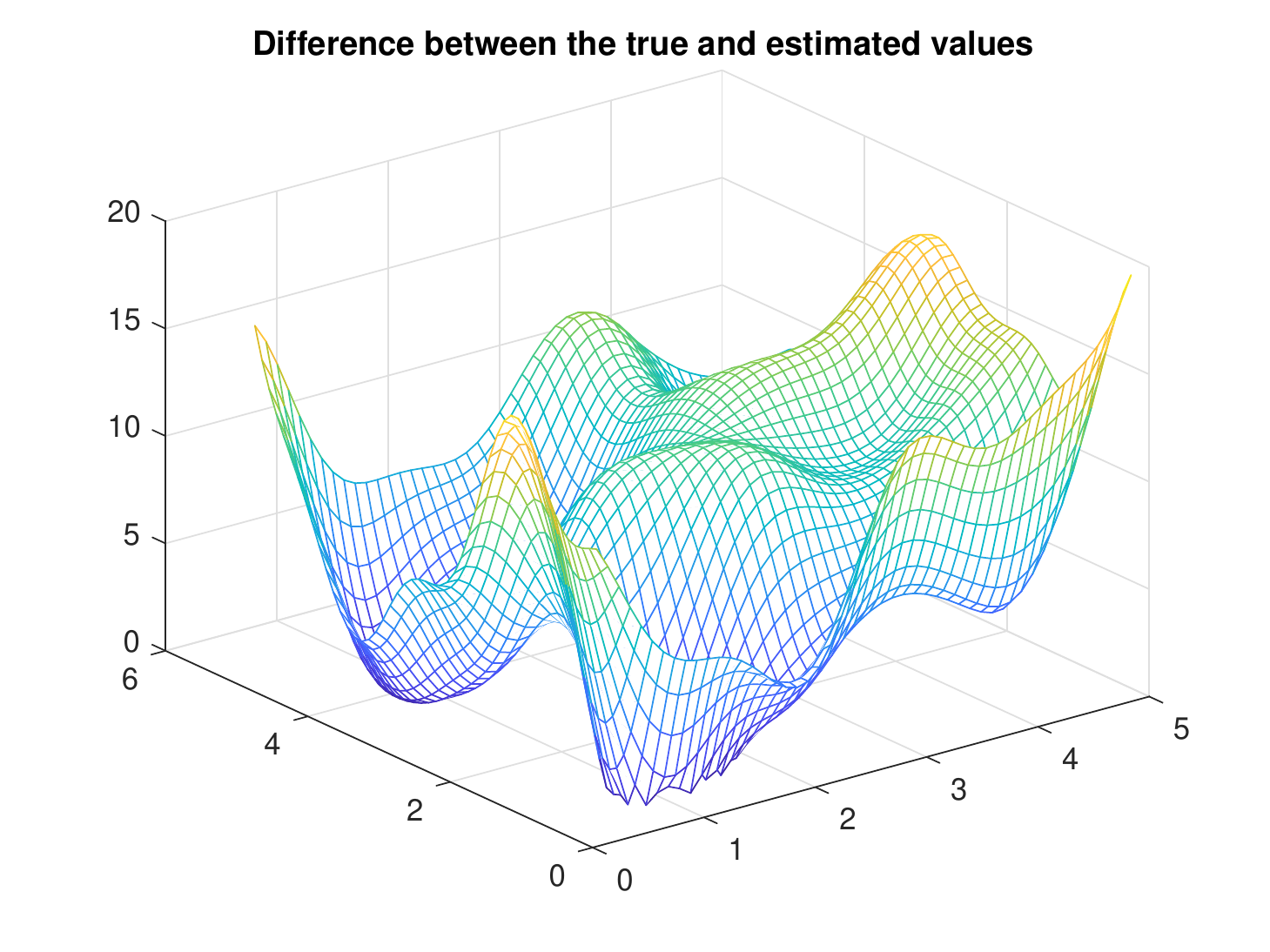}\label{fig5b}}
\caption{Comparison of EM Field Reconstruction based on 9 sensors using the Proposed Mean function and Zero Mean.} 
\label{fig_meanVSzeromean}
\end{figure*}
\begin{figure*}[!htb]
\centering
\subfloat[Proposed Mean]{\includegraphics[scale=0.3]{Re_P_Pred_Mean9sensors-eps-converted-to.pdf}\label{fig6a}}
\subfloat[Zero Mean]{\includegraphics[scale=0.3]{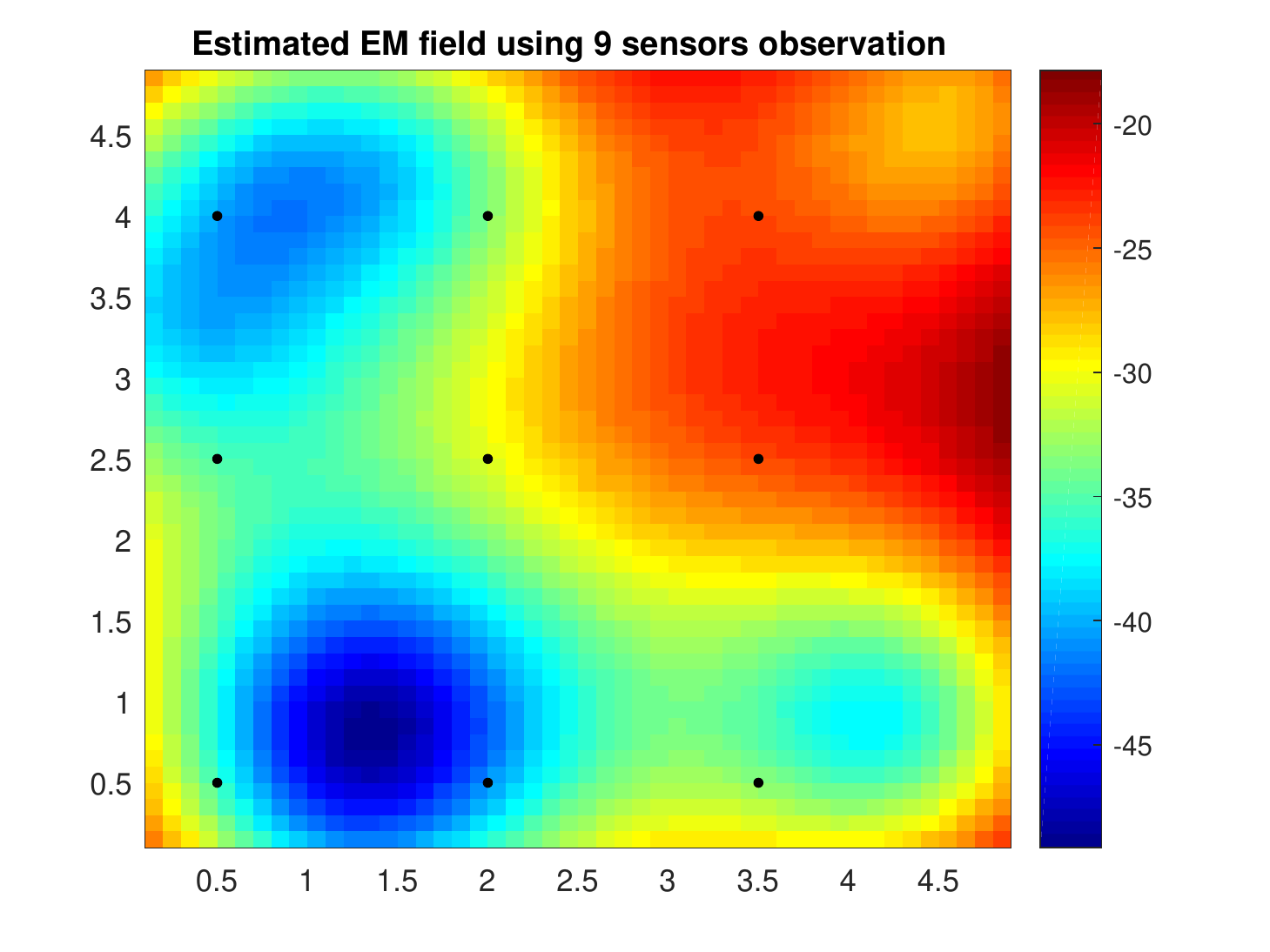}\label{fig6b}}
\subfloat[Difference Proposed Mean]{\includegraphics[scale=0.3]{Re_P_DiffMean-eps-converted-to.pdf}\label{fig7a}}
\subfloat[Difference Zero Mean]{\includegraphics[scale=0.3]{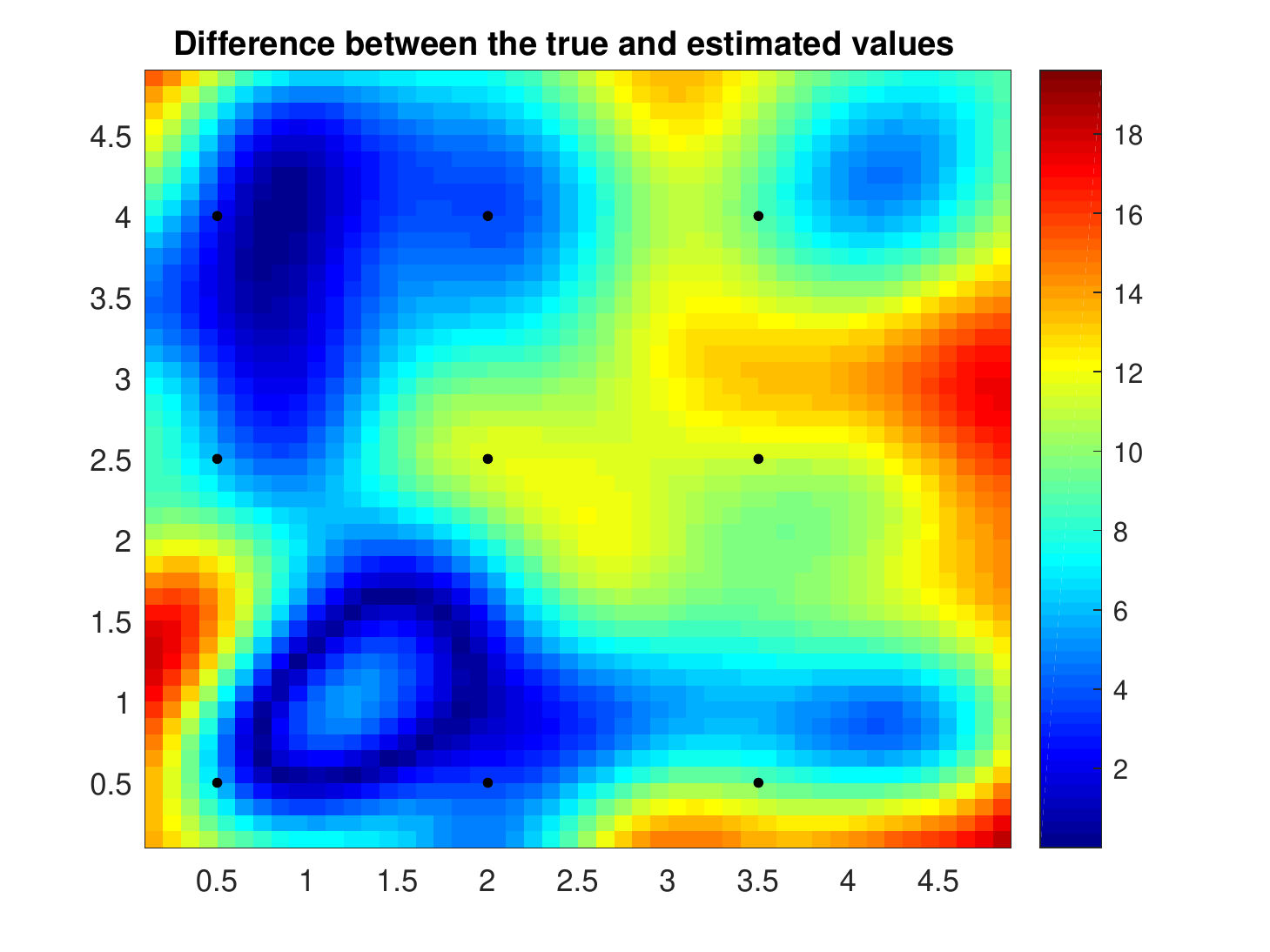}\label{fig7b}}
\caption{Comparison of Pseudocolor EM Field Reconstruction based on 9 sensors using the Proposed Mean function and Zero Mean.} 
\label{fig_meanVSzeromean_pplot}
\end{figure*}
 \begin{table}
 	\centering
 	\caption{Comparison between Spatial Reconstruction of EM field Using Proposed Mean Function and Zero Mean.    }
 	\label{table4} 
 	\resizebox{\columnwidth}{!} {
 	 	\begin{tabular}{| >{\centering\arraybackslash}m{0.8in} || >{\centering\arraybackslash}m{0.8in} || >{\centering\arraybackslash}m{0.8in} |N}
 		\hline
 		\textbf{Criteria}    & \textbf{With Proposed Mean Function} & \textbf{With Zero Mean}  & \\  [10pt]\hline
 		NMSE &  0.3763  & 0.4780      &\\  [10pt]\hline
 		Correlation &  0.8141  &   0.7781   &\\ [10pt]\hline   
 		\end{tabular}}
 \end{table}
 As it is shown in Table \ref{table4} the estimation performance of electromagnetic field intensities using the proposed mean function is better than the case where the mean is zero, and this is because of the included mean function that is the prior information about the behavior of electromagnetic field propagation in the considered room. We have used 2401 training points to train our algorithm and we also trained our algorithm using different number of training points to observe and compare the effect of the number of training samples on the prediction performance. In this case we vary the number of training points to set the hyperparameters and we performed the prediction based on only 9 sensors, then compare the results. 
\begin{figure*}[!htb]
\centering
\subfloat[True value]{\includegraphics[scale=0.3]{Re_TrueValue-eps-converted-to.pdf}\label{fig111}}
\subfloat[Estimated with 9 sensors]{\includegraphics[scale=0.3]{Re_Pred_Mean9sensors-eps-converted-to.pdf}\label{fig100}}
\subfloat[Estimated with 30 sensors]{\includegraphics[scale=0.3]{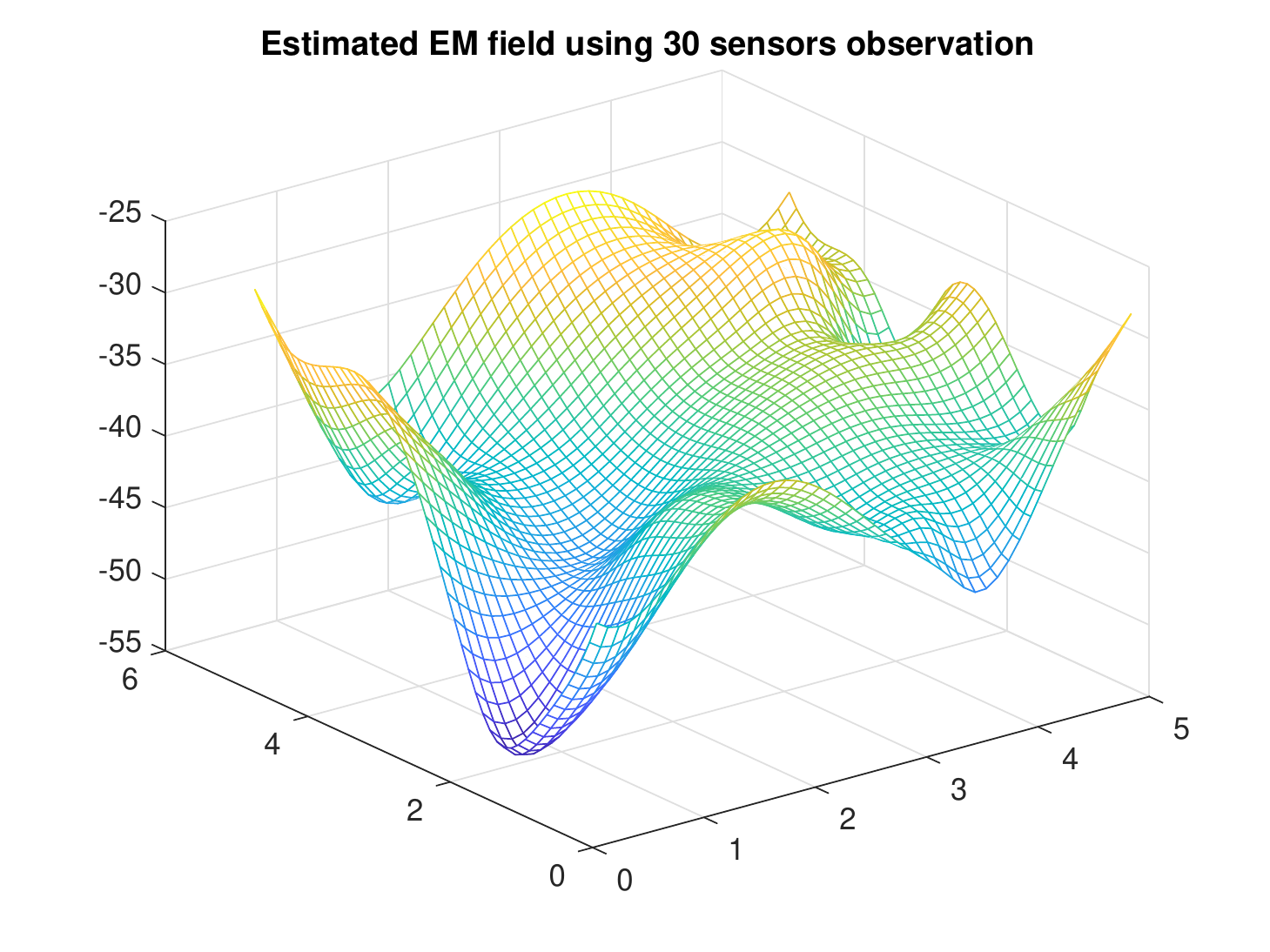}\label{fig8a}}
\subfloat[Estimated with 100 sensors]{\includegraphics[scale=0.3]{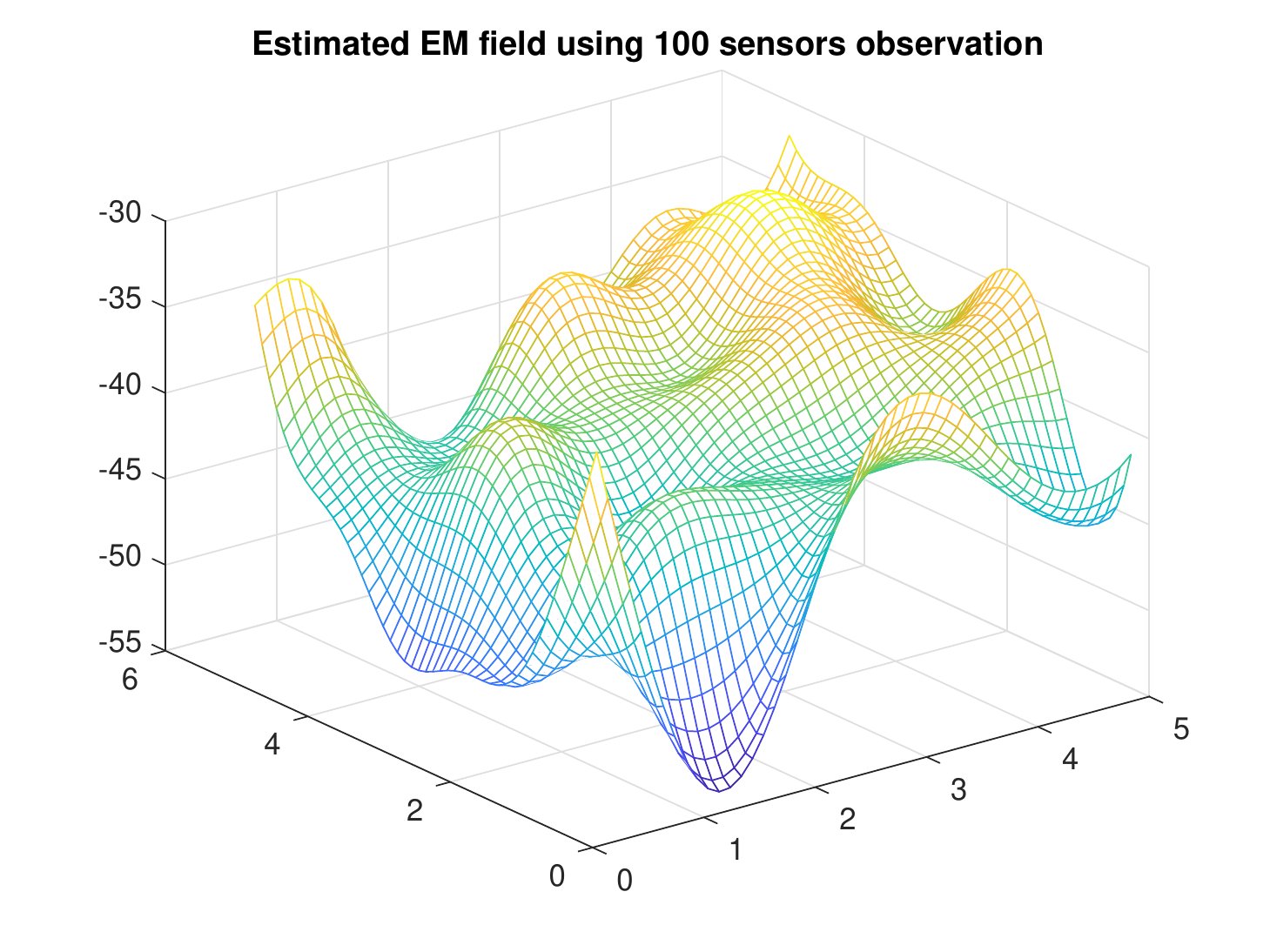}\label{fig8b}}
\caption{Spatial Reconstruction of EM field intensities based on 9, 30 and 100 sensors observation points.} \label{varysensors1}
\end{figure*}
\begin{figure*}[!htb]
\centering
\subfloat[True value]{\includegraphics[scale=0.3]{Re_P_TrueValue-eps-converted-to.pdf}\label{fig1111}}
\subfloat[Estimated with 9 sensors]{\includegraphics[scale=0.3]{Re_P_Pred_Mean9sensors-eps-converted-to.pdf}\label{fig1000}}
\subfloat[Estimated with 30 sensors]{\includegraphics[scale=0.3]{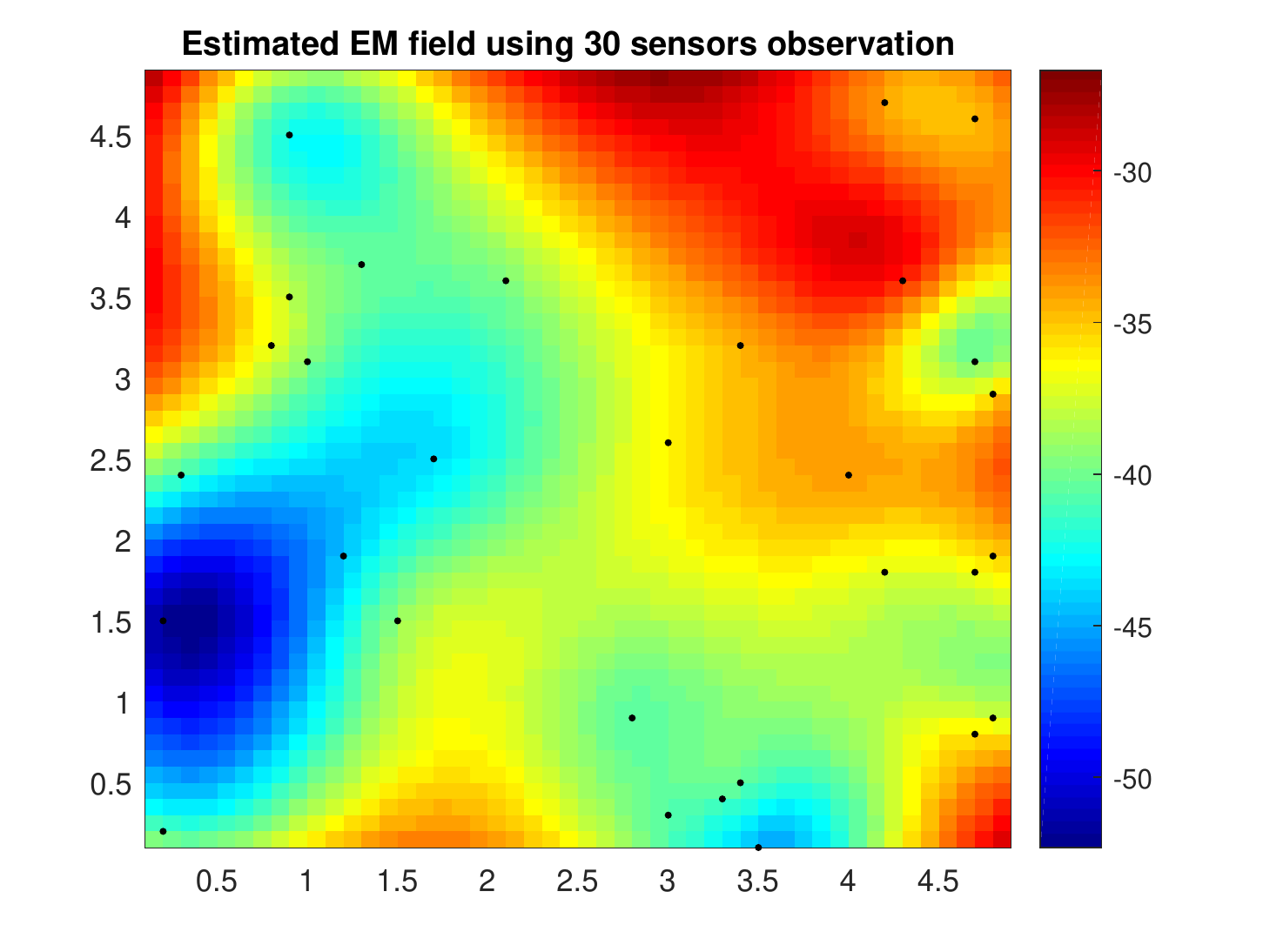}\label{fig9a}}
\subfloat[Estimated with 100 sensors]{\includegraphics[scale=0.3]{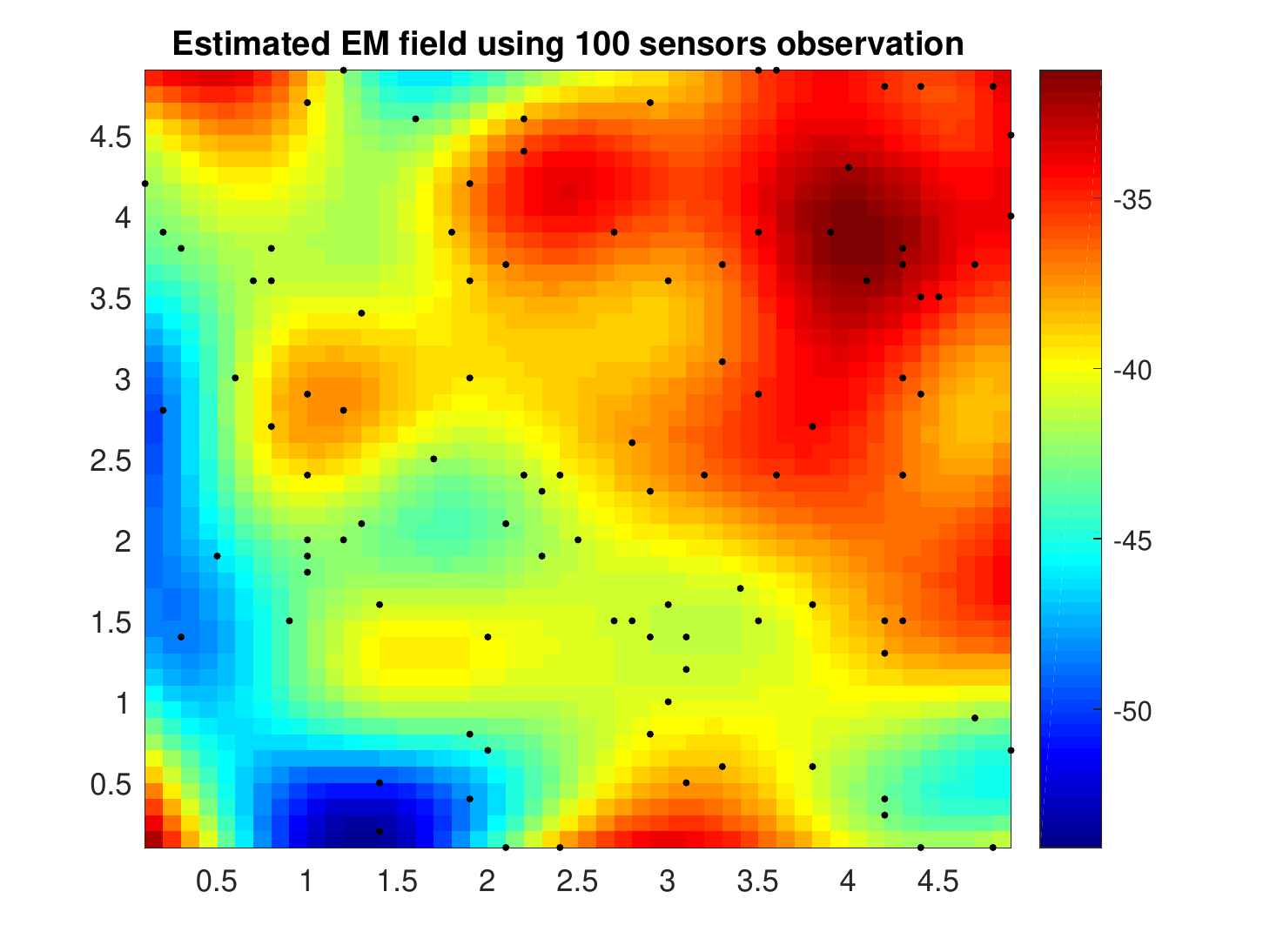}\label{fig9b}}
\caption{Pseudocolor of Spatial Reconstruction of EM field intensities based on 9, 30 and 100 sensors observation points.} \label{varysensors2}
\end{figure*}
\begin{table}
\caption{Comparison of the spatial reconstruction based on different number of sensors observation. }	\label{table6} 
\centering
\resizebox{\columnwidth}{!} {
\begin{tabular}{|| >{\centering\arraybackslash}m{0.6in} ||>{\centering\arraybackslash}m{0.6in}| >{\centering\arraybackslash}m{0.6in} |  >{\centering\arraybackslash}m{0.6in} ||N}
		\hline
		& \multicolumn{3}{|c|}{\textbf{Number of sensors observation}} &\\ \hline
			\textbf{Dataset}     & \textbf{9} & \textbf{30} &\textbf{100}  &\\  [10pt]\hline
			\textbf{NMSE}        &  0.3763 &  0.3073   & 0.1177   &\\  [10pt]\hline
			\textbf{Correlation} &  0.8141  &   0.8214   & 0.8723   &\\ [10pt]\hline   
		\end{tabular}}
	\end{table}
Finally, we perform spatial reconstruction of electromagnetic filed based on 30, 100 sensors observation and compared the performance obtained with our case where only 9 sensors observation were used. The results are shown in figures \ref{varysensors1} and \ref{varysensors2} respectively and we present the performance evaluation of these predictions in Table \ref{table6}.
\section{Conclusion and Future Work}
\label{cha:5}
In this thesis work we developed an algorithm for spatial reconstruction of electromagnetic field which is based on only 9 sensor observation points  using Gaussian process. We started our work by first collecting different set of training data points to train our algorithm. The measurement for training sets has been carried out in IRCICA using a modeling tool called SIMUEM3D, which is available for research work in the company. We modeled the measurement room first and then by fixing the necessary parameters in the tool, we conducted the exact measurement of the electromagnetic field intensity at the specified locations. The positions of the observations are set by grid location, which is a grid coordinates starting from $0.1$ up-to $4.9$ and by stepping $0.1$ in both x and y directions. We collected total of $2401$ spatial data points to represent the space of the real measurement room and we used only 9 sensors measurement to perform the estimation. 

We structured our algorithm in two portions, training and prediction phases.
In the training phase, we used the training points to perform model selection, this means, we selected suitable covariance function for our model and set the hyperparameters by  computing the optimal hyperparameters using the training points. We implemented the marginal likelihood approach to compare between covariance functions that could fit the model and based on the performance comparison we performed, the Mat\'{e}rn covariance function was found to be a better choice for our model. 
To include the properties of electromagnetic fields in the algorithm, we also considered a mean function which is defined by weighted sum of fixed basis function. Since the intensity of electromagnetic fields observed by the sensors is inversely proportional to the square of the separation between the positions of the source and sensors, we defined the basis function of the mean function as proportional to the inverse of the squared distance between the source and sensors. 

In prediction phase, we used the optimized hyperparameters and only 9 sensors observation points to make real time estimation not only at the location of the sensors but also at any location in the considered room. The spatial electromagnetic field reconstruction contains 2401 spatial points that represents the positions in the room where the real observation is made by the sensors network. 
We studied the performance of the algorithm that can be obtained when we use a mean function which represents the properties of the electromagnetic fields and when use zero mean. To conclude, the results shows that the proposed algorithm has a better performance with the considerer mean function rather than the case where the mean is zero, and we observe that the number of training points has an impact on the performance. We also compare the prediction performance obtained by varying the number of sensor observations and increasing the sensor observation points improves the performance. 
In this thesis we fixed the location of the sensors and observed the electromagnetic field intensities, but this thesis work can also be extended further by considering to locate the sensors in different positions to study and achieve better performance. It is possible to observe the EM fields by the sensors in different locations and compare the positions of the sensors that gives a better performance.   

\bibliographystyle{IEEEtran}
\bibliography{biblio}

\end{document}